\documentclass[preprint,12pt,authoryear]{elsarticle}
\usepackage{subfigure}
\usepackage{graphicx}
\usepackage{epstopdf,epsfig}
\usepackage{newtxtext}
\usepackage{multirow}
\usepackage{multicol}
\usepackage{newtxmath}
\usepackage{amsmath}
\usepackage{natbib}
\usepackage{hyperref}
\usepackage{mathtools}
\usepackage{relsize}
\graphicspath{ {./Fig/} }
\usepackage{comment}
\hypersetup{
    colorlinks = true,
    urlcolor   = blue,
    citecolor  = black,
}

\newcommand{\RomanNumeralCaps}[1]
\linenumbers
\usepackage{float}
\usepackage[margin=2.5cm]{geometry}

\newcommand\blfootnote[1]{%
  \begingroup
  \renewcommand\thefootnote{}\footnote{#1}%
  \addtocounter{footnote}{-1}%
  \endgroup
}

\begin{document}

\begin{frontmatter}

\title{Unsteady Cylinder Wakes from Arbitrary Bodies with Differentiable Physics-Assisted Neural Network}

\author[inst1]{Shuvayan Brahmachary}

\affiliation[inst1]{organization={Department of Mathematics and Informatics},
            addressline={Boltzmannstrasse 3}, 
            city={Garching bei Muenchen},
            postcode={85748}, 
            state={Munich},
            country={Germany}}

\author[inst1]{Nils Thuerey}

\begin{abstract}
This work delineates a hybrid predictive framework configured as a coarse-grained surrogate for reconstructing unsteady fluid flows around multiple cylinders of diverse configurations. The presence of cylinders of arbitrary nature causes abrupt changes in the local flow profile while globally exhibiting a wide spectrum of dynamical wakes fluctuating in either a periodic or chaotic manner. Consequently, the focal point of the present study is to establish predictive frameworks that accurately reconstruct the overall fluid velocity flowfield such that the local boundary layer profile, as well as the wake dynamics, are both preserved for long time horizons. The hybrid framework is realized using a base differentiable flow solver combined with a neural network, yielding a differentiable physics-assisted neural network (DPNN). The framework is trained using bodies with arbitrary shapes, and then it is tested and further assessed on out-of-distribution samples. Our results indicate that the neural network acts as a forcing function to correct the local boundary layer profile while also remarkably improving the dissipative nature of the flowfields. It is found that the DPNN framework clearly outperforms the supervised learning approach while respecting the reduced feature space dynamics. The model predictions for arbitrary bodies indicate that the Strouhal number distribution with respect to spacing ratio exhibits similar patterns with existing literature. In addition, our model predictions also enable us to discover similar wake categories for flow past arbitrary bodies. For the chaotic wakes, the present approach predicts the chaotic switch in gap flows up to the mid-time range. 
\end{abstract}


\end{frontmatter}

\blfootnote{\hspace{-0.62cm}codes to follow shortly: https://github.com/tum-pbs/DiffPhys-CylinderWakeFlow/}

\section{Introduction}

Machine learning has drawn significant attention towards the field of fluid simulations \citep{kutz2017deep,brunton2020machine}, particularly in the last two decades. Owing to its successful inception in the field of turbulence modeling \citep{ling2016reynolds,wu2018physics,duraisamy2019turbulence,srinivasan2019predictions}, it has spawned multiple avenues such as super-resolution of fluid flows \citep{xie2018tempogan,fukami2019super}, detection of turbulent interface \citep{li2020using}, active flow control \citep{pino2023comparative}, fluid-particle interaction \citep{davydzenka2022high}, reduced-order modeling (ROM) for predicting the dynamical state of a fluid system \citep{xiao2015non,pawar2019deep}, aerodynamic shape optimization \citep{chen2021numerical}, among others. To incorporate human knowledge in the form of physics models, methods were introduced that integrate them into the training procedure, \textit{e.g.,} physics-informed neural network \citep{raissi2019physics, sun2020surrogate}, graph neural network \citep{belbute2020combining,brandstetter2022message}, 
and generative adversarial network \citep{cheng2020data} with physical loss functions \citep{lee2019data}. While custom loss functions satisfy the boundary conditions and minimize the residual of the underlying governing partial differential equations, they only allows for partial coupling of the flow solver with the neural network. Another alternative that allows full integration of the flow solver within the training loop is via differentiable models, \textit{e.g.,} differentiable physics for turbulence modeling \citep{list2022learned}. This approach has demonstrated its effectiveness as a cost-efficient strategy for turbulence modeling compared to the cost-intensive direct numerical simulations.  Such approaches have also shown the potential to serve as a coarse-grained surrogate for unsteady fluid flows. For instance, the works of \cite{um2020solver} and \cite{kochkov2021machine} demonstrated that differentiable flow solvers coupled with a neural network can yield satisfactory reconstruction of flowfields for representative test cases. However, the full merits of the hybrid strategy as a generic and accurate approach are far from established.

A primary task when evaluating the merits of a predictive framework is to identify a fluid test problem that is of significant practical interest and complexity. One such problem that has been of significant interest to the fluid dynamics community is that of flow past body \citep{hasegawa2020machine} or multiple bodies \cite{wan2018machine}. Typical factors and challenges that plague the analysis of such fluid flows may include computational time or the turn-around time, low to large temporal scales in the flow, unsteadiness in the flow, the presence of multiple bodies, a large spectrum of wake dynamics, etc. While coarse-grained surrogates reduce the computational burden \citep{stachenfeld2021learned}, the effectiveness of such a strategy to reliably reconstruct the flowfields for multiple wake regimes, besides ensuring accuracy in local boundary profiles and global wake dynamics, is yet to be examined. For instance, authors in work \cite{morimoto2022generalization} examine flow past two side-by-side cylinders to qualitatively evaluate the generalizability aspect of their models without considering the local fluid behavior around the body boundary. In addition, the work by \cite{hasegawa2020machine} examined flow past a single bluff-body of various shapes using a convolutional neural network-based auto-encoder (CNN-AE) model with an emphasis on global fluid dynamics. In a similar vein, the investigation by \cite{lee2019data} points out that their model predictions for flow past a single cylinder exhibited the largest errors in either the body boundary or the wake region. Clearly, this presents a significant limitation that should be addressed by advancing performance measuring metrics based on both the local as well as global fluid properties.



Flow past objects are of significant interest to the engineering community from the hydrodynamics as well as structural design point of view. The potential implications of the underlying instabilities \citep{mckinley1993wake}, wake dynamics \citep{williamson1996vortex}, wake induced vibration \citep{bearman2011circular} and its control \citep{choi2008control} have led to establishing a significant body of research. One of the focal points concerning this research is understanding the spectrum of wake flow exhibited by multiple cylinders placed relative to each other. Some of the investigations that present a comprehensive study for two cylinders are \citep{zravkovich1987effects, papaioannou2006three, sumner2010two} etc. and \citep{lam1988phenomena, guillaume1999investigation, bao2010numerical, zheng2016numerical} for three or more cylinders. These investigations point towards the direct dependency of the resulting wake regimes with the spacing ratio ($\textit{i.e.,}$, length $L$ to diameter $D$ ratio) $L/D$. More recently, the investigation by \citep{chen2020numerical} showed the presence of up to nine distinct flow regimes for cylinders arranged in an equilateral-triangle position at various spacing ratios, 1 $\leq$ $L/D$ $\leq$ 6 for multiple Reynolds numbers 50 $\leq$ $Re$ $\leq$ 175. While these investigations provide fundamental insights about wake flow regimes and flow states, their analysis is limited to the use of equi-diameter circular cylinders or rectangular cylinders with varying aspect ratios. Consequently, this begs the question if the observations related to the universal non-dimensional parameters such as Strouhal number and critical spacing ratio (as reported by \citep{chen2020numerical}) still hold for arbitrarily shaped cylinders. 






Analyzing large datasets of wake flow exhibited by arbitrarily shaped complex configurations offers potential for novel physical insights; however, employing traditional body-conforming continuum computational fluid dynamics (CFD) flow solvers can be time-consuming and warrants special treatment for catering to changing geometries via the underlying computational grid. While non-conformal immersed boundary (IB) approaches \citep{mittal2005immersed, brahmachary2019finite} allow for the use of a fixed Cartesian mesh, exploring a large design space of possible spacing ratios could significantly ramp up the turn-around time. Consequently, neural network-based coarse-grained surrogate models for unsteady fluid flows provide a cost-effective alternative approach \citep{stachenfeld2021learned}. 

In this work, we build on the ideas by \cite{um2020solver} and present a hybrid Differentiable Physics-Assisted Neural Network (or DPNN) framework that acts as a coarse-grained surrogate for accurate unsteady fluid simulations for low Reynolds number flows past multiple bodies. This framework requires a base solver integrated into the neural network (NN) along with ground truth data to train the resulting hybrid framework. We employ a Cartesian grid-based immersed boundary method (\textit{i.e.,} FoamExtend) as the \textit{Reference} solver. In addition, we employ an in-house developed differentiable flow solver (\textit{i.e.,} PhiFlow) as the \textit{Source} solver. For faster training and inference, the \textit{Source} intentionally resorts to a first-order spatial accuracy along with masked stair-step representation of the underlying body boundary. This inherently reduces the computational burden of the base differentiable flow solver. In contrast, however, the \textit{Reference} solver utilizes an accurate local algebraic reconstruction for fluid flow around the body boundary with second order spatial accuracy. Given this stark contrast between the \textit{Reference} and \textit{Source} solvers, besides the highly multi-modal nature of the present problem, this naturally poses a very challenging task for NNs \textit{i.e.,} to act as a forcing function to learn the non-trivial corrections. 
We comprehensively evaluate the hybrid approach for multiple wake categories while drawing necessary comparisons with various benchmarks. Moreover, our approach of utilizing arbitrarily shaped bodies in a staggered equilateral triangle position allows for a large data diversity and sheds light on the physics of the wake dynamics resulting from the arbitrary cylindrical configurations. 

Our focus is thus twofold. First, we train and test the DPNN framework on a dataset comprising unsteady flow past arbitrary bodies placed in a staggered equilateral triangle position, each at a unique spacing ratio. These tests were performed once the \textit{Reference} solver was throughly validated and verified to generate accurate solutions. Second, after establishing the effectiveness of our approach using multiple performance measuring metrics, we evaluate the out-of-distribution generalization capabilities of our framework. These evaluations is based on configurations that align closely with the current state of existing literature. Throughout the study, we strive to understand the physical implications of the solutions derived from the predictive framework, examining both local boundary layer profiles and global wake dynamics. We also highlight the framework's performance in both physical and reduced feature spaces to demonstrate its long-term temporal stability.

\section{Numerical setup and validation}

In this section, we illustrate the computational setup underpinning the entire hybrid framework. This is followed by validation test cases undertaken to affirm the correctness of the setup as well as verify the ability to render accurate solutions. 

\begin{figure}
\centering
\includegraphics[scale=0.45, angle=0]{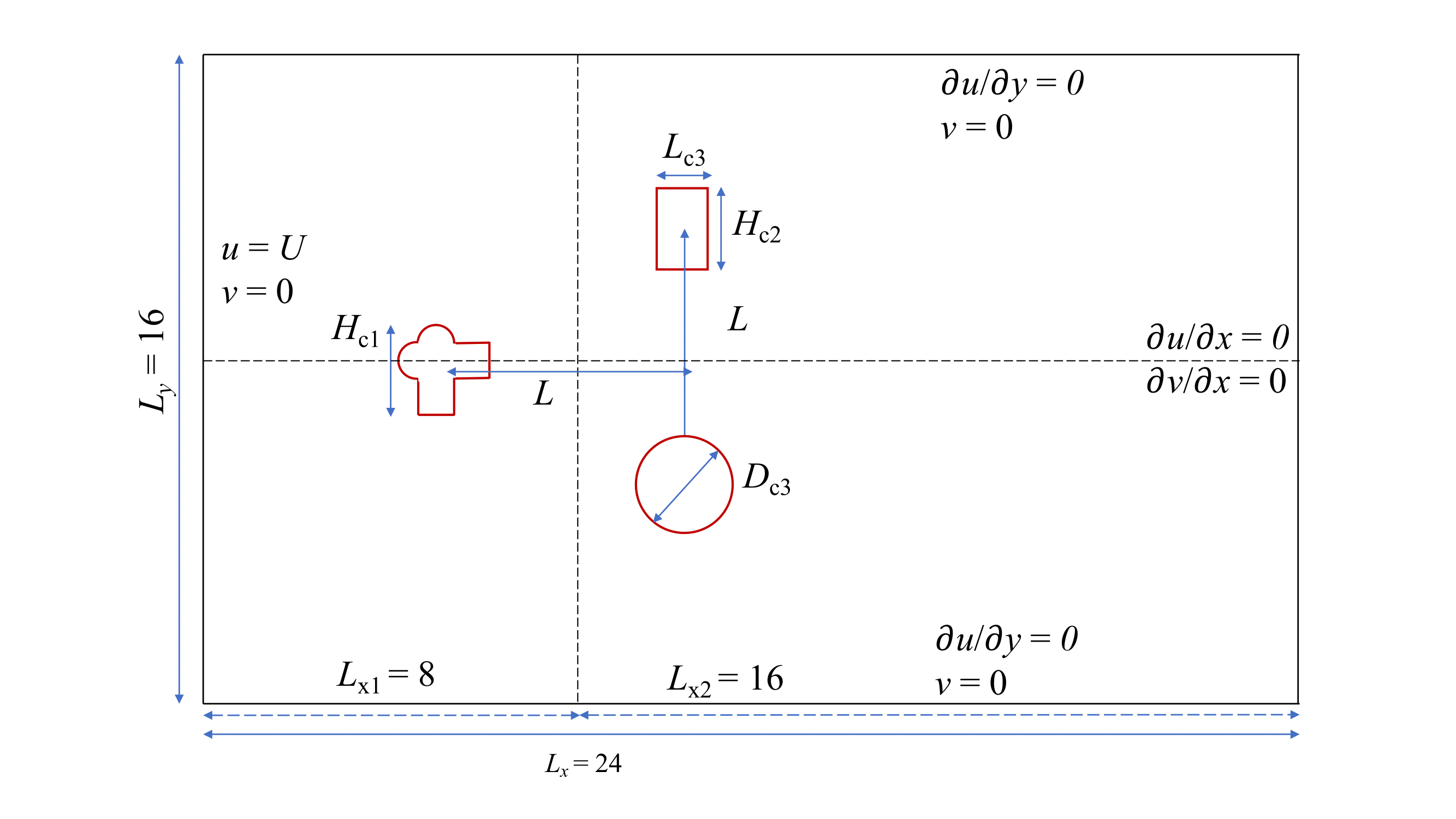}
\caption{Computational domain including the three bodies placed in an equilateral triangle position (not to scale)}
\label{cyl_layout}
\end{figure}

\subsection{Reference solver}

The present study employs the open source \texttt{FoamExtend}  V 4.0 immersed boundary (IB) approach as the \textit{Reference} solver. As mentioned earlier, the IB-based non-conformal approach serves a dual purpose. Firstly, it greatly simplifies the use of a fixed Cartesian mesh for all underlying configurations, irrespective of the geometric complexity. This methodology circumvents the necessity for labor-intensive re-meshing for each geometric configuration and the corresponding coordinate transformation from physical to computational space. Secondly, the use of a fixed Cartesian mesh ensures seamless integration to the deep-learning pipeline wherein one can leverage the benefits of the Eulerian (\textit{i.e.}, fixed cell in space) viewpoint. \texttt{FoamExtend} utilizes a discrete forcing approached treatment strategy for enforcing the boundary condition for the immersed bodies and has been built on top of the well-established OpenFoam \citep{jasak2015immersed} framework. 

In the present work, we consider the 2D Navier Stokes equations as mentioned below:

\begin{equation}
 \
    \begin{cases}
      \frac{\partial \mathbf{u}}{\partial t} + (\mathbf{u} .\nabla)\mathbf{u} = -\frac{1}{\rho}\nabla p + \nu \nabla^2\mathbf{u} \\
      \nabla.\mathbf{u} = 0\\
    \end{cases}       
\end{equation}

where, $\mathbf{u} = (u,v)$ is the velocity field, $\rho$ is the density, $p$ is the pressure, and $\nu$ is the dynamic viscosity. The first equation represents the momentum conservation equations, whereas the second equation represents the continuity equation, which also serves as a constraint on the velocity fields, otherwise known as the ``divergence-free'' constraint. 

The flow scenario describes the incompressible flow past cylinders at moderate Reynolds number Re $\approx 100$. For the present case, the Reynolds number is evaluated based on the cylinder height of the primary upstream cylinder as $Re = \frac{U H}{\nu}$, where $U$ and $H$ represent the freestream velocity and height of the upstream cylinder respectively. For the present setup, $\nu$=0.01 Ns/$m^2$ and $U$=1m/s are used. A second-order accurate Crank Nicolson scheme is used to march the solutions in time, while a second-order limited linear (central-differencing) scheme is used for convective flux discretization. Pressure-Implicit with Splitting of Operators ($\texttt{PISO}$) \citep{issa1986solution} is employed to solve for the discretized momentum and continuity equations in 1 predictor and 2 corrector steps. In particular, preconditioned conjugate gradient (PCG) based solver is used for $p$ linear equations along with a diagonal-based incomplete Cholesky (DIC) precondition for symmetric matrices. In addition, for $U$ linear equations, BiCGSTAB is used along with diagonal-based incomplete LU (DILU) preconditioned for asymmetric matrices. 

\subsection{Validation: Strouhal number and force characteristics}
\label{val}

\begin{table}
  \begin{center}
  \def~{\hphantom{0}}
\begin{tabular}{ccccc} \\ 
   $Re$=100 (single cylinder)           &  & $\bar{C}_d$ & $C^{'}_1$ & $St$ \\  \\
\cite{constant2017immersed} ($\Delta x=\Delta y = 0.02D$)   &  & 1.38        & -    & 0.165 \\
\cite{constant2017immersed} ($\Delta x=\Delta y = 0.010D$)   &  & 1.37        & -    & 0.165 \\
Present ($\Delta x=\Delta y = 0.0312D$)            &  & 1.412       & 0.276   & 0.156 \\ 
Present ($\Delta x=\Delta y = 0.02D$)            &  & 1.403       & 0.285   & 0.156 \\ 
\end{tabular}
\caption{Coefficient of lift $C_l$, drag $C_d$, and Strouhal number $St$ for flow past a single cylinder at $Re$=100}
\label{1_cyl}
\end{center}
\end{table}

We execute a comprehensive grid independence study to corroborate the \textit{Reference} solver used to acquire the ground truth data. We initiate our numerical experiments with flow past a single cylinder at Reynolds number $Re$=100. A computational domain of size [0,24D] $\times$ [0,16D] is chosen using a uniformly spaced Cartesian grid. In particular, we choose medium and fine spatial resolutions to undertake a grid-independence study, \textit{i.e.,} $\Delta x/D$ = $\Delta  y/D \in$ \{1/32, 1/50\}, where $D$ represents the diameter of the cylinder. This leads to a blockage ratio, $B$=$D/H$ = 0.0625. Further, we use a constant time-step $\Delta t$ = 0.1s for this validation test case and perform the simulations until a total time $T$ = 200s. 

\begin{table}
  \begin{center}
\begin{tabular}{cccccccc} \\
$Re$=100 (three cylinders, $L/D=2.5$)
&  & $\bar{C}_{d,1}$ & $\bar{C}_{l,1}$ & $C^{'}_{l,1}$ & $\bar{C}_{d,2}$ & $\bar{C}_{l,2}$ & $C^{'}_{l,2}$ \\
\\ 
\cite{zheng2016numerical}  &  & 1.23    & 0.0 &  -0.002 &  1.53    & -0.087 &  0.335    \\
Present ($\Delta x=\Delta y = 0.0312D$)             &  & 1.249    & 0.0 &  0.0  &  1.561    & 0.166 &  0.433    \\ 
\end{tabular}
\caption{Coefficient of lift $C_l$, drag $C_d$, and Strouhal number $St$ for flow past three cylinders at various time-step sizes}
\label{3_cyl}
\end{center}
\end{table}

\begin{table}
  \begin{center}
\begin{tabular}{ccccccccc} \\
$Re$=100 (three cylinders, $L/D=2.5$)
&  & $\bar{C}_{d,1}$ & $\bar{C}_{l,1}$ & $C^{'}_{l,1}$ & $\bar{C}_{d,2}$ & $\bar{C}_{l,2}$ & $C^{'}_{l,2}$ & $St$ \\
\\ 
Present ($\Delta t$ = 0.25)     &  & 1.254    & 0.0 &  0.0  &  1.584    & 0.172 &  0.49 & 0.148  \\ 
Present ($\Delta t$ = 0.1)      &  & 1.249    & 0.0 &  0.0  &  1.561    & 0.166 &  0.433  & 0.156  \\ 
Present ($\Delta t$ = 0.05)     &  & 1.244    & 0.0 &  0.0  &  1.545    & 0.166 &  0.359  & 0.156  \\ 
\end{tabular}
\caption{Coefficient of lift $C_l$, drag $C_d$, and Strouhal number $St$ for flow past three cylinders}
\label{3_cyl_dt}
\end{center}
\end{table}

The time average drag coefficient $\bar{C_d}$ as well as the root mean square (r.m.s) lift coefficient $C^{'}_1$ obtained from the \textit{Reference} flow solver is compared with existing results. Specifically, Table \ref{1_cyl} shows the $\bar{C_d}$, $C^{'}_1$, and Strouhal number $St$ values for flow past a single cylinder at various grid resolutions. It is found that both $\bar{C_d}$ and $C^{'}_l$ obtained from the medium as well as the fine spatial resolutions compare really well with the existing solutions of \cite{constant2017immersed}, who also employ a non-conformal immersed boundary-based numerical solver for the computations. It must be noted that a change in the underlying grid from medium to fine spatial resolution results in an increase in the total number of control volumes $n_c$ from 3,93,216 to 9,60,000. Consequently, to allow for a reasonable computational time while retaining accuracy, we choose the medium grid for the remainder of the computations in this section as well as for obtaining the ground truth solution. 

Table \ref{3_cyl} presents the mean as well as the r.m.s values for the coefficient of drag and lift for the upstream and downstream cylinder (only one) for flow past three cylinders placed in an equilateral triangle position at spacing ratio $L/D$=2.5. It is found that the present solutions auger really well with the ones obtained by \cite{zheng2016numerical}, who employ a body-conformal finite-volume approach using a time step size $\Delta t$=0.05s. The influence of the time-step size is also analyzed in Table \ref{3_cyl_dt}. It is found that the choice of time step size results in a marginal difference in force quantities as well as the Strouhal number. Consequently, a moderate value of $\Delta t$=0.1 s is adopted for the present study. It can now be remarked that the grid resolution and time-step size chosen for the present study allow for the accurate computation of flow past multiple bodies. In addition, we also qualitatively verify the wake regimes produced using the flow solver, as highlighted in Appendix \ref{wake_regime}. 

\subsection{Differentiable flow solver} 

The present study uses the open source simulation toolkit \texttt{Phiflow} ($\phi_{\mathrm{flow}}$) as the base differentiable flow solver (or \textit{Source}). It leverages automatic differentiation to enable an end-to-end recurrent training paradigm. While it is crucial that the base solver (\textit{i.e.,} Source) be differentiable, the \textit{Reference} solver (\textit{i.e.,} FoamExtend in this case) need not be differentiable.  We employ the projection method to decouple the momentum and continuity equation in a two predictor-corrector step. Firstly, we use the operator splitting of the diffusion, advection, and pressure terms in the predictor phase. Specifically, the MacCormack advection step for advecting the initial velocity fields. Secondly, in the corrector phase, the spatial gradients of pressure are used to compute the divergence-free velocity field \cite{chorin1968numerical}. 

\begin{figure}
    \centering
    \subfigure[]{\includegraphics[width=0.575\textwidth]{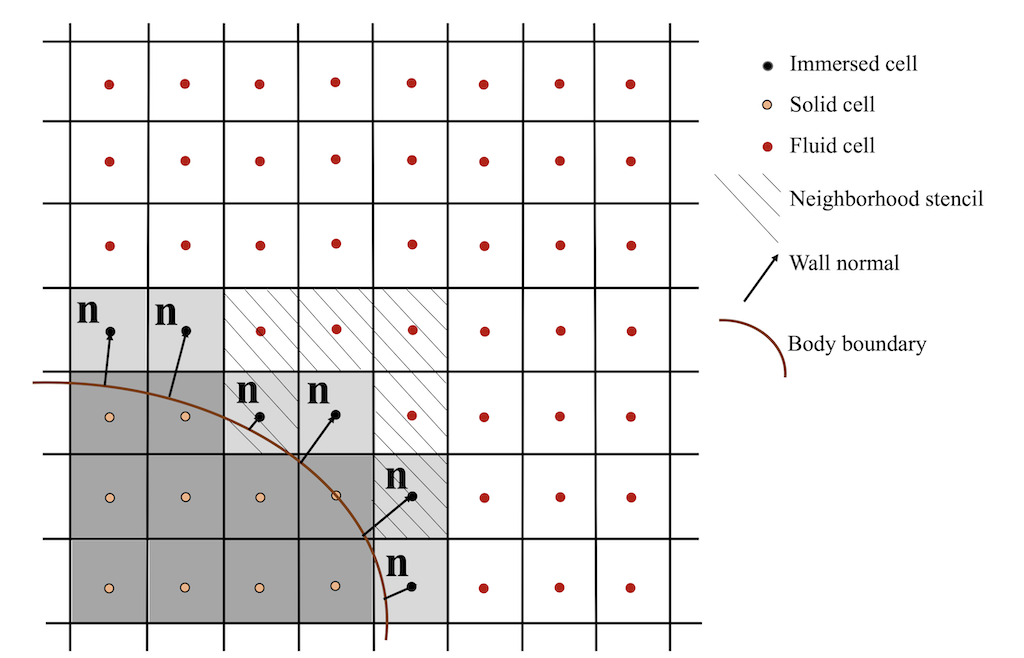}} 
    \subfigure[]{\includegraphics[width=0.4\textwidth]{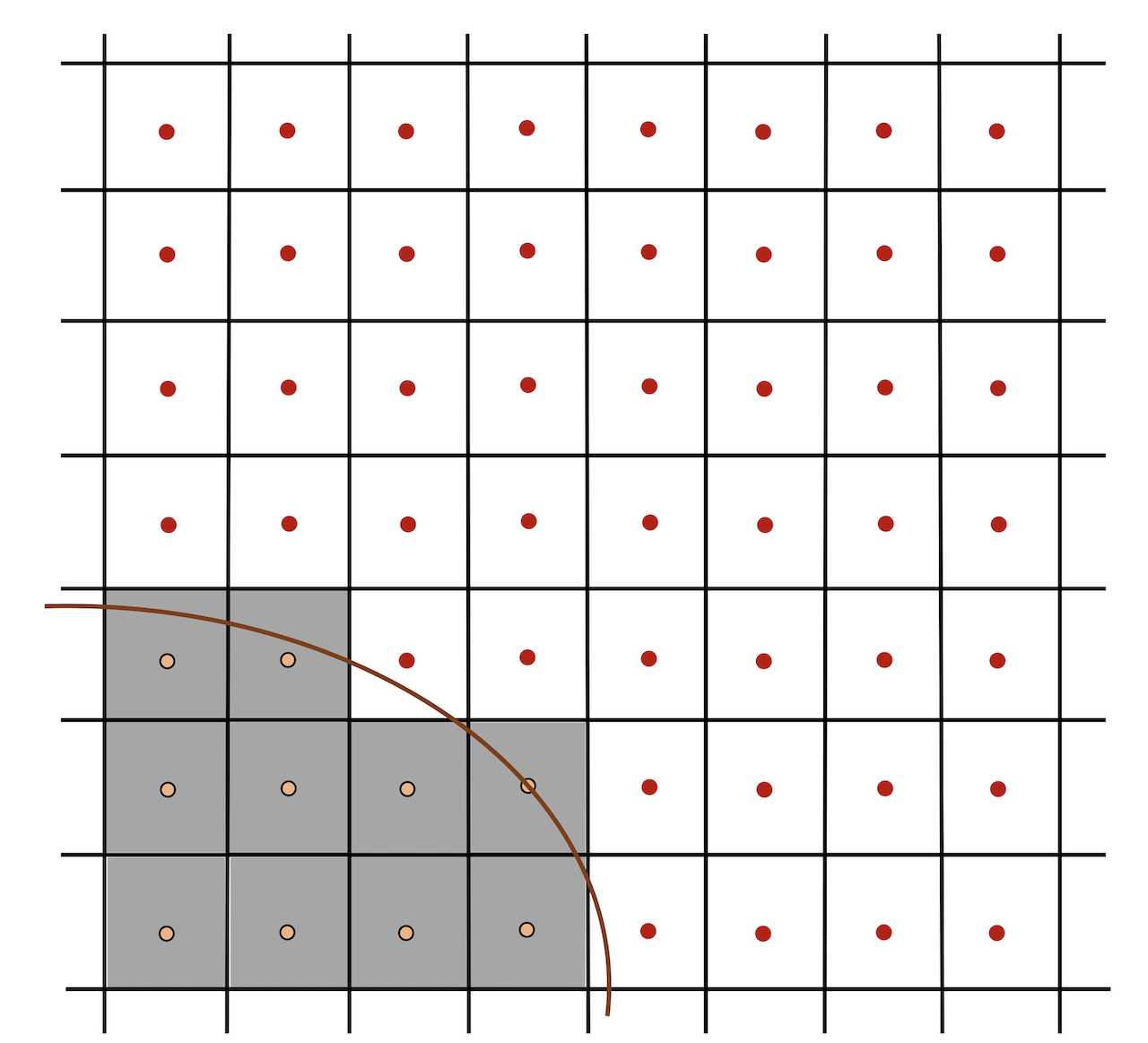}} 
    \caption{Boundary reconstruction approach used by (a) \texttt{FoamExtend} (\textit{Reference} solver) (b) \texttt{Phiflow} (\textit{Source} solver)}
    \label{reconstruction}
\end{figure}

Finally, the choice of Cartesian grids in both \textit{Reference} and \textit{Source} solver allows for a one-to-one mapping between each control cell. This is especially important given that the use of masked stair-step representation of the body boundary by the \textit{Source} results in a very approximate reconstruction of near-body fluid properties. This allows for a training setup wherein the network learns to faithfully reconstruct the local boundary flow profile based on the accurate \textit{Reference} solver. 

\subsection{Cylinder layout}
\label{cylinder_setup}

We carefully choose a setup for cylinder cluster arrangement that allows us to mimic the traditional approach adopted in open literature. However, as we seek to generate and exploit a multitude of complexities in the training data to harness the full capabilities of the present hybrid learning strategy, we introduce some randomness to the setup. Some specifications related to the cylinder arrangements are as follows:


\begin{enumerate}
    \item \textbf{Cylinder arrangement}: We place three cylinders in an equilateral triangle position, \textit{i.e.,} one upstream cylinder followed by two downstream cylinders, placed apart from each other via the spacing ratio $L/D$. This setup mimics the ``regular triangular" cluster setup used in \cite{zravkovich1987effects} and has also been investigated recently by \cite{chen2020numerical}. Additionally, this arrangement allows us to investigate \textit{proximity} interference (P), \textit{wake} interference (W), a combination of (P+W) interference as well as no-interference. 

    \item \textbf{Upstream composite cylinder}: The upstream cylinder (cylinder $C_\mathrm{1}$) is framed as a composite body consisting of a base rectangle (at its center) and secondary rectangle or semi-circle on its sides (see Fig. \ref{cyl_layout}). The selection between a secondary rectangle or a secondary semi-circle as one of the sides of the base rectangle is random. We define a function $f_{\mathrm{1}}:R_{\mathrm{u,i}} \rightarrow S$, such that it maps a random integer $R_{\mathrm{u,i}}$ $\in$ \{0,1\} to geometric shapes, S $\in$ \{semi-circle, rectangle\}, where $i \in$ \{1,2,3,4\}. The outcome can be represented as follows:

    \[
    R_{\mathrm{u,i}}  = \begin{dcases*}
    0 & $f_{\mathrm{i}}$ = $\mathrm{semi circle}$ \\
    1 & $f_{\mathrm{i}}$ = $\mathrm{rectangle}$ \\
    \end{dcases*}
    \]

    The length of the base rectangle, as well as the secondary rectangles, are randomly chosen between fixed upper and lower limits (see Table \ref{up_lower}). The diameter of the semi-circle is equal to the edge length of the base rectangle to which it is attached.

    \item \textbf{Downstream cylinders}: The choice of the downstream cylinders is either a rectangular cylinder of a certain length and height or a circular cylinder of a certain diameter. This selection is partially random. We define a function $f_{\mathrm{2,3}}:R_{\mathrm{d}} \rightarrow S$, such that it maps a random integer $R_{\mathrm{d}}$ $\in$ \{0,1\} to geometric shapes, S $\in$ \{circle, rectangle\}. Once the upper downstream cylinder (cylinder $C_\mathrm{2}$) is chosen (say a rectangular cylinder), the lower downstream cylinder (cylinder $C_\mathrm{3}$) gets fixed (a circular cylinder). This outcome can be represented as shown below. 

    \[
    R_{\mathrm{d}} = \begin{dcases*}
    0 & $f_{\mathrm{2}}$ = $\mathrm{circle}$; \ $f_{\mathrm{3}}$ = $\mathrm{rectangle}$\\
    1 & $f_{\mathrm{2}}$ = $\mathrm{rectangle}$; \ $f_{\mathrm{3}}$ = $\mathrm{circle}$\\
    \end{dcases*}
    \]
    
    This forces the training dataset to always contain a rectangle and a circular cylinder as the two downstream cylinders. In addition, once the height of the downstream rectangular cylinder, $H_{\mathrm{C2}}$ (or the diameter of the downstream circular cylinder, $D_{\mathrm{C3}}$) is chosen, the dimension (\textit{i.e.,} diameter or cylinder height) of the other downstream cylinder is fixed to satisfy the following criterion. 

    \begin{equation}
        H_{\mathrm{C2}}+D_{\mathrm{C3}} = 2 H_{\mathrm{C1}}
    \end{equation}

    This restricts the cylinders from overlapping. The dimensions of the rectangle and the circular cylinder are randomly chosen between a fixed upper and lower limit (see Table \ref{up_lower}). 

    \item \textbf{Spacing ratio}: The distance between centre-to-centre distance between the downstream cylinders, as well as the downstream and upstream cylinders, is controlled by the spacing ratio $L/D$. The $L/D$ values are chosen such that it covers a wide spectrum of possible wake flow regimes (\textit{i.e.,} 1.2 $\leq$ $L/D \leq$ 5.5). Care is taken to avoid overlapping bodies for low $L/D$ values. Finally, to ensure an unbiased selection of the spacing ratio, the Latin Hypercube Sampling (LHS) is chosen as a design of the experiment method.
    
\end{enumerate}

Points 2 and 3 above intentionally introduce arbitrariness to the training data to serve a dual purpose. Firstly, we analyze the role of arbitrariness in the wake flow regime to offer new insight into the present body of knowledge around flow past equi-diameter cylinders. Secondly, this introduces data diversity while also allowing us to evaluate the generalizability of the hybrid learning framework to more representative problems. Further, point 4 allows us to uncover various wake categories that have been introduced recently by \cite{chen2020numerical} for similar spacing ratios.

\subsection{Wake categories}
\label{cylinder_wakes}

This section introduces the various wake categories obtained from the \textit{Reference} solver at multiple spacing ratios $L/D$ for equi-diameter cylinders placed in an equilateral triangle position. These representative spacing ratios for equi-diameter cylinders are for illustration purposes and are not part of the training data. We purposely choose equi-diameter circular cylinders so that fair comparisons can be drawn with \cite{chen2020numerical} who also evaluate it on identical conditions.

\textbf{Single bluff-body wake:} At extremely small spacing ratio \textit{i.e.,} $L/D$ = 1.2, the three cylinders are close to each other such that they exhibit vortices as if it were a ``single bluff-body". This is further demonstrated pictorially in Fig. \ref{LD_12} in the Appendix \ref{wake_regime}, along with other representative cases. This kind of wake is periodic in nature.

\textbf{Deflected gap wake:} At slightly larger spacing ratio \textit{i.e.,} $L/D$ = 1.5, the fluid in between the two downstream cylinders (or gap flow) switches direction to either the upper cylinder (\textit{i.e.,} cylinder 2) or the lower cylinder (\textit{i.e.,} cylinder 3), hence the name ``deflected gap". This switch is constant with time. 
The kind of wake is periodic, with some modulation over time.

\textbf{Flip-flopping wake:} For marginally larger spacing ratio \textit{i.e.,} $L/D$ = 2.25, the gap flow in between the two downstream cylinders switches direction erratically between the upper cylinder (\textit{i.e.,} cylinder 2) or the lower cylinder (\textit{i.e.,} cylinder 3), hence the name ``flip-flopping". This chaotic switch is a challenging case for any predictive framework, 
and the kind of wake is irregular in nature.

\textbf{Anti-phase wake:} For moderately larger spacing ratio \textit{i.e.,} $L/D$ = 3.5, the vortices shed by the two downstream cylinders rotate in opposing directions, hence the name ``anti-phase". This wake category is periodic in nature and exhibits a symmetry about the centerline streamwise direction. For arbitrarily shaped bodies, this symmetry is lost and is consequently referred to in this work as ``quasi anti-phase" wake. 

\textbf{Fully developed in-phase wake:} For larger spacing ratio \textit{i.e.,} $L/D$ = 5, the vortices shed by the two downstream cylinders are in same direction; hence the name ``in-phase". Further, the upstream cylinder also sheds vortices which merges with the vortices shed by the downstream cylinders, hence the term ``fully-developed". 
This kind of wake is periodic in nature. For arbitrarily shaped bodies, this is referred to in this work as "quasi in-phase" wake.

\begin{table}
\begin{center}
\begin{tabular}{llcccccc}
\multicolumn{2}{c}{\multirow{2}{*}{Component}}                                                                                             & \multicolumn{2}{c}{Length ($L$)}   & \multicolumn{2}{r}{Height ($H$)}    & \multicolumn{2}{c}{Radius ($R$)}     \\ \\
\multicolumn{2}{l}{}     & \begin{tabular}[c]{@{}c@{}}lower \\ limit\end{tabular} & \begin{tabular}[c]{@{}c@{}}upper \\ limit\end{tabular} & \begin{tabular}[c]{@{}c@{}}lower \\ limit\end{tabular} & \begin{tabular}[c]{@{}c@{}}upper \\ limit\end{tabular} & \begin{tabular}[c]{@{}c@{}}lower \\ limit\end{tabular} & \begin{tabular}[c]{@{}c@{}}upper \\ limit\end{tabular} \\ 
\multirow{2}{*}{\begin{tabular}[c]{@{}l@{}} \\ Upstream \\ Cylinder\end{tabular}}   & \begin{tabular}[c]{@{}l@{}}(primary)\\ Base rectangle\end{tabular} & 0.4                                                    & 0.5                                                    & 0.4                                                    & 0.5                                                    & -                                                      & -                                                      \\ \\
                                                                                & \begin{tabular}[c]{@{}l@{}}(secondary)\\  Rectangle\end{tabular}   & 0.15                                                   & 0.25                                                   & 0.15                                                   & 0.25                                                   & -                                                      & -                                                      \\ \\
\multirow{2}{*}{\begin{tabular}[c]{@{}l@{}} \\Downstream \\ Cylinder\end{tabular}} & Rectangle                                                          & 0.4                                                    & 0.5                                                    & 0.7                                                    & 1.0                                                    & -                                                      & -                                                      \\ \\
                                                                                & \begin{tabular}[c]{@{}l@{}}Circular \\ cylinder\end{tabular}       & -                                                      & -                                                      & -                                                      & -                                                      & 0.4                                                    & 0.5           
\end{tabular}
\caption{Upper and lower limits for the dimensions of the cylinders}
\label{up_lower}
\end{center}
\end{table}

\section{Hybrid predictive framework}

\begin{figure}
\centering
\includegraphics[scale=1.0, angle=0]{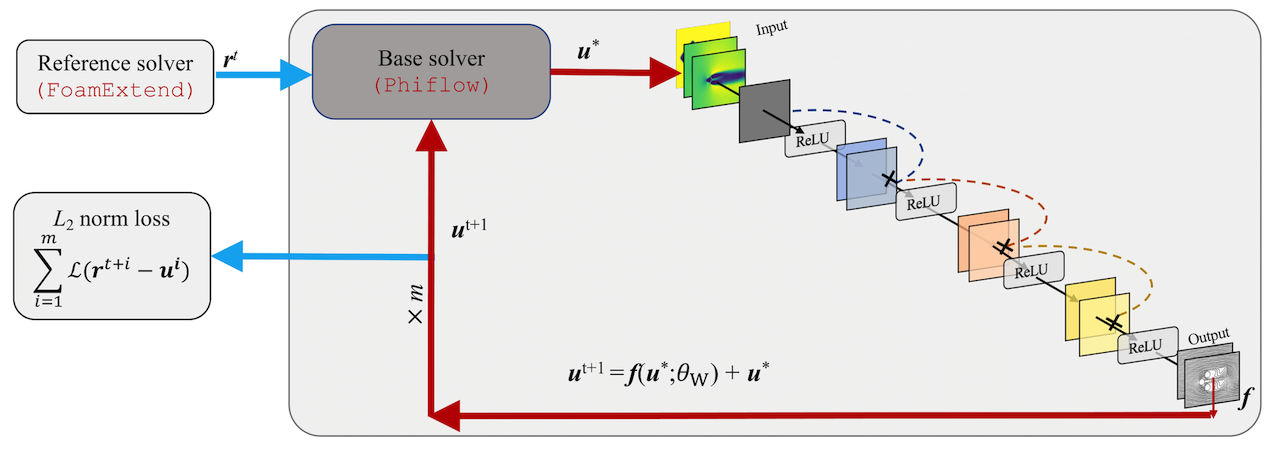}
\caption{DPNN-based hybrid predictive framework}
\label{framework}
\end{figure}

This section introduces the methodology behind the DPNN-based hybrid framework used in the present work. As alluded earlier, such a strategy embeds the base differentiable flow solver \texttt{Phiflow} with a neural network (NN) architecture, mimicking the \textit{solver-in-the-loop} strategy previously employed by \cite{um2020solver}. Such a strategy reaps the benefit of long-term training feedback while enabling an end-to-end training paradigm. We build on the above by enabling a robust and efficient coarse-grained surrogate model for unsteady flows for investigating one of the core fluid flow problems, \textit{i.e.,} wake flows caused by arbitrarily shaped objects. This serves a dual purpose, \textit{i.e.,} firstly, the arbitrarily shaped bodies allow for a rich and diverse set of datasets. Secondly, while it is well known that the shape of the body is responsible for the resulting wake characteristics, such an issue has not prompted an investigation into a large spectrum of spacing ratios to the best of our knowledge.

\subsection{Neural network architecture}

In this section, we highlight the neural network (NN) architectures that were explored in the present study. 
We explore a range of popular architectures while keeping the total number of parameters in each network architecture within reasonable limits to facilitate a fair comparison. 

The network takes as input of channel size = 3, \textit{i.e.,} velocity fields, $\mathbf{u} = u, v$ along with the shape mask or marker field $m_f  \in R^{n_x \times n_y}$, where $n_x$ and $n_y$ represents the number of grid points along the longitudinal and lateral directions, respectively. The shape mask $m_f$ uniquely identifies the shape of the enclosed body and is fixed for a given test configuration. While the ground truth data is obtained on a high-resolution grid of size $n_x, n_y$ = (768 $\times$ 512), the network only receives a downsampled (3x) velocity and shape mask as inputs (\textit{i.e.,} 96 $\times$ 64), for faster training. For a training sample, there are a total of $t_f$ = 3000 frames or snapshots (of which the first 1500 are discarded to only consider statically stable snapshots), each representing velocity fields generated sequentially using a constant time-step of $\Delta t$ = 0.1s (or, total time $t$ = 300s). The entire dataset consists of 100 different experiments each at a unique spacing ratio, resulting in 100 unique shape masks $m_f$. As mentioned in the earlier Section \ref{cylinder_setup} and Fig. \ref{cyl_layout}, to allow for a fair comparison, we employ the same cylinder cluster arrangement as also employed by \cite{chen2020numerical}, with the equi-diameter cylinders employed by \cite{chen2020numerical} being replaced by arbitrarily shaped cylinders in the present work. Among the entire dataset, 50 are used for training, and the remaining 50 are used for testing. The training is executed using a batch size of 50 for 50 epochs using Adam \citep{kingma2014adam} as the optimizer. A variable learning rate $\alpha$ is employed, which is sequentially reduced with epochs. The output from the network is constrained to be of channel size = 2, \textit{i.e.,} the velocity field at the next time step. In this work, we predominantly employ residual neural network (ResNet)-based architecture besides also using other popular designs. Appendix \ref{NN_arch} provides the details related to each architecture. 

\subsection{Present hybrid framework}

The DPNN-based hybrid predictive framework is a \textit{recursive} learning strategy wherein the base flow solver \texttt{PhiFlow} ($\phi_{\mathrm{flow}}$) is coupled with the neural network as shown in Fig. \ref{framework}. 
In the present study, the \textit{Reference} solver (\textit{i.e.,} \texttt{FoamExtend}) provides the pre-computed ground truth data. 
The hybrid framework receives an initial guess $\mathbf{r}^{t}$ from the \textit{Reference} solver and feeds it to the base solver (\textit{i.e.,} \texttt{PhiFlow}). The base solver then updates the velocity field to the next time-step $\mathbf{u}^{*}$. At this stage, no learning is involved. The output from \texttt{PhiFlow} is then fed to the NN architecture as input, \textit{i.e.,} velocity fields $u, v$, and shape marker field $m_f$. The output from the NN, $\mathbf{f} (\mathbf{u^*};\theta_W)$ acts as the forcing function which is used to correct the output from the base solver \texttt{PhiFlow} ($\theta_w$ represents the network weights). This step is made recursive for $m$ solver unrollment steps signifying as many forward steps by base solver in time by feeding the cumulative output back to the base solver, \textit{i.e.,} $\mathbf{u}^{t+1}$ = $\mathbf{u}^{*}$ + $\mathbf{f}(\mathbf{u^*};\theta_W)$. 

We remark that both the \textit{Reference} and the base solvers are subject to identical values for the dynamic viscosity $\nu$, time step size $\Delta t$, and freestream conditions, besides making sure to incorporate the same object under consideration. The critical difference here lies in their respective solution strategies and the techniques to handle body boundary conditions, \textit{i.e.,} while the former is spatially second-order accurate and algebraically reconstructs the fluid properties along the body boundary to locally satisfy the no-slip boundary conditions, the latter is spatially first-order accurate and resorts to a masked stair-step representation for computational efficiency. This fact serves as the key motivation to learn the global wake dynamics as well as the local boundary representation via the loss formulation, as will be discussed in the next Section. 

\subsection{Loss formulation}

The loss formulation provides the central goal of the learning process. In this study, this means recovering the ground truth velocity flowfields as accurately as possible. The  trainable parameters of the model are iteratively updated by minimizing the loss function. This allows the optimizer to navigate the non-convex energy landscape with the goal to arrive at a global/local minima. Typical choices of a loss function $\mathcal{L}$ use some error norm to compare network prediction $\mathbf{u}$ with the ground truth $\mathbf{r}$. For instance, the $\mathcal{L}_2$ norm for a purely supervised training is given as:

\begin{equation}
    \mathcal{L} = \mathcal{L}_2 (\mathbf{r}- \mathbf{u})
\end{equation}

Such a formulation minimizes the Euclidean distance between $\mathbf{u}$ and $\mathbf{r}$. Although popular, such a loss is devoid of additional information from the physics point of view. For instance, \cite{um2020solver} showed that one can formulate physics-based loss formulations by taking advantage of the temporal unrolling during the training. This was followed by an independent investigation  \citep{list2022learned} wherein they built a custom loss function to accurately predict turbulent flows. Consequently, a greater emphasis is placed on rendering a loss yielding a physically consistent solution. Such a physics-based loss could take the form:

\begin{equation}
    \mathcal{L}  = \frac{1}{m} \sum_{i=1}^{m} \mathcal{L}_2(\mathbf{r}^ {t+i}- \mathbf{u}^i)
\end{equation}

where $m$ represents the number of unrollment steps used during training and \textit{t} represents the instant of time. Interested readers are referred to the webbook \cite{thuerey2021pbdl} for further details. 

\subsection{Benchmarks}

In addition to our present DPNN framework, we also compare it to existing benchmarks.
As part of this comparative evaluation, we employ the data-driven supervised learning (SL) method. In addition, we evaluate results produced by only running the base solver \textit{i.e., Source}a. For the SL approach, the network directly receives the input $\mathbf{r}^t$ from the \textit{Reference} solver and produces the output as the flowfield for the subsequent time step \textit{i.e.,} $\mathbf{f}(\mathbf{r}^t;\theta_W)$. As this SL framework does not contain an integrated \textit{Source} solver, the output generated by the network serves as the final prediction for the next time step (\textit{i.e.,} $\mathbf{u}^{t+1}$ = $\mathbf{f}(\mathbf{r}^t;\theta_W)$). This makes the SL approach faster. Notably, to ensure a fair comparison, the outputs from both DPNN and SL frameworks utilize identical underlying networks and receive the same input during training/testing. For the output produced from the \textit{Source} solver, the next time step predictions can be written as $\mathbf{u}^{t+1}$ = $\mathbf{u}^*$. Finally, For reasons of completeness and validation, we also juxtapose our predictions with those from the \textit{Reference} solver. 
Comparisons with the \textit{Source} solver highlight the starting point on which the DPNN solver learns to apply its corrections.
%
%
Further details related to the network architecture may be found in Appendix \ref{NN_arch}. 

\section{Results and Discussions}

In this section, we present a detailed analysis of the results obtained from the present hybrid learning framework as well as other benchmarks. As mentioned earlier, we employ multiple neural network architectures within the hybrid learning approach. Henceforth, we begin by undertaking an analysis that compares various network architectures for a common problem in the following section. 

\subsection{Role of Network Parameters in Model Performance}

We start by presenting the statistical evaluations of models builds using different network architectures for all the testing samples. An effective representation of the distribution of the model performance across a large spectrum of testing samples is done via kernel density estimation (or KDE). Besides highlighting the model confidence, it further allows for the potential identification of outliers. The KDE is evaluated based on the mean absolute error, $\mu$ of the model predictions compared to the ground truth data, for the 100th testing frame. The mean absolute error, $\mu$ is evaluated as follows:

\begin{table}
  \begin{center}
\begin{tabular}{ccccc} \\
Architecture & No. of parameters & Mean, $\mu_N$ & Standard deviation, $\sigma_N$ &  \\
CNN          & 546,478          & 0.03124     & 0.0082                       &  \\
ResNet       & 516,674          & \textbf{0.02626}     & 0.00657                      &  \\
Unet (with DS)        & 520,342          & 0.07693     & 0.0138                      &  \\
Unet          & 520,342          & 0.06067     & 0.0152                      &  \\
Dil-ResNet (with DS)  & 548,066          & 0.07266     & 0.0211                       &  \\
Dil-ResNet   & 548,066          & 0.04303     & 0.01101                       &  \\
ResNeXt      & 524,322          & 0.03871     & 0.00787                      &  \\
DenseNet     & 566,578          & 0.04258     & \textbf{0.0047}                       & 
\end{tabular}
\caption{Comparative performance among various NN architectures in terms of mean $\mu_N$ and standard deviation $\sigma_N$ (across 50 testing samples, at the end of 100 testing frame)}
\label{net_perf}
\end{center}
\end{table}

\begin{table}
  \begin{center}
\begin{tabular}{cccc} \\
Random seed number & Mean, $\mu$ & Standard deviation, $\sigma$ &  \\
1           & 0.03029     & 0.00818                      &  \\
2           & 0.02643     & 0.00747                      &  \\
3           & \textbf{0.02483}     & \textbf{0.00706}                      &  \\
4           & 0.02991     & 0.00896                      &  \\
42          & 0.02626     & 0.0082                       & 
\end{tabular}
\caption{Comparative performance of ResNet architecture for different random seeds}
\label{resnet_rand}
\end{center}
\end{table}

\begin{figure}
\centering
\includegraphics[scale=2.2, angle=0]{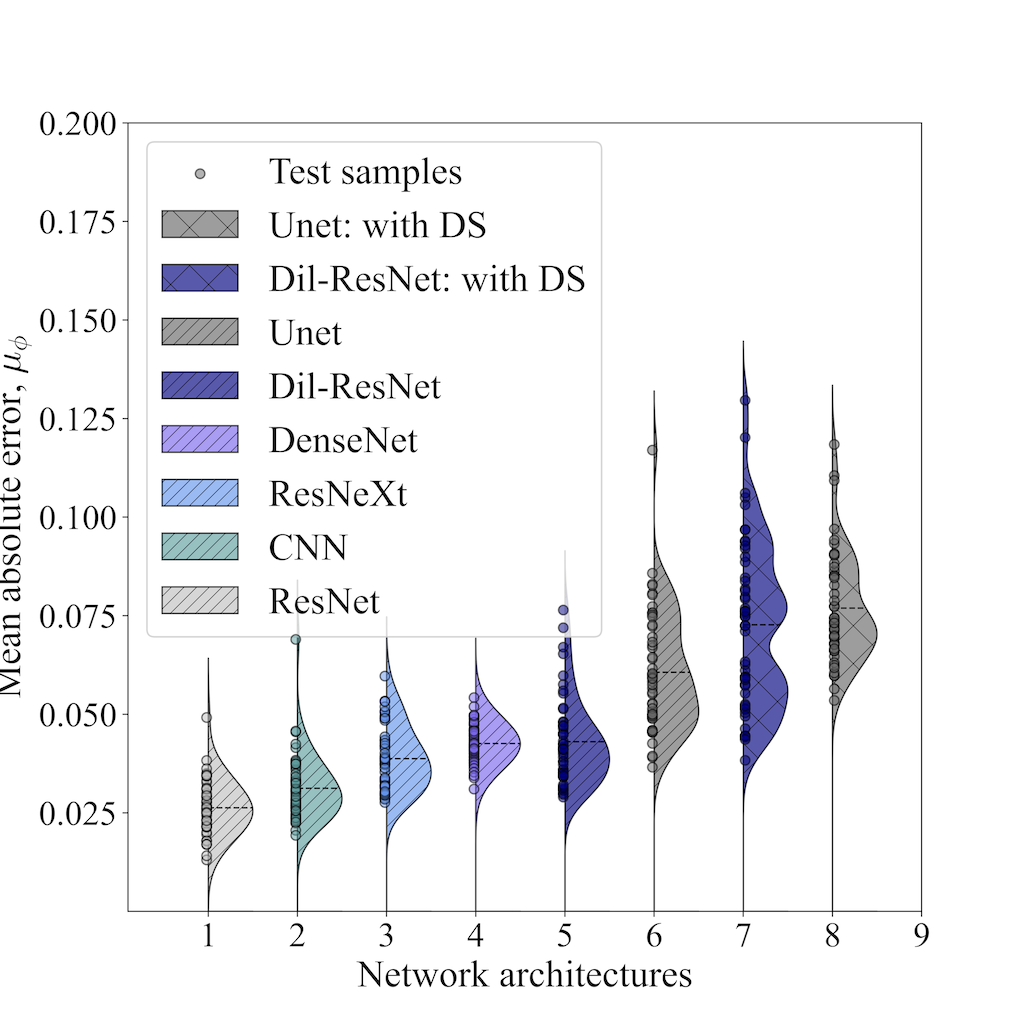}
\caption{Comparison of the mean absolute error $\mu$ at the end of 100th testing frame for 50 testing samples obtained from different NN architectures}
\label{All_models}
\end{figure}

\begin{equation}
\label{mu_1}
\mu = \frac{1}{n_c}\sum_{i=1}^{n_c} (|u_{\mathrm{ref,i}}-u_{\mathrm{baseline,i}}|) + (|v_{\mathrm{ref,i}}-v_{\mathrm{baseline,i}}|) 
\end{equation}

Figure \ref{All_models} presents the approximate probability distribution via the KDE for the 50 testing samples from different network architectures. It is found that most of the approximate probability distribution is indicative of a Gaussian-like distribution. 
These include the KDE obtained from the convolution neural network (CNN) and the residual network-based models \textit{i.e.,} ResNet \citep{he2016deep}, together with the newer variants ResNeXt \citep{xie2017aggregated}, DenseNet \citep{huang2017densely}, and diluted residual network  (Dil-ResNet) from  \cite{stachenfeld2021learned}. On the contrary, however, the Unet \citep{ronneberger2015u} exhibits multiple peaks in the distribution while also incurring greater sample mean error $\mu_N$ (see Table \ref{net_perf}). While attributing higher error to the architectures is not straightforward, strikingly, both the Unet with downsample (with DS) and the Dil-ResNet (with DS) architectures employ encoder blocks that compress the input data to a reduced dimension via the \texttt{stride} operation. This is not entirely surprising as the model struggles to retain local features of the input upon downsampling. We remark that identical choice of activation function, optimizer, learning rate $\eta$, initial weights $\theta_{\mathrm{W}}$, loss function, etc. have been used for all the architectures. Consequently, it can be mentioned that the models without any compression or convolution downsample serve as a better alternative for the present choice of problems. 

\begin{figure}
\centering
\includegraphics[scale=0.35, angle=0]{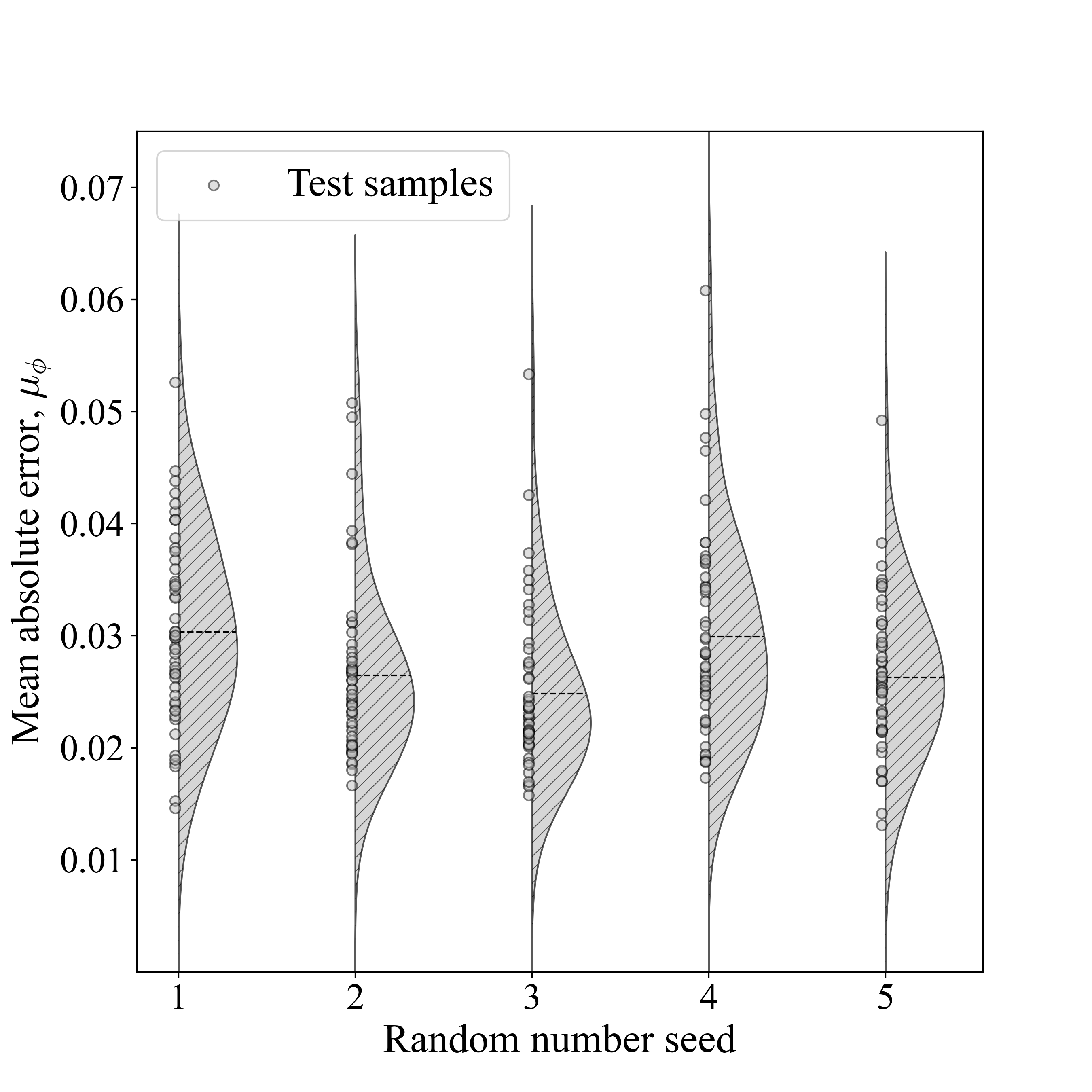}
\caption{Influence of random number seed on the performance of residual network (ResNet) architecture}
\label{rng_seeds}
\end{figure}

Neural network-based learning is a stochastic, non-linear optimization process that aims to iteratively improve the model predictions on training data while managing uncertainty and navigating through a complex, multidimensional solution space. In this context, using a gradient-based optimizer like Adam may result in a locally-optimal solution with some sensitivity to the initial guess. To check the robustness of the model, the influence of the random number seed must be evaluated. We investigate the performance of the ResNet-based architecture for multiple seeds. Figure \ref{rng_seeds} shows the approximate probability density based on the mean absolute error $\mu$ obtained using ResNet-based architectures for different values of random number seed for all the testing samples. It can be observed that the Gaussian-like distribution of the error $\mu$ is retained for different random seed values, besides showing consistency in the model performance in terms of the sample mean $\mu_N$ (see Table \ref{resnet_rand}). In light of these discussions, it can be remarked that the performance is independent of the random number seed. Finally, we also determine the influence of the adaptive learning rate $\eta$ that determines the convergence and consequently the nature of solutions obtained. This is briefly discussed in Appendix \ref{LRate}. The remainder of the study is based on the ResNet architecture. 


\subsection{Qualitative comparison of predictive models across wake categories}
\label{qual}

In this section, we compare various benchmark models, assessing their capability to accurately reproduce the dynamic wake characteristics for new, unseen testing samples with different spacing ratios. This analysis is fundamentally important because, intriguingly, different spacing ratios between the cylinders yield diverse wake behaviors, each of which necessitates faithful reproduction. For instance, the deflected gap wake should exhibit a narrow and a wide wake immediately downstream of the two subsequent cylinders, with the gap flow between these cylinders being deflected towards the narrow wake region. This deflection should remain constant with time. In contrast, for the flip-flopping or the chaotic wakes, the gap flow switches direction in an erratic manner. As such, while some wake patterns exhibit a periodic, uniform repetition, others veer into the realm of chaos. It is, therefore, a fascinating problem that tests the predictive capabilities of the models.

\begin{figure}
\centering
\includegraphics[scale=2.575, angle=0]{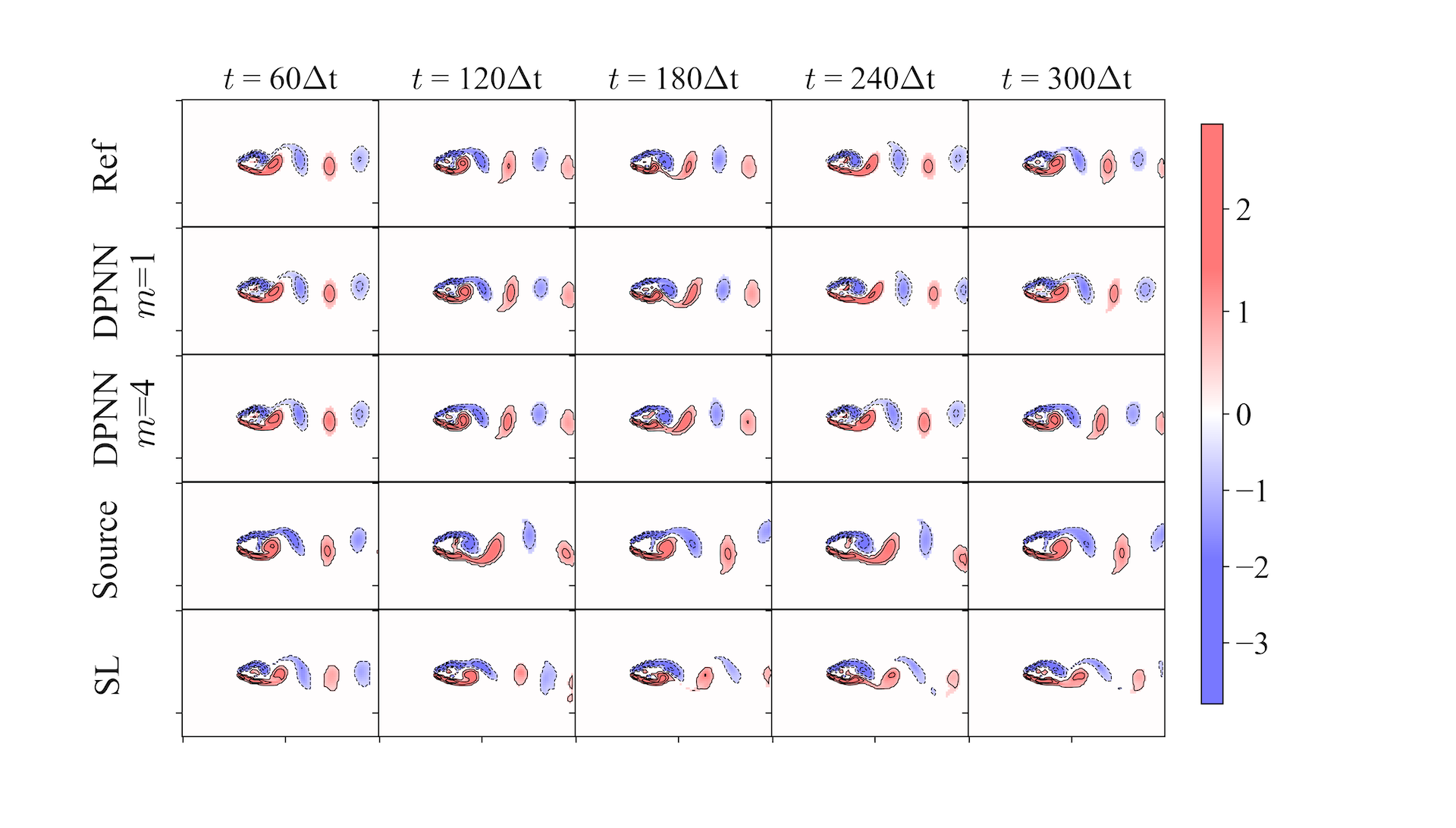}
\caption{Comparison of velocity flowfields obtained using various benchmarks for different time instances for the spacing ratio $L/D$ = 1.26 (single bluff-body wake, 17th testing sample)}
\label{common_omega_sin}
\end{figure}

Five representative testing samples are chosen from the available 50 experiments, each exhibiting a unique spacing ratio and the corresponding wake dynamics. We present vorticity flowfields generated entirely from the network's output. This allows us to highlight the regions of interest, \textit{i.e.,} cylinder wake and near body boundary, besides also indicating the direction of rotation of the vortices. Figure \ref{common_omega_sin} compares the vorticity flowfield obtained from the various benchmarks, including the \textit{Reference} solver at the spacing ratio $L/D$ = 1.26. The vorticity flowfield obtained from the \textit{Reference} exhibits a single bluff-body wake with a certain shedding frequency. It is found that while all the baselines can reproduce the correct category of the wake, the \textit{Source} and the supervised learning (SL) approach very quickly deviate from the correct vortex shedding cycle. However, the present DPNN approach compares remarkably well with the \textit{Reference} data even for long time horizons.

\begin{figure}
\centering
\includegraphics[scale=2.575, angle=0]{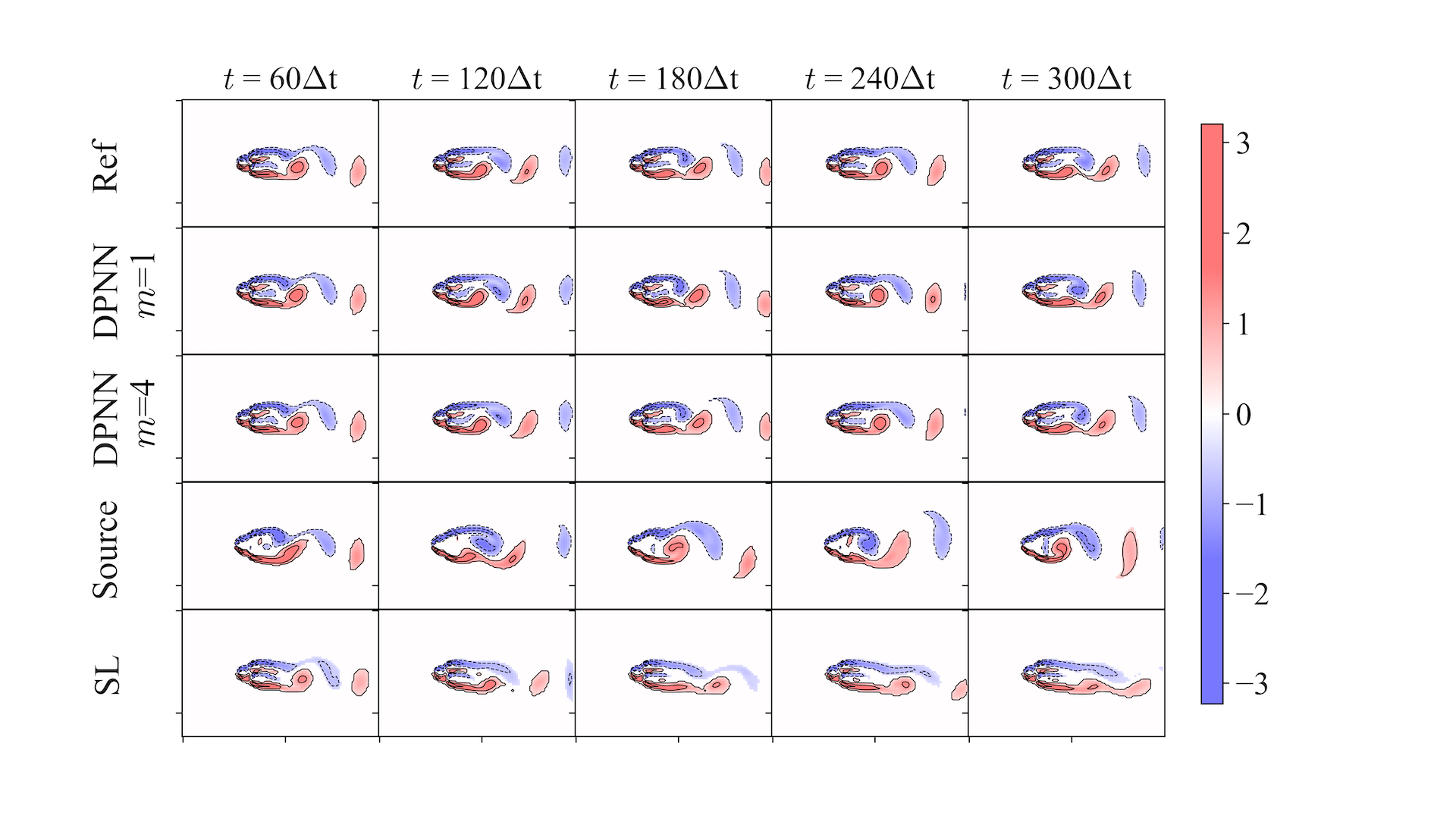}
\caption{Comparison of vorticity flowfields obtained using various benchmarks for different time instances for the spacing ratio $L/D$ = 1.57 (deflected gap wake, 39th testing sample)}
\label{common_omega_def}
\end{figure}

Figure \ref{common_omega_def} presents the comparison of the vorticity flowfield obtained at the spacing ratio $L/D$ = 1.57 for which the \textit{Reference} solver portrays deflected gap wake. As alluded to earlier, the deflected gap flow shifts toward the narrower wake region (\textit{i.e.,} lower downstream cylinder in this case), with the direction of the gap flow remaining stable over time. Consequently, the predictive models should preserve this phenomenon while accurately predicting the vortex-shedding cycle. It is found from Fig. \ref{common_omega_def} that the predictions from \textit{Source} quickly transitions to a single bluff-body wake, whereas the vorticity contours obtained from the SL approach transform into a hybrid quasi in-phase wake. In contrast, however, the current DPNN approach aligns impressively with the \textit{Reference} data across extended time horizons. This observation holds true for the deflected gap flow direction and the overall vortex shedding cycle.

\begin{figure}
\centering
\includegraphics[scale=2.575, angle=0]{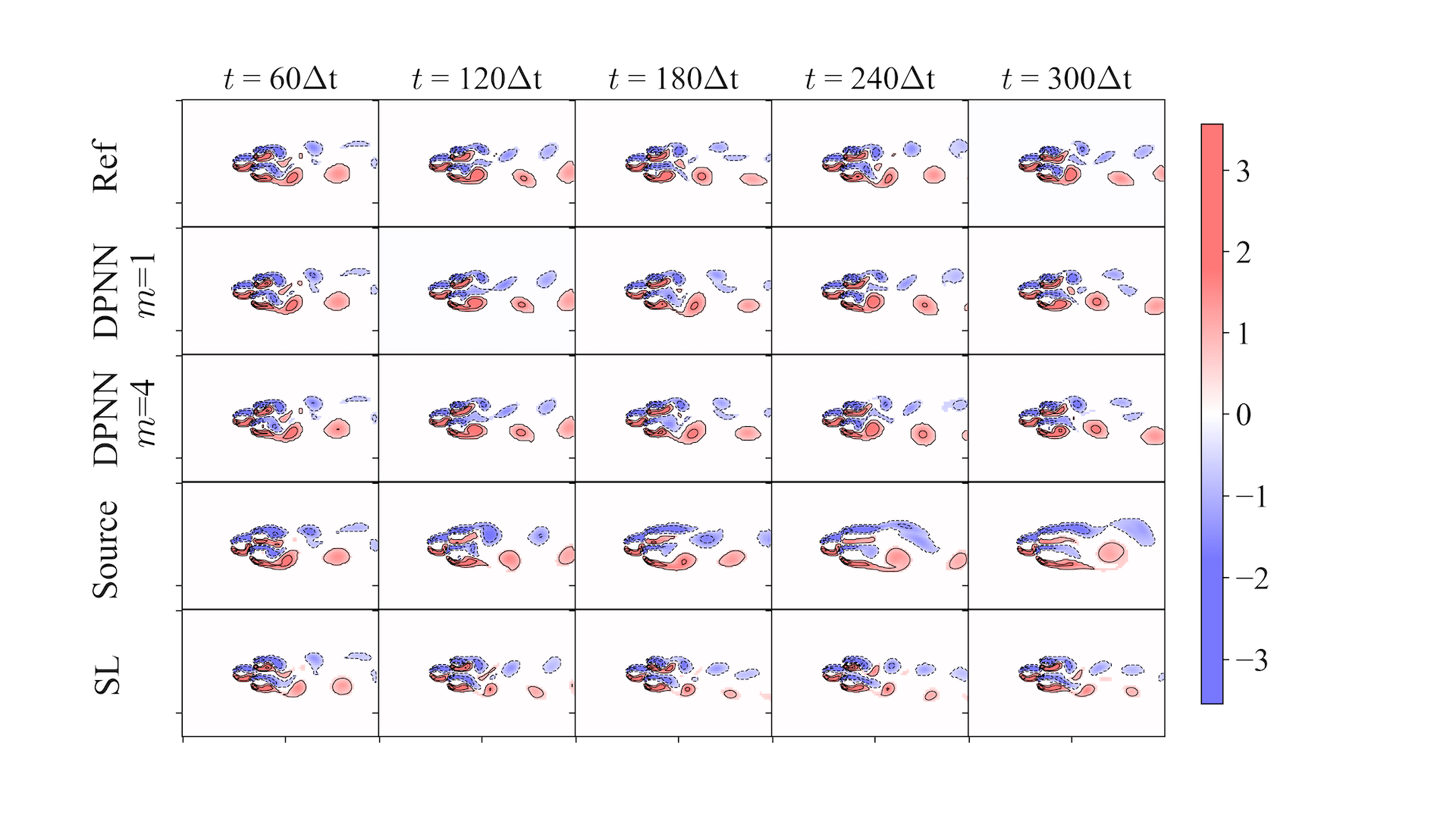}
\caption{Comparison of vorticity flowfields obtained using various benchmarks for different time instances for the spacing ratio $L/D$ = 2.34 (chaotic wake, 41st testing sample)}
\label{common_omega_cha}
\end{figure}

Figure \ref{common_omega_cha} showcases a comparison of the vorticity flowfield at a spacing ratio $L/D$ = 2.34, where the \textit{Reference} solver displays a flip-flopping (or chaotic) wake, \textit{i.e.,} the gap flow in between the two downstream cylinders chaotically switches direction. This is evident in Fig. \ref{common_omega_cha} wherein the gap flow pointing downwards from the frame at $t = 60 \Delta t$ has transitioned to the top at $t = 120 \Delta t$. The predictions from DPNN seem to capture this switch correctly up to frame $t = 180 \Delta t$, whereas, for long temporal horizons, both the gap flow direction and the vortex shedding cycles have visible differences. These differences further augments when compared to the vorticity flowfields obtained from \textit{Source} and SL approaches.

\begin{figure}
\centering
\includegraphics[scale=2.5, angle=0]{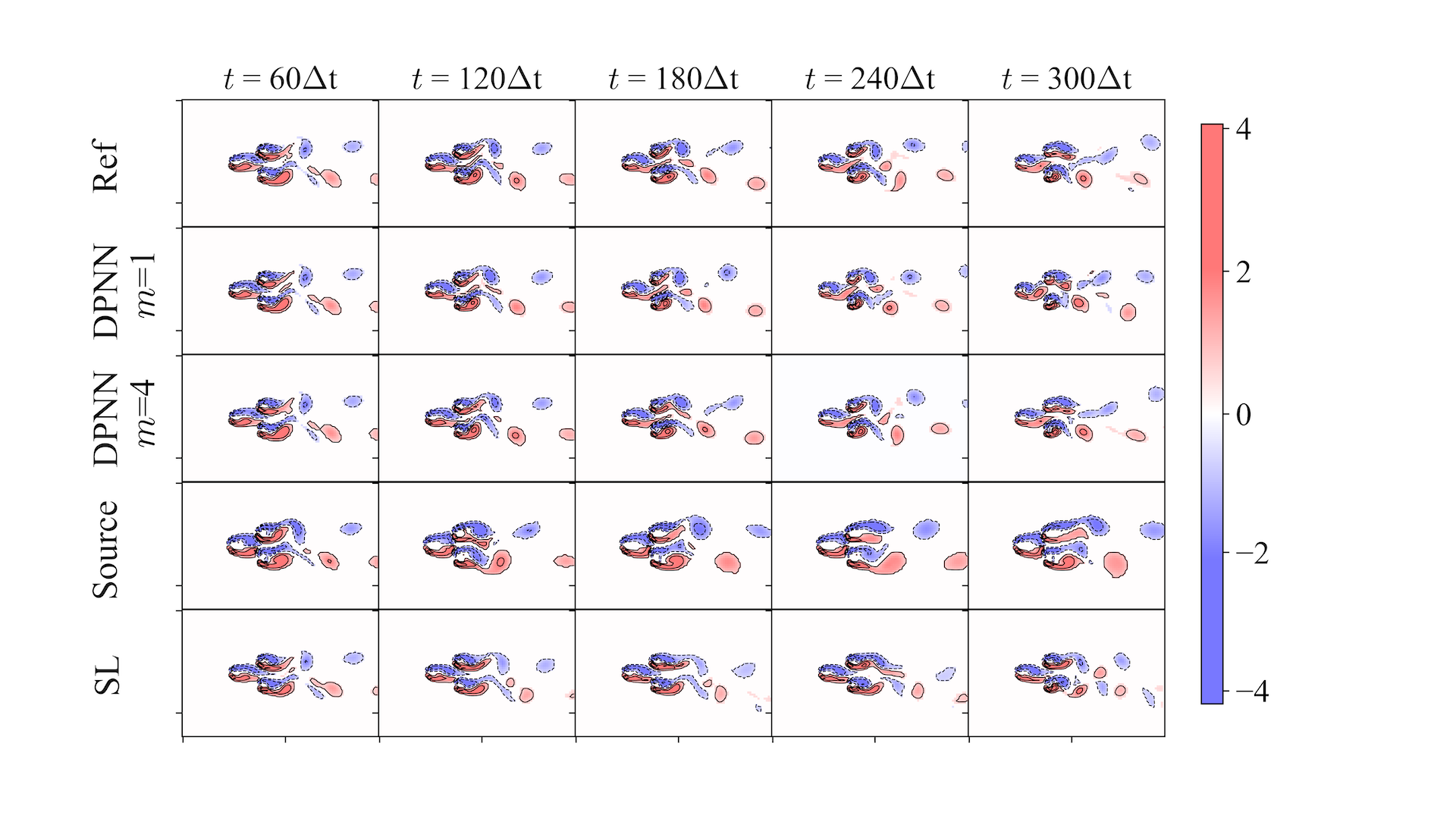}
\caption{Comparison of vorticity $\omega$ flowfields obtained using various benchmarks for different time instances for the spacing ratio $L/D$ = 3.37 (fully developed quasi in-phase wake, 26th testing sample)}
\label{common_omega_qip}
\end{figure}

Figure \ref{common_omega_qip} exhibits a comparative analysis of vorticity flowfields at a spacing ratio of $L/D$ = 3.37, which, as per the \textit{Reference} solver, corresponds to a fully developed quasi in-phase wake. The clockwise rotating vortices (highlighted in blue) emitted from the freestream side of the upper cylinder and the gap side of the lower cylinder, appear to be in quasi in-phase. The same can be mentioned for the anti-clockwise rotating vortices (highlighted in red) shed from the freestream side of the lower cylinder and the gap side of the upper cylinder. It is found that among all the baselines, both DPNN and SL approaches perform well, with the former preserving the vortex structure quite remarkably for long-time horizons. On the other hand, the \textit{Source} depicts a chaotic wake and fails to reproduce the actual wake dynamics for the given spacing ratio and geometric configuration. 

\begin{figure}
\centering
\includegraphics[scale=2.5, angle=0]{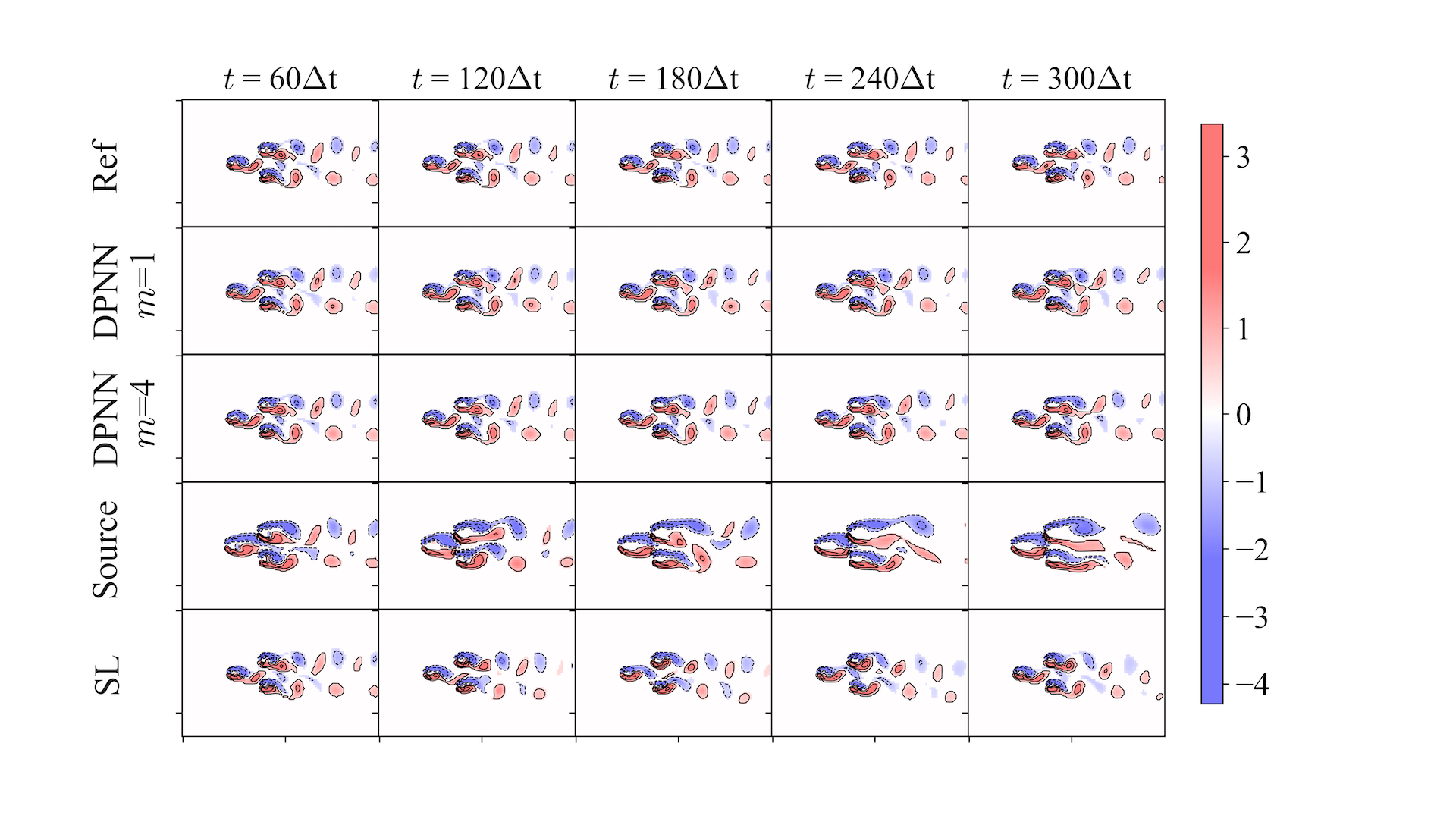}
\caption{Comparison of vorticity $\omega$ flowfields obtained using various benchmarks for different time instances for the spacing ratio $L/D$ = 3.89 (fully developed quasi anti-phase wake, 32nd testing sample)}
\label{common_omega_qap}
\end{figure}

Figure \ref{common_omega_qap} highlights the vorticity flowfields obtained at the spacing ratio of $L/D$ = 3.89 from various benchmarks. The flowfield obtained from the \textit{Reference} solver indicates a fully-developed quasi anti-phase wake. It is observed that the vortices emitted from the freestream side of the two downstream cylinders shed vortices in the opposite direction. This dynamic nature of the wake is accurately retained by the present DPNN approach. On the contrary, however, the SL approach produces a hybrid wake, \textit{i.e.,} wake transitions into a hybrid wake, whereas the \textit{Source} yields a chaotic wake.

The exercise underscored the crucial need for developing robust and precise models. These models should accurately identify the underlying wake category and deliver sustained temporal fidelity over extended periods. While the current section provides a qualitative comparison of individual baselines across multiple wake categories, the subsequent section will present a quantitative analysis of the key variables of interest.

\subsection{Quantitative comparisons in physical and reduced feature space}
\label{quan}

In the preceding section, we employed velocity and vorticity flowfields as a qualitative measure of performance accuracy. The network used in our study is trained purely on physical space, specifically the velocity flowfields as a collection of control volumes with staggered faces. Through the actions of the kernels, convolutions transform these physical quantities into an equivalent dimensional feature space, achieved through zero-padding. This process poses an engaging question - does the network retain its performance in Fourier space or the reduced feature space offered by the first three principal components?

\begin{figure}
\centering
\includegraphics[scale=1.75, angle=0]{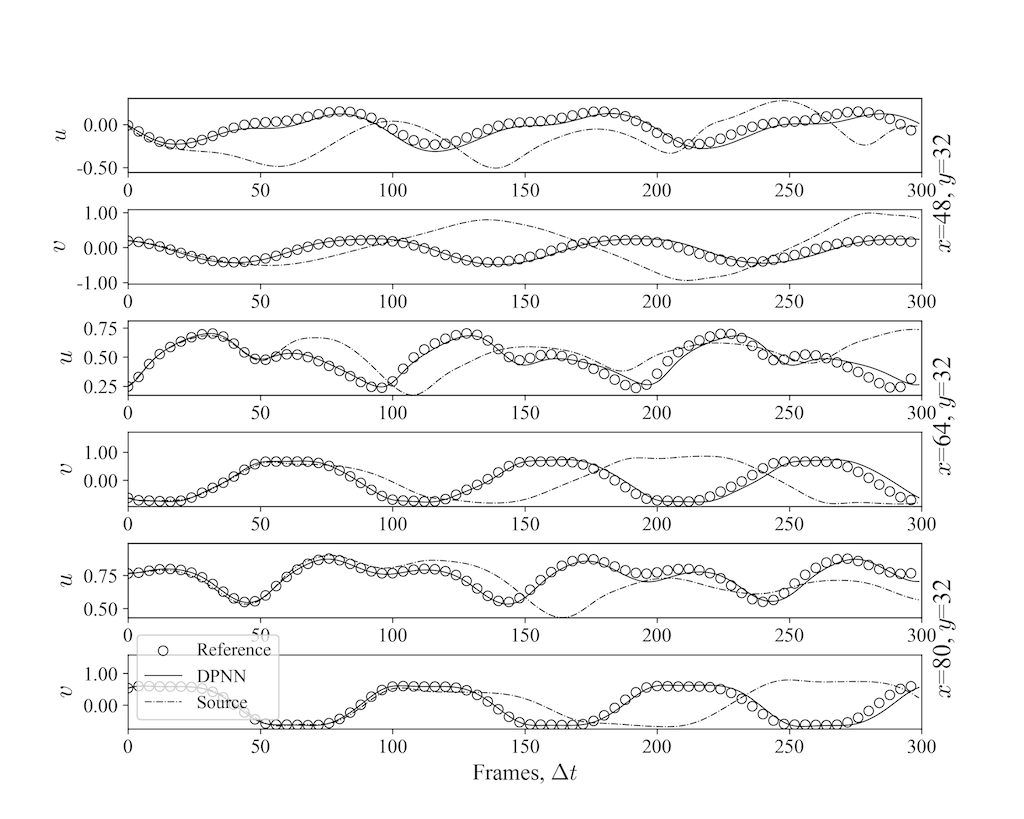}
\caption{Comparison of velocity downstream of the bodies and its variation with time (39th testing sample)}
\label{common_uv_probe_89}
\end{figure}

We begin our investigation by adopting the 39th testing sample, which depicts a deflected gap wake, as discussed in Section \ref{qual}. In Fig. \ref{common_uv_probe_89}, we present the temporal variation of the velocity components $u, v$ at multiple spatial probes using different baselines. These probes are placed at the wake of the downstream cylinders along the centerline axis, \textit{i.e.,} ($n_{\mathrm{px}}, n_{\mathrm{py}}$) = (48,32), (64,32), (80,32), where $n_{\mathrm{px}}, n_{\mathrm{py}}$ represents the nodal positioning in the $x$ and $y$ directions, respectively. It must be noted here that the present DPNN approach from hereon refers to $m$=4 steps of unrolling during training. Thus, an evaluation period up to $t=300\Delta t$ is a substantial period wherein multiple vortex-shedding cycles are captured. It is found that the present DPNN framework results in very good agreement with the modulated signals obtained for both $u, v$ velocity components, with the agreement seemingly better for probes placed further away from the bodies. The minor offset with the data obtained from the \textit{Reference} solver tends to happen for higher time values $t$. On the contrary, however, this offset is rather early for the predictions obtained from the \textit{Source} solver. Moreover, for higher values of time $t$, the predictions significantly differ from the \textit{Reference}. 

\begin{figure}
\centering
\includegraphics[scale=1.75, angle=0]{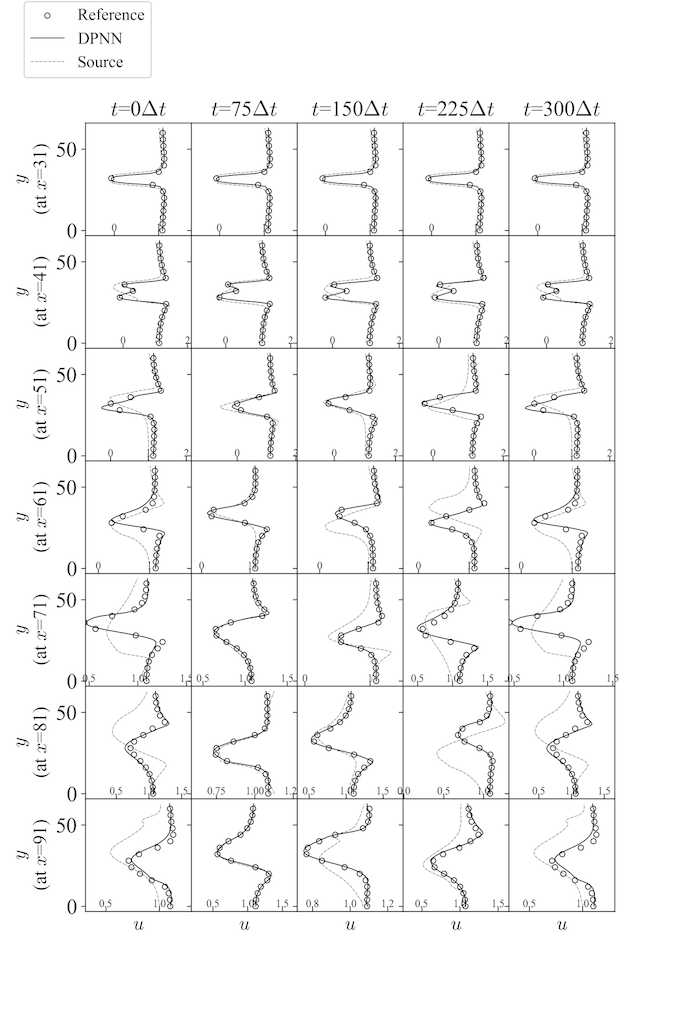}
\caption{Comparison of the $u$-velocity at multiple vertical cross-sections (39th testing sample)}
\label{uvel_xsections_89}
\end{figure}

\begin{figure}
\centering
\includegraphics[scale=1.75, angle=0]{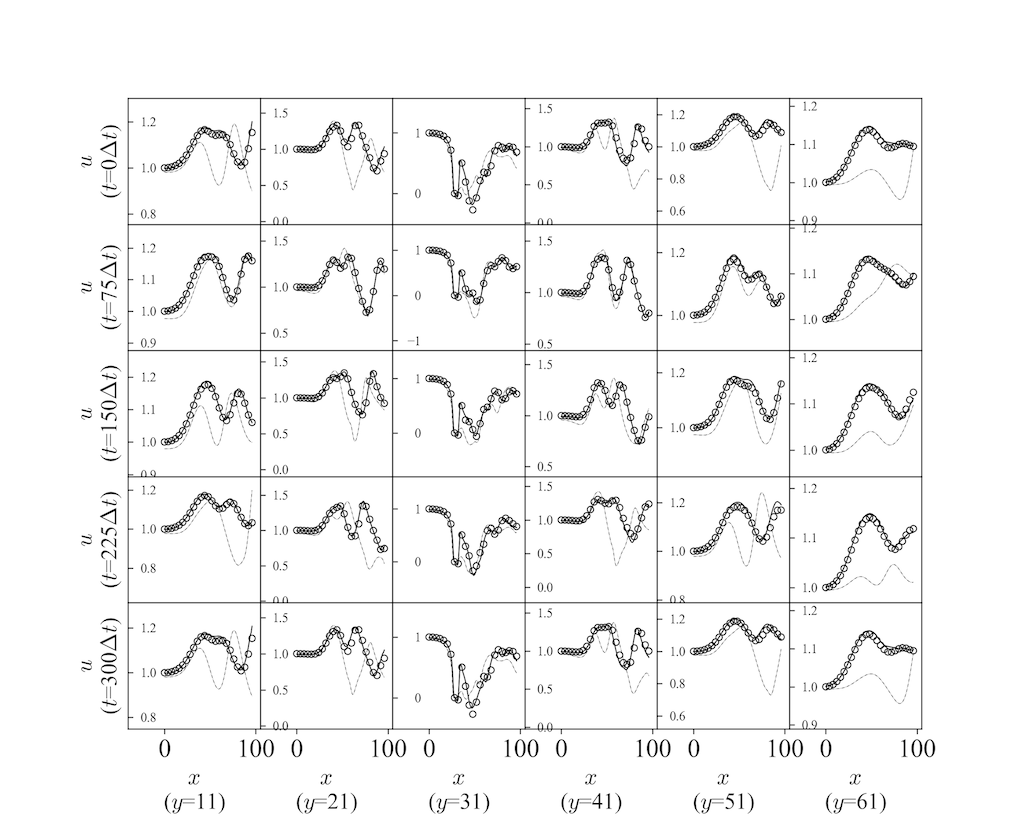}
\caption{Comparison of the $u$-velocity at multiple horizontal cross-sections (39th testing sample)}
\label{vvel_xsections_89}
\end{figure}

\begin{figure}
\centering
\includegraphics[scale=0.6, angle=0]{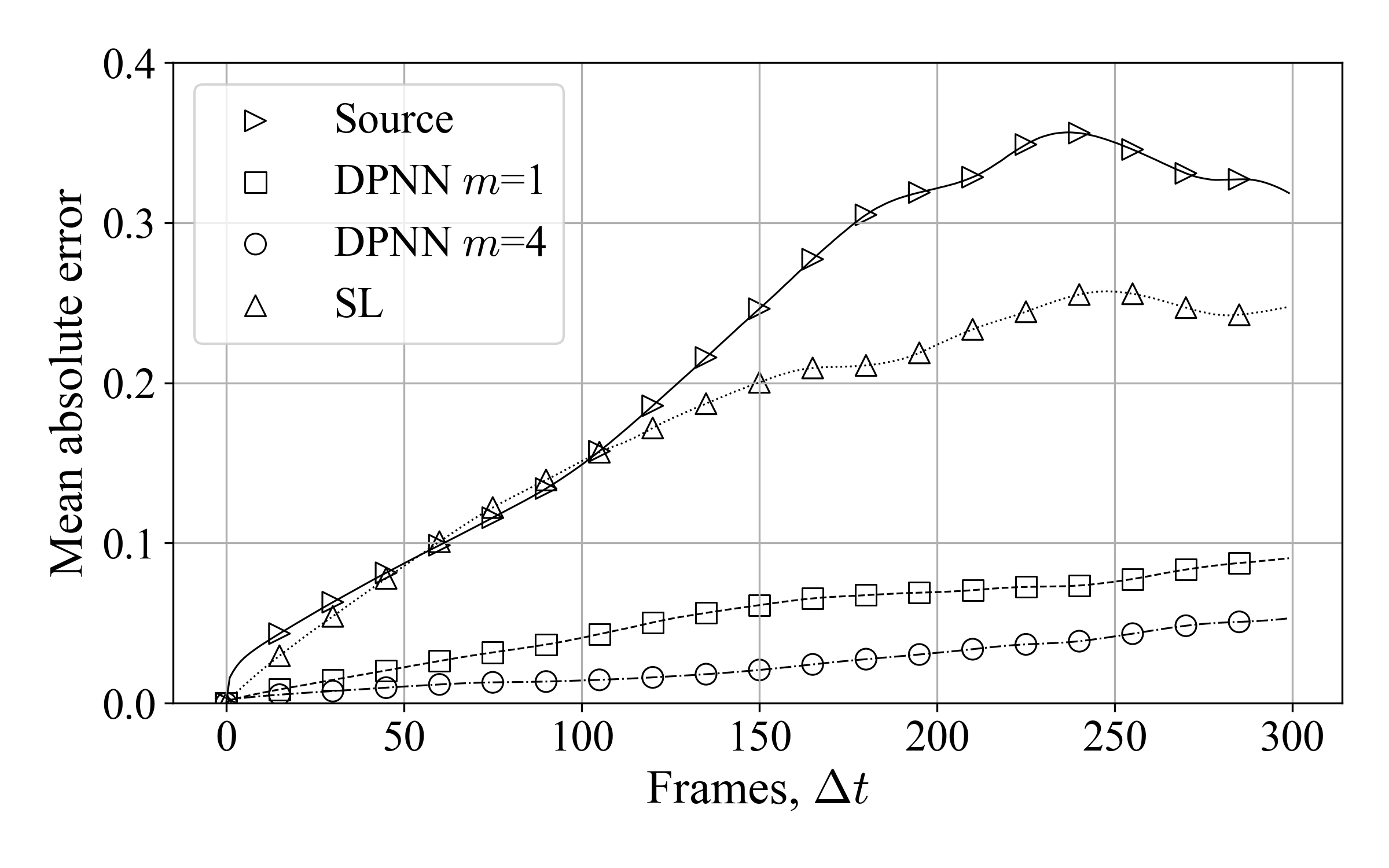}
\caption{Variation of the mean absolute error from various frameworks with testing frames for the 39th testing sample}
\label{common_mae_89}
\end{figure}

Figure \ref{uvel_xsections_89} displays the distributions of $u$-velocity across several cross-sections aligned in the longitudinal direction (along the rows) at multiple instances of time (along the columns), obtained from multiple benchmarks. Barring some minor differences, we found that the current framework aligns exceptionally well with data obtained from the \textit{Reference} solver across various spatial locations and at different time instances. Similar observations can also be drawn from Fig. \ref{vvel_xsections_89}, where the $u$-velocity distributions are plotted along multiple horizontal cross-sections at different instances of time. Naturally, these observations are also reflected in Fig. \ref{common_mae_89} where the mean absolute error, $\mu$ (calculated based on Eq. \ref{mu_1}) for the recursive predictions with respect to the ground truth data are shown. It is found that the present DPNN-based approach yields the lowest cumulative error over time, whereas the data from the \textit{Source} results in the highest growth of error over time. As such, it can be mentioned that the DPNN approach has successfully enabled a hybrid framework that can faithfully reproduce the dynamical behavior of the physical quantity (\textit{i.e.,} velocity flowfields).  

\begin{figure}
\centering
\includegraphics[scale=0.625, angle=0]{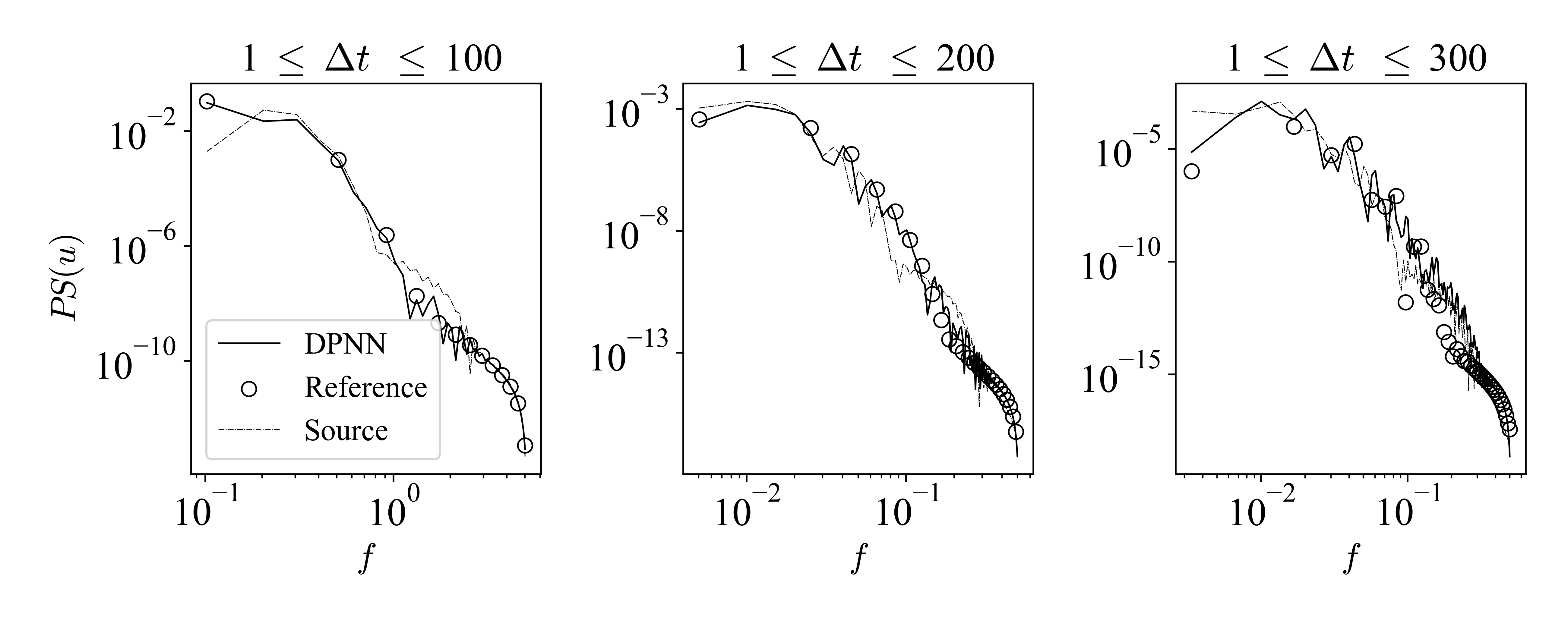}
\caption{Comparison of the power spectral density of the streamwise velocity for multiple time ranges (39th testing sample)}
\label{psd_89}
\end{figure}

We now compute the power spectral density (or PSD) based on the velocity flowfields obtained from the predictive frameworks. 
Figure \ref{psd_89}(a)-(c) presents the PSD obtained from the velocity measurements at the $n_{\mathrm{px}},n_{\mathrm{py}}$ = (64,32) probe location with respect to different time ranges. It is found that for a relatively smaller time range of up to 100 $\Delta t$, the PSD vs. frequency obtained from the DPNN aligns exceptionally well with the data obtained from the \textit{Reference} solver signifying that the present framework is able to faithfully reconstruct small as well as large scale structures in the velocity flowfield. However, minor differences in the high-frequency range are found for large recursive predictions (\textit{i.e.,} up to 300 $\Delta t$), indicating small-scale differences in the velocity flowfields. This bodes well with the observations in Fig. \ref{common_uv_probe_89}, where observable differences are found for larger recursive predictions.

\begin{figure}
\centering
\includegraphics[scale=1.75, angle=0]{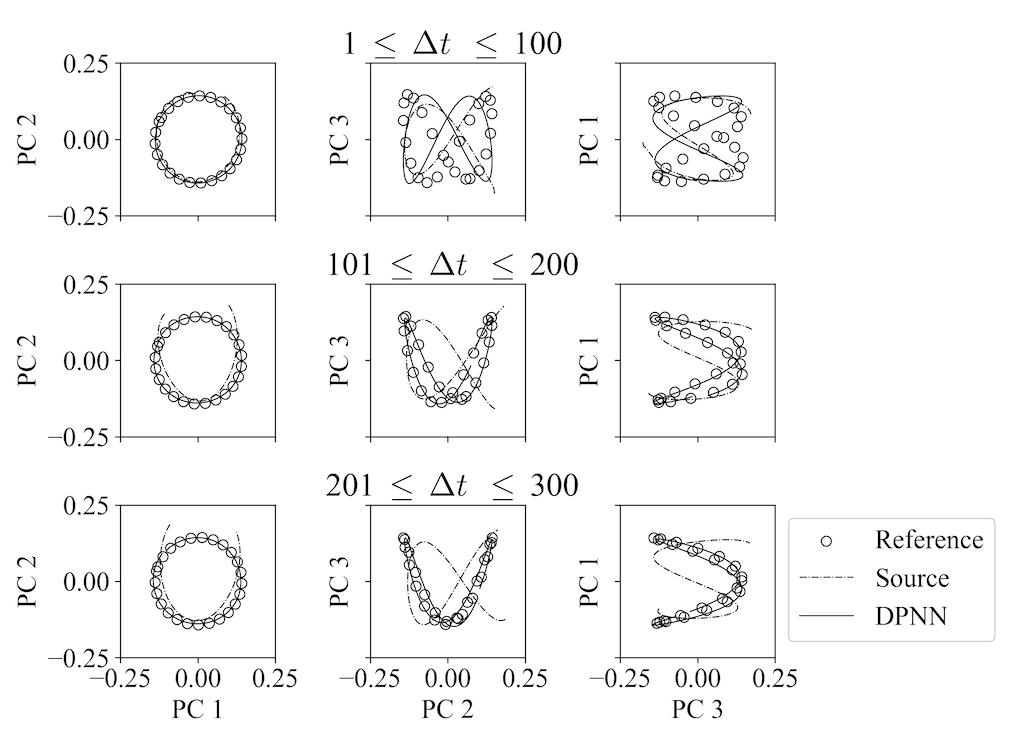}
\caption{Comparison of the first three principal components and its variation with time (39th testing sample)}
\label{pca_89}
\end{figure}

We perform principal component analysis (or PCA) on the predicted velocity fields obtained from the baselines to evaluate the principal components (or PC) and their variation with respect to time $t$. Figure \ref{pca_89} presents the first three PC's and their distribution with time. It is found that the temporal variation of PC 1 and PC 2 indicates a periodicity in the data obtained from the \textit{Reference} solver, which seems to be captured correctly in predictions obtained by DPNN. Moreover, the temporal variation in the PC 2, PC 3, and PC 3, PC 1 components are also in good overall agreement with the \textit{Reference} solver. The protracted vortex shedding cycle obtained from \textit{Source} observed in Fig. \ref{common_uv_probe_89} is also reflected in Fig. \ref{pca_89} in terms of the open-ended arc indicating a longer vortex shedding time-period. As a result of these investigations, it can be remarked that the performance of the DPNN approach demonstrates consistency in both the physical and reduced feature space.

\subsection{Reconstructing flowfields: global and local perspectives}
\label{local}

As we delve deeper into investigating the present coarse-grained surrogate model, it becomes increasingly clear that both local and global fluid structures are crucial for the overall representation of flow physics. Specifically, precisely reproducing local boundary layer phenomenon and global wake dynamics becomes essential. The intersection of these scales - the local and the global - is where true accuracy in understanding and predicting flow dynamics lies. To that extent, we investigate the ability of the learned models to correctly reconstruct the local mean velocity boundary layer profile (with  momentum thickness, $\theta$) as well as the mean gap flow $\bar{U_\mathrm{G}}$, while also evaluating the global fluid variables in terms of kinetic energy $KE$ and enstrophy $\Omega$. This step is not trivial given that the \textit{Source} solver employs a masked stair-step representation for the underlying body boundary for computational efficiency. 

It is also crucial to remark here that while three level coarsening will naturally influence the local resolution of the boundary layer profile it still retains a parabolic profile for the upstream body, as will be shown later on. Thus at a given level of grid resolution, the goal is to mitigate the difference between the under-resolved boundary layer of the \textit{Source} and the resolved boundary layer of the \textit{Reference} transferred to the \textit{Source} mesh. Hence, our goal is not a super-resolution task, i.e. transferring a low-resolution solution to a higher resolution, but instead to improve the solution given a fixed computational mesh.

The momentum thickness, $\theta$ is evaluated as below,

\begin{equation}
  \theta(x) = (\int_{0}^{\delta_\mathrm{max}} \frac{\bar{u}(x,y)}{\bar{u}_e(x)} \Big( 1-\frac{\bar{u}(x,y)}{\bar{u}_e(x)} \Big) \,dy) 
\end{equation}

The time and space averaged streamwise velocity at the gap \citet{chen2020numerical}, $U_G/U$ is evaluated as follows,

\begin{equation}
  \frac{U_\mathrm{G}}{U_{\mathrm{\infty}}}= \frac{1}{U_{\mathrm{\infty}}} \int_{-0.5}^{0.5} \bar{u} \,d\Big(\frac{y}{G}\Big) 
\end{equation}

where, $\bar{u}$ is the time-mean streamwise velocity at the gap. The kinetic energy $KE$ and enstrophy $\Omega$ are calculated as follows,

\begin{equation}
KE(t,U) = \frac{1}{2} \int_{L_{2(A)}}^{} {U(t)^2} dA
\end{equation}

\begin{equation}
\Omega(t,U) = \frac{1}{2} \int_{L_{2(A)}}^{} {\omega(t)^2} dA
\end{equation}

While KE is a popular choice for statistical evaluation of fluid flows, enstrophy represents the rotational energy of the flow and corresponds to the dissipative effects of the flow \citep{tong2015numerical}. \\

\begin{figure}
    \centering
    \subfigure[]{\includegraphics[width=0.31\textwidth]{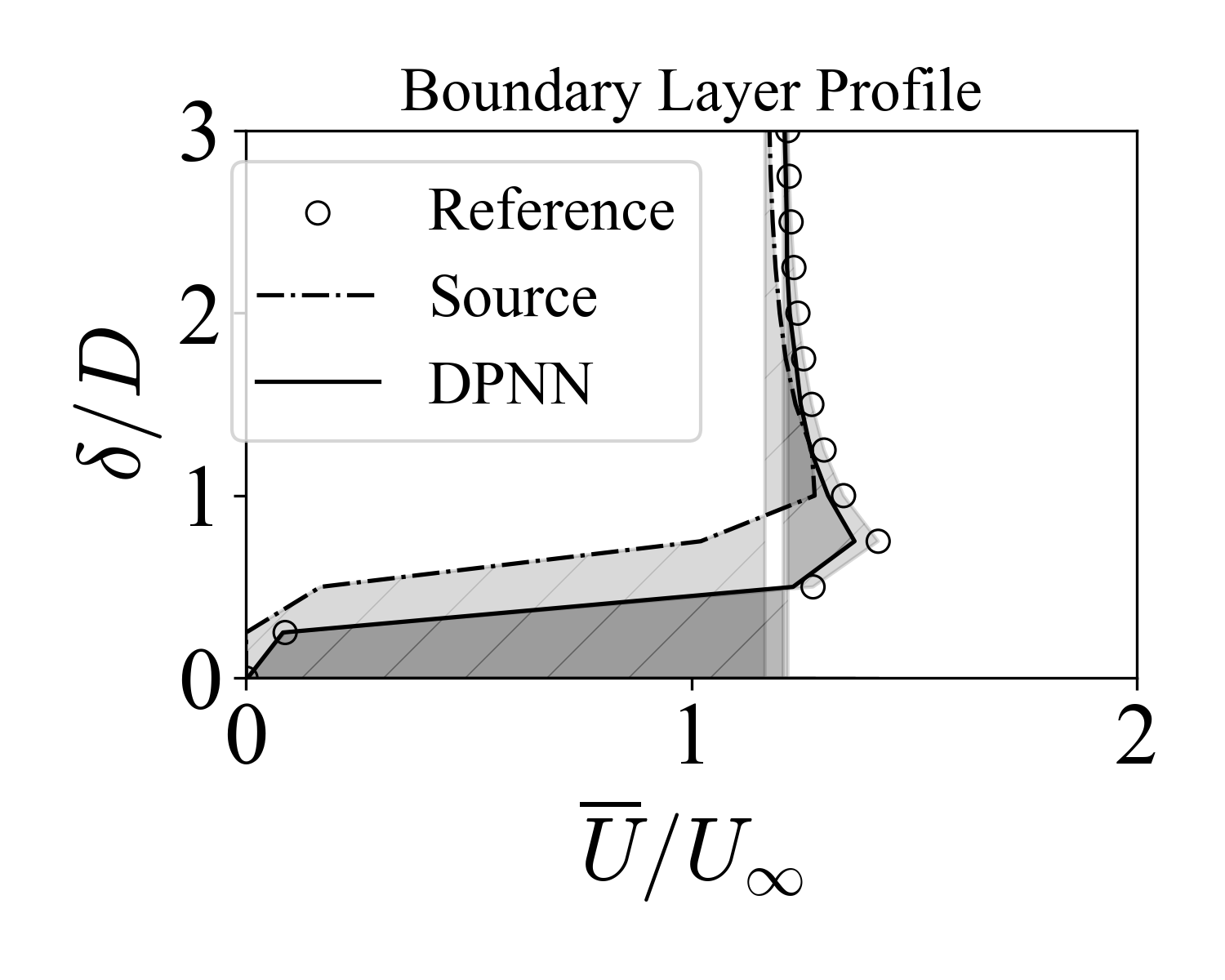}} 
    \subfigure[]{\includegraphics[width=0.31\textwidth]{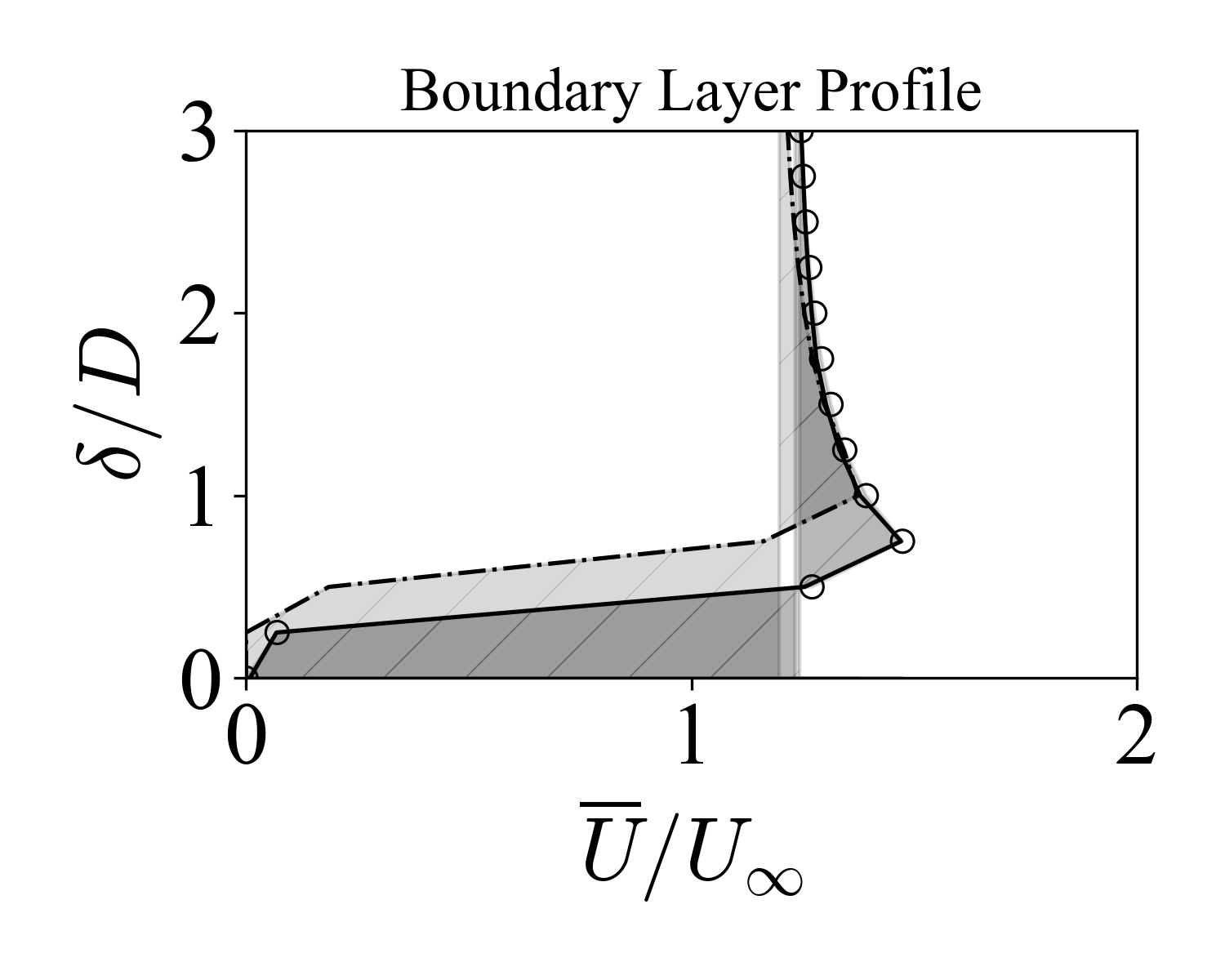}} 
    \subfigure[]{\includegraphics[width=0.31\textwidth]{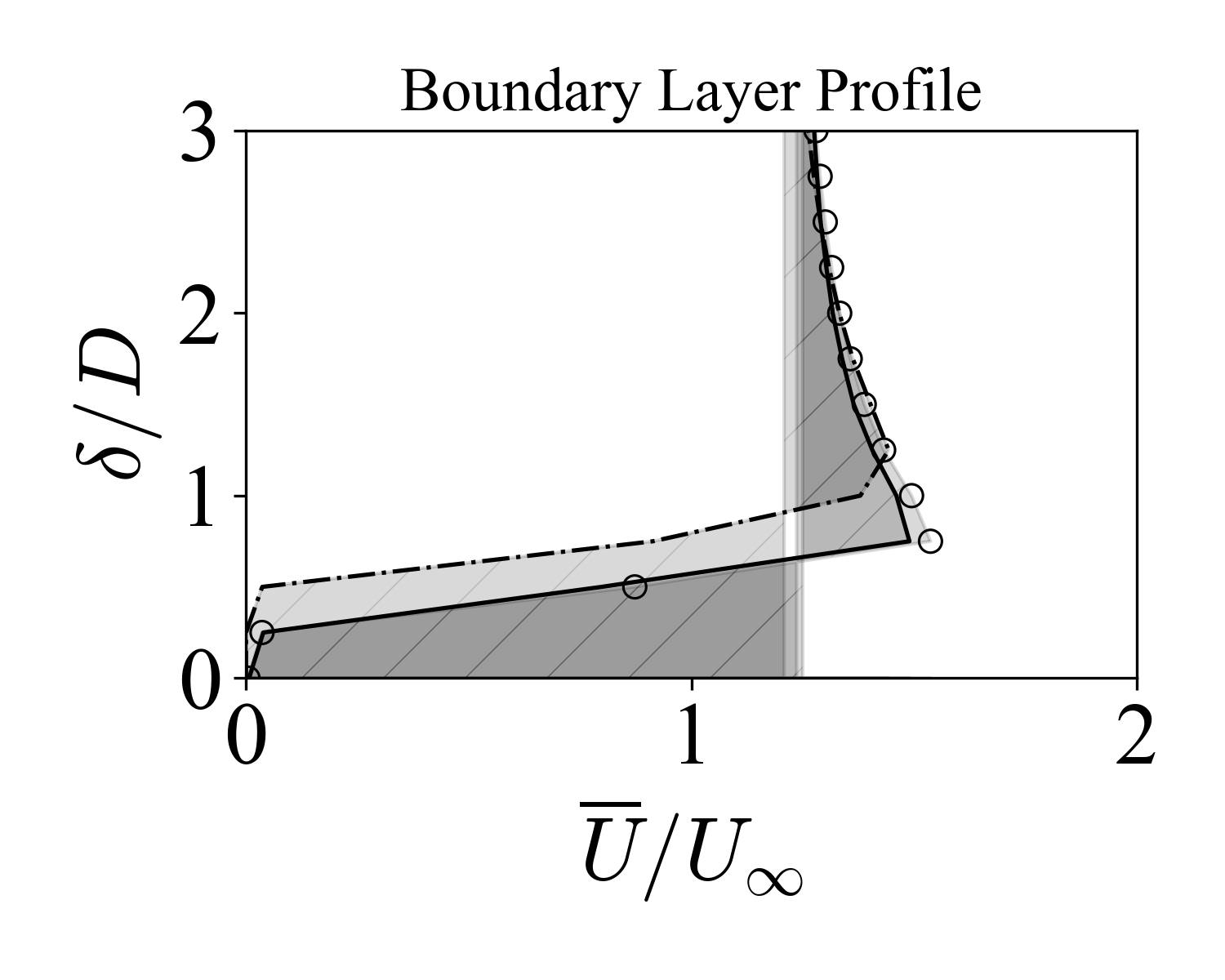}}
    \caption{Comparison of the time-averaged boundary layer profile along the lateral $y$ direction of the upstream body for (a) 28th (b) 40th (c) 46th testing samples}
    \label{BLProfile}
\end{figure}

\begin{table}
\begin{center}
\begin{tabular}{cccc} \\
                & \multicolumn{3}{l}{Boundary layer momentum thickness, $\theta/D$} \\
Testing samples no & \textit{Reference}              & \textit{Source}              & DPNN               \\
28               & 0.852                  & 0.527               & 0.697              \\
40               & 0.91                   & 0.723               & 0.902              \\
46               & 1.01                   & 0.812               & 0.895             
\end{tabular}
\caption{Comparison of the boundary layer momentum thickness $\theta/D$ for multiple testing samples}
\label{mom_thickness}
\end{center}
\end{table}

Figure \ref{BLProfile} presents the boundary layer profile in the lateral \textit{y} direction just above the body boundary along the cross-section drawn at the upstream cylinder for three distinct testing samples. Despite the fact that downsampling high-resolution data obtained from the \textit{Reference} solver can distort the actual boundary profile, reconstructing the steep rise in velocity remains a critical criterion to be preserved. It is evident that the inadequate treatment of the body boundary results in a much thicker boundary layer profile obtained from the \textit{Source}. This results in an enlarged effective diameter of the body, subsequently changing the equivalent effective spacing ratio. Such alteration points to a potential difference in the wake category (single bluff-body wake instead of deflected gap wake from \textit{Source} in Fig. \ref{common_omega_def}). On the contrary, the boundary layer profile obtained from the DPNN approach agrees quite well with the \textit{Reference} data. This is also observed in Table \ref{mom_thickness}, where the momentum thickness $\theta/D$ has been reported for three representative testing samples.  

\begin{figure}
    \centering
    \subfigure[]{\includegraphics[width=0.31\textwidth]{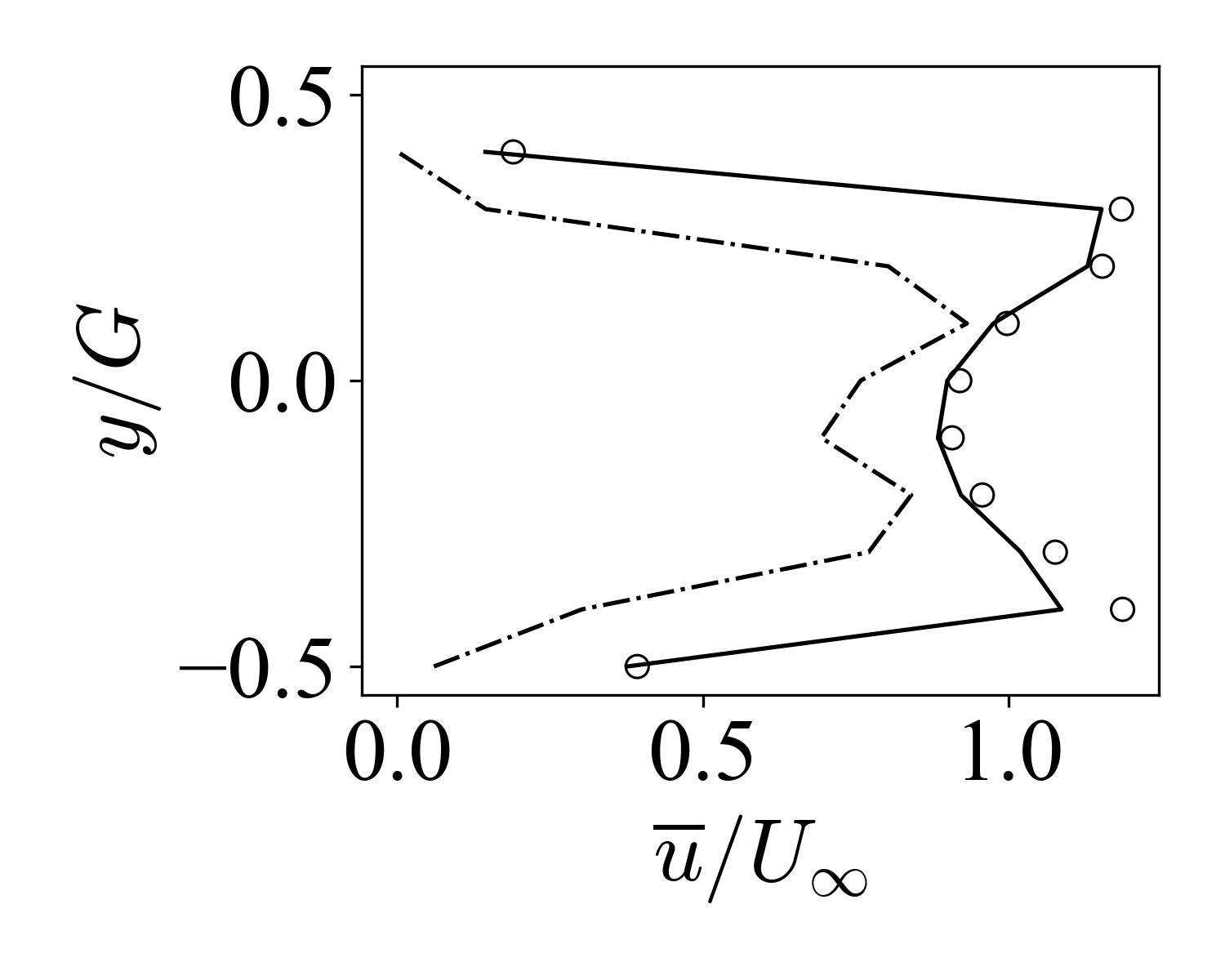}} 
    \subfigure[]{\includegraphics[width=0.31\textwidth]{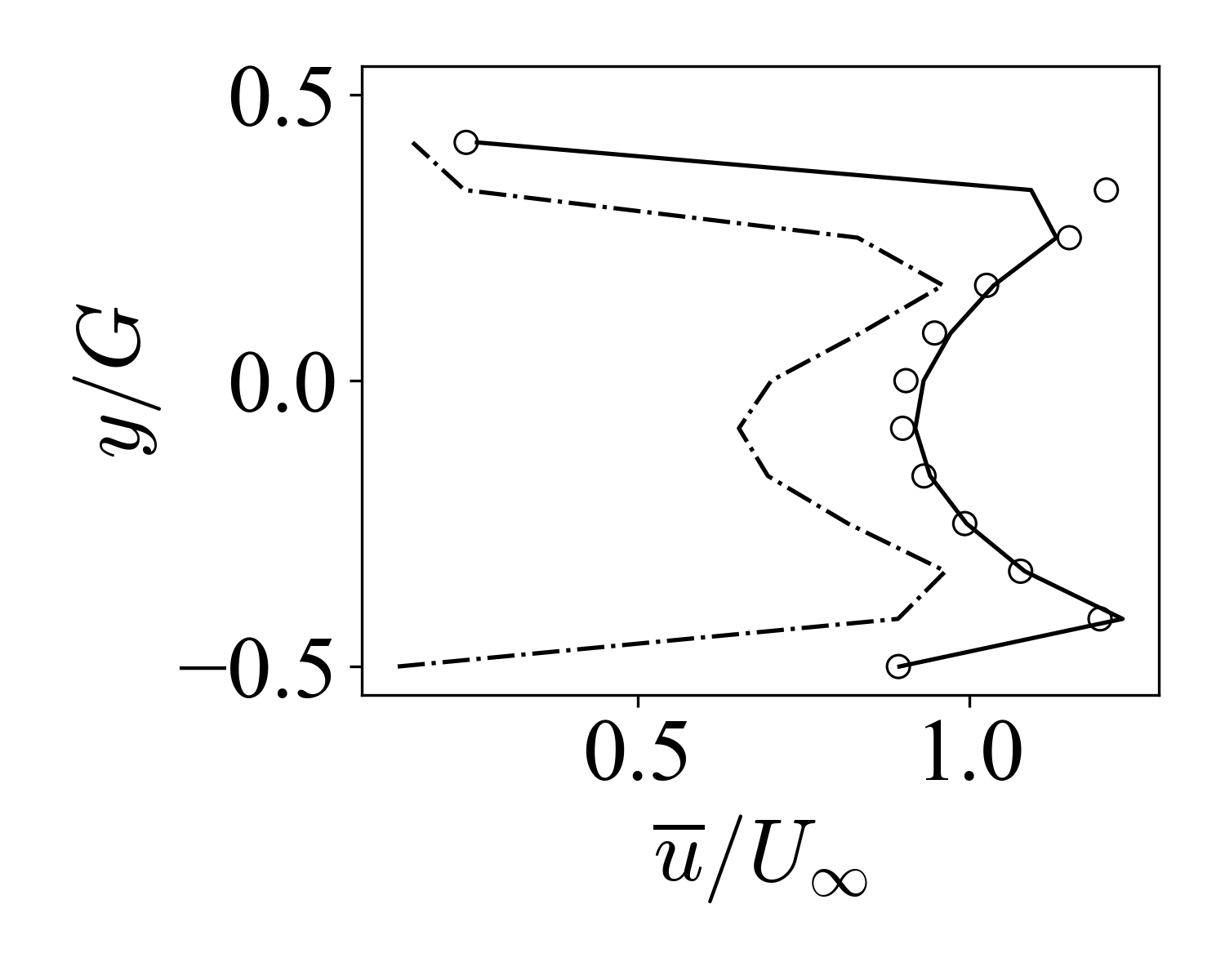}} 
    \subfigure[]{\includegraphics[width=0.31\textwidth]{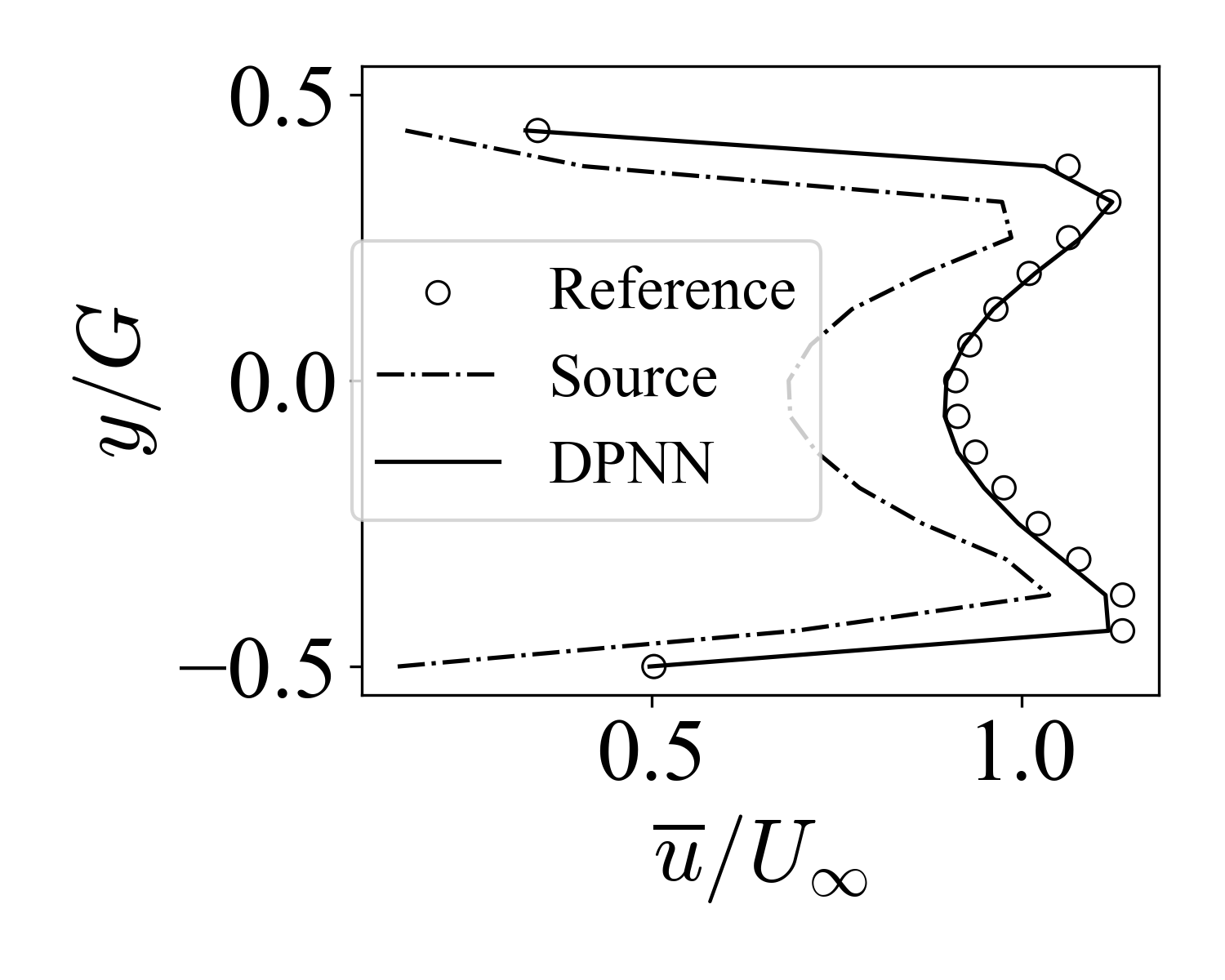}}
    \caption{Comparison of the time-averaged streamwise gap velocity profile in between the two downstream bodies for (a) 28th (b) 40th (c) 46th testing samples}
    \label{gapMvel}
\end{figure}

\begin{table}
\begin{center}
\begin{tabular}{cccc} \\
                & \multicolumn{3}{l}{Mean streamwise velocity at the gap, $\frac{U_\mathrm{G}}{U_{\mathrm{\infty}}}$} \\
Testing samples no & \textit{Reference}            & \textit{Source}             & DPNN               \\
28               & 0.896                  & 0.530               & 0.857              \\
40               & 0.956                  & 0.657              & 0.958              \\
46               & 0.958                  & 0.728              & 0.946             
\end{tabular}
\caption{Comparison of the time and space averaged streamwise velocity at the gap $\frac{U_\mathrm{G}}{U_{\mathrm{\infty}}}$ for multiple testing samples}
\label{meanVelgap}
\end{center}
\end{table}

Figure \ref{gapMvel} shows the streamwise mean gap velocity profile distribution between the two downstream regions. The prominent peaks adjacent to the parabolic profile near the center are caused by the interaction of the separated shear layer from the upstream cylinder with the gap side shear layers of the downstream cylinders. This interaction penetrates the gap flow. There seems to be a very good agreement of the mean streamwise velocity profile at this gap obtained from the DPNN approach with that of the \textit{Reference}, whereas the thicker boundary layer profile exhibited by the \textit{Source} results in a narrower gap profile. Finally, we also quantify the time and space averaged streamwise velocity at the gap $\frac{U_\mathrm{G}}{U_{\mathrm{\infty}}}$ in Table \ref{meanVelgap}, where a very good agreement between DPNN and \textit{Reference} data has been noted.

\begin{table}
\begin{center}
\begin{tabular}{cccc} \\
                & \multicolumn{3}{l}{Mean enstrophy, ${\Omega/ \Omega_0} $} \\
Testing samples no & \textit{Reference}            & \textit{Source}             & DPNN               \\
28               & 1.001                  & 0.926              & 0.985              \\
40               & 0.998                 & 0.944              & 0.979              \\
46               & 0.999                  & 0.988              & 0.997              
\end{tabular}
\caption{Comparison of the normalized time averaged enstrophy ${\Omega / \Omega_0}$ for multiple testing samples}
\label{mean_ens}
\end{center}
\end{table}

\begin{table}
\begin{center}
\begin{tabular}{cccc} \\
                & \multicolumn{3}{l}{Mean kinetic energy, ${ KE / KE_0}$ } \\
Testing samples no & \textit{Reference}            & \textit{Source}             & DPNN               \\
28               & 1.000                  & 0.924              & 0.991              \\
40               & 1.000                  & 0.929              & 0.995              \\
46               & 1.000                  & 0.927              & 0.996              
\end{tabular}
\caption{Comparison of the normalized time averaged kinetic energy ${ KE / KE_0}$ for multiple testing samples}
\label{mean_KE}
\end{center}
\end{table}

To enable a comparison between global variables that represent interesting fluid phenomena, we focus on assessing the enstrophy, symbolized by $\Omega$, and the kinetic energy, represented by $KE$. Tables \ref{mean_ens} and \ref{mean_KE} present the time-averaged enstrophy $\Omega$ and kinetic energy $KE$ obtained from multiple benchmarks. It is found that the flowfields obtained from \textit{Source} exhibit considerably lower values of $\Omega$ and $KE$, indicating loss of rotational and kinetic energy over time. In other words, the flowfields obtained from \textit{Source} result in significant decay of kinetic and rotational energy due to stronger dissipation effects resulting from the coarse grid simulations. In contrast, however, the time-averaged enstrophy and kinetic energy obtained from DPNN are in excellent agreement with the data obtained from the \textit{Reference} solver. Given the fact that the same numerical schemes are employed in the base solver for both the \textit{Source} and the DPNN (i.e., the solver embedded in the loop), the network thus acts as a forcing function that counters the strong numerical dissipation introduced by the \textit{Source} solver. This exercise, thus, clearly points towards the ability of the DPNN approach to yield predictions that preserve the local boundary layer profile along with the global wake dynamics in a way that results in lower numerical dissipation, as achieved by the \textit{Reference} solver. So far, we have demonstrated the reliability of the model at an individual prediction level. 
Additional evaluations for all the 50 previously unseen test samples have also been performed and are shown in Appendix \ref{compre}, supporting the conclusions drawn above.

\subsection{Insights through model predictions}

\begin{figure}
\centering
\includegraphics[scale=0.6, angle=0]{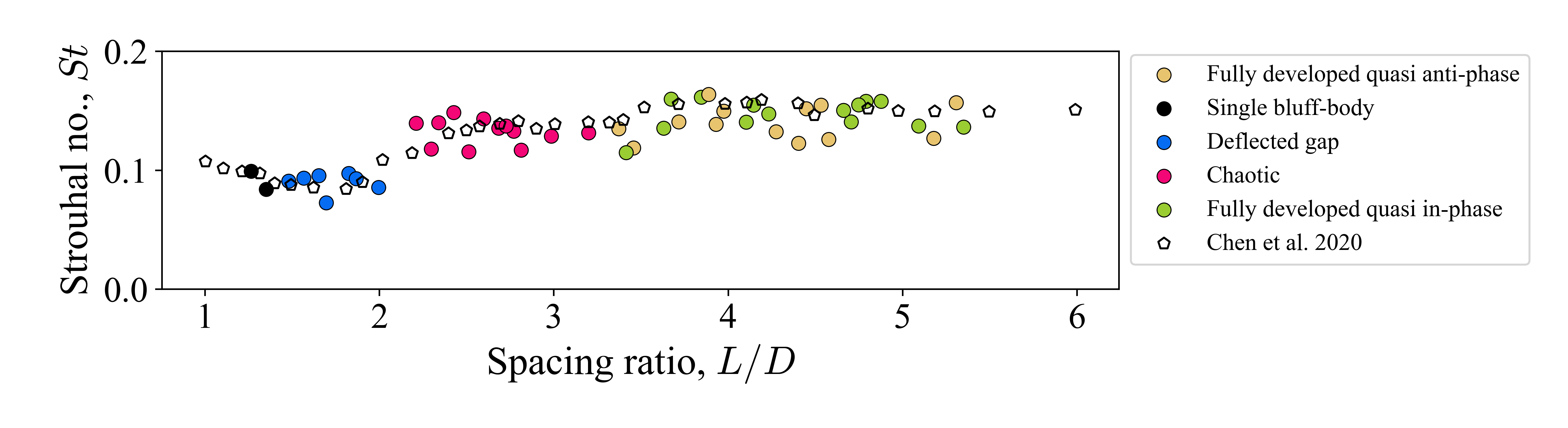}
\caption{Strouhal number along with the spacing ratio for the arbitrarily shaped cylinders and its comparison with Chen et al. 2020. Strouhal number is evaluated based on the dominant frequency of the downstream (upper) cylinder obtained for 300 frames}
\label{st_compare}
\end{figure}

A key aspect of our research is the model's ability to produce not only robust but also practical and valuable predictions. As we advance, we will undertake a comprehensive evaluation to further substantiate the model's reliability in producing results comparable to established literature. Specifically, we will compare non-dimensional parameters such as the Strouhal number ($St$) and spacing ratio ($L/D$), which are typically associated with the flow around two or more cylinders. This broadens our evaluation beyond individual testing samples, as was conducted in previous sections. To our knowledge, this marks the first occasion when such an investigation has been carried out for arbitrarily shaped bodies. 


Despite the fact that the present dataset corresponds to flowfields obtained for flow past arbitrary configurations, one can still compare the distributions of the Strouhal number $St$ based on the dominant frequency of one of the two downstream cylinders with the spacing ratio $L/D$. The diameter in the spacing ratio $L/D$ is simply the projected height of the upstream cylinder. This results in the distribution of the non-dimensional number (\textit{i.e.,} Strouhal number and spacing ratio), which allows for comparison with \citep{chen2020numerical}, who report the data obtained for flow past equi-diameter cylinders computed using an immersed boundary-based flow solver. Figure \ref{st_compare} compares the Strouhal number based on the dominant frequency of the downstream cylinder (upper) as a function of the spacing ratio $L/D$. 
Figure \ref{st_compare} offers multiple insights, viz., besides the observable oscillations in the Strouhal number, a good overall fit between the solutions from the two different methodologies is noted. This illustrates that despite the arbitrary nature of the embedded bodies, the underlying frequency of vortex shedding (non-dimensionalized by the corresponding diameter and freestream velocity) for a certain body at a given spacing ratio exhibits similar patterns. Besides, it is also noted that the transition from low-frequency single bluff-body wake to deflected gap wake and subsequently to high-frequency, chaotic wake happens at nearly the same spacing ratio (as compared to \cite{chen2020numerical}). This observation is especially important given that the Strouhal number accuracy is crucially dependent on long-term recursive predictions, and deviations cause significant differences in the vortex shedding frequency \citep{hasegawa2020machine} \textit{e.g.,} from \textit{Source} (not shown here). Consequently, it can be remarked that the present DPNN-based predictive framework allow for useful and reliable predictions enabling long-term temporal accuracy. While multiple factors, such as shear layer growth, instabilities, gap flow, etc., contribute to the frequency of vortex shedding, further investigations are needed to probe the cause of the fluctuations. 

\begin{figure}
\centering
\includegraphics[scale=0.6, angle=0]{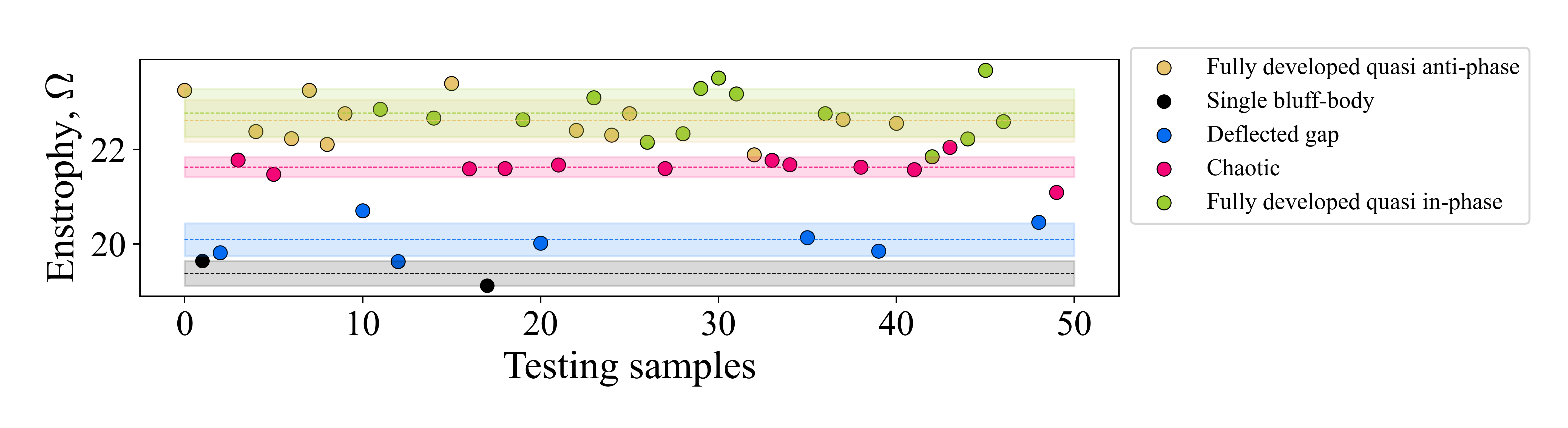}
\caption{Comparison of the enstrophy $\Omega$ of the predictions obtained from DPNN color-coded by the wake category. The color bars represent the region within the maximum and the minimum $\Omega$ for each wake category. }
\label{category_ens}
\end{figure}

In prior sections, we emphasized the necessity of evaluating the performance of the predictive model using test samples representative of different wake categories. In this section, we expand on this by presenting additional metrics of interest. 
This exercise allows for inferring greater physical insights from the predictions obtained so far, thus offering a unique perspective on solutions obtained from the predictive framework. Figure \ref{category_ens} represents the time-averaged enstrophy $\Omega$ evaluated over the spacial domain for 300 testing frames for all testing samples, color marked by the category of the wake. It is found that there exist clear strata or bands of regions belonging to certain wake categories, \textit{e.g.,} single bluff-body, deflected gap, and chaotic wake, whereas these bands overlap for the fully developed quasi in/anti-wake categories. This indicates that the mean rotational energy of the flowfields is lowest for the single bluff-body wake and highest for the fully-developed quasi in-phase wake. This is not entirely surprising as the vortices shed from the upstream cylinder in the fully developed wake category penetrate the gap region while merging with vortices shed from the downstream cylinder, thereby increasing the rotational energy of the flow. 

\begin{figure}
\centering
\includegraphics[scale=0.6, angle=0]{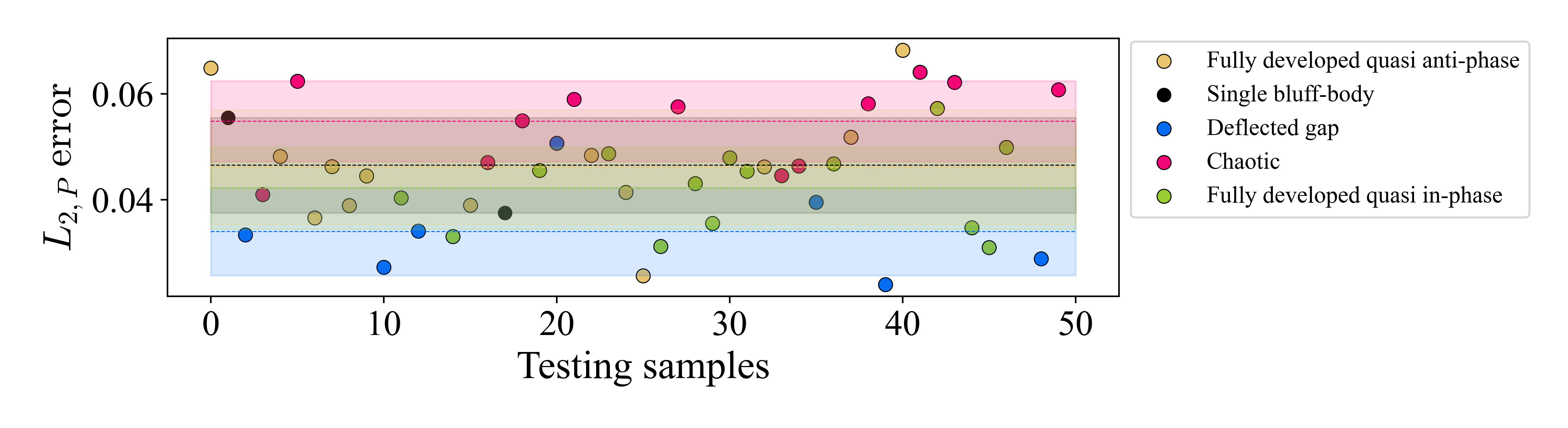}
\caption{Comparison of the $l_2$ norm error of the predictions obtained from DPNN}
\label{category_errP}
\end{figure}

\begin{figure}
\hspace{-0.2cm}
\includegraphics[scale=0.6, angle=0]{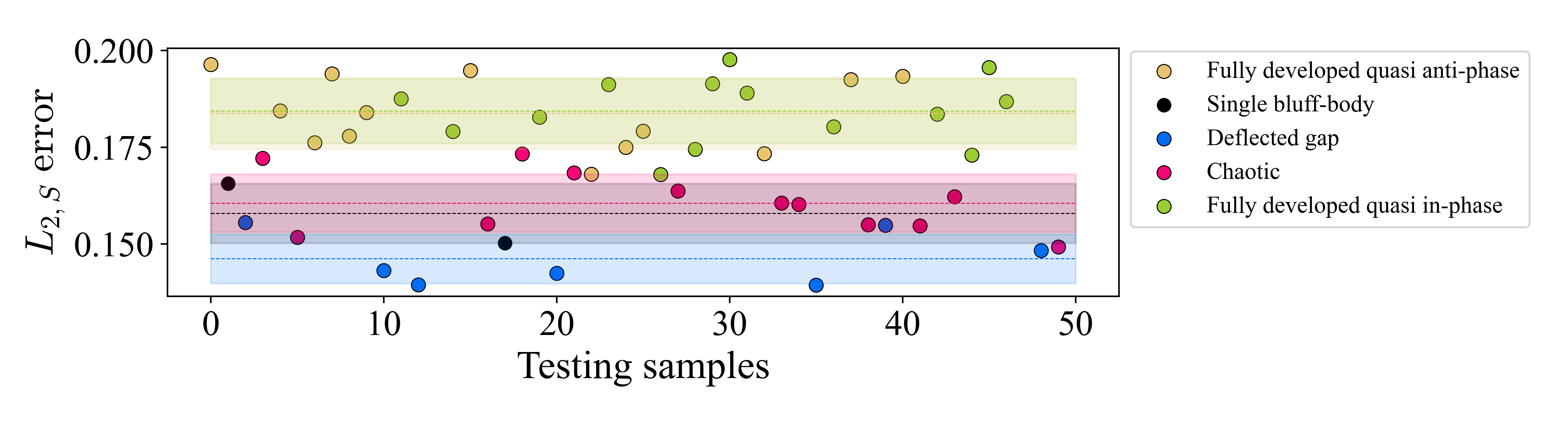}
\caption{Comparison of the $l_2$ norm error of the predictions obtained from \textit{Source}}
\label{category_errS}
\end{figure}

\begin{equation}
\label{l2_err}
L_{2,\phi} = \sqrt{  \frac{ \sum_{i=1}^{nc}  (\phi_{\mathrm{Ref}, i}^j - \phi_{\mathrm{baseline}, i}^j)^2 } {n_x \times n_y}}
\end{equation}

\begin{figure}
\centering
	\includegraphics[width=0.75\textwidth]{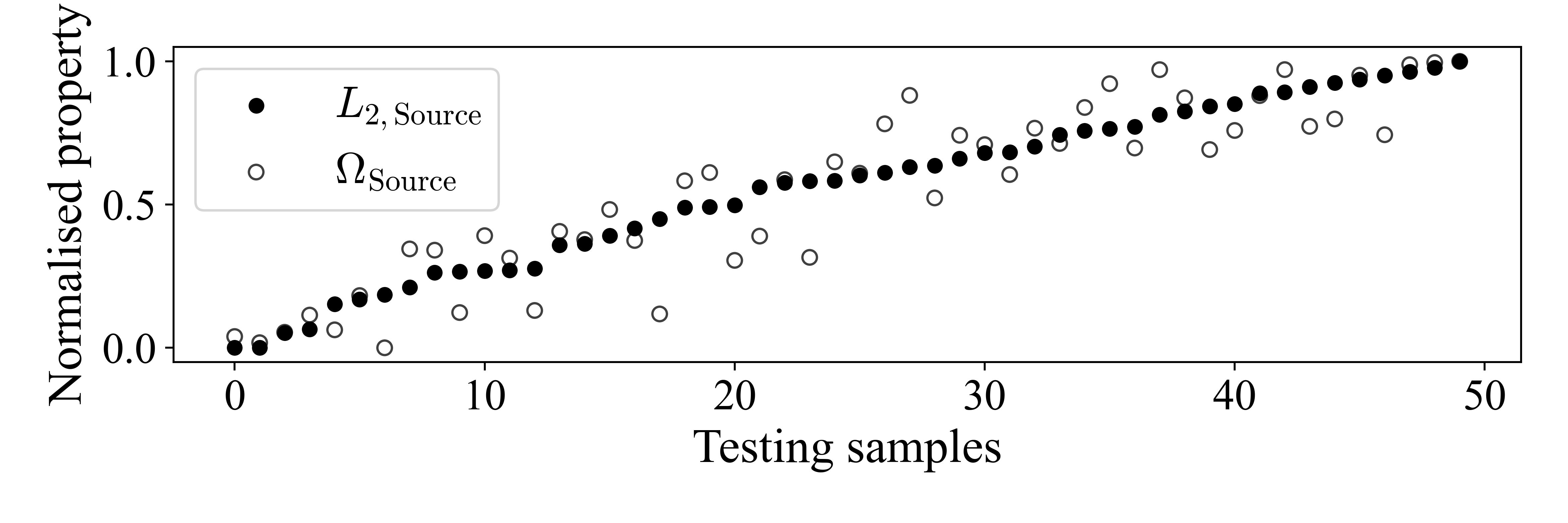}
    	\caption{Comparison of the enstrophy with $L_2$ norm of the predictions from \textit{Source} solver, sorted in ascending manner}
\label{category_ens_l2}
\end{figure}

Figure \ref{category_errP} illustrates the time-averaged $L_2$ norm error computed using Eq. \ref{l2_err} for each testing sample based on the velocity field up to $t_f$=300 frames. The variables $n_x$, $n_y$ are the numbers of computational cells in the streamwise and lateral directions, and $\phi$ represents the quantity of interest.  The samples are color-marked according to the wake category they exhibit. Unsurprisingly, it is found that the testing samples belonging to the chaotic wakes incur the highest error. Additionally, the testing samples corresponding to the deflected gap wake result in the lowest error from both DPNN and Source, which can be attributed to the stable gap flow (see Fig. \ref{category_errS}). On the contrary, the testing samples exhibiting fully developed quasi in-phase wake result in the highest error from the \textit{Source}. To probe further, in Fig. \ref{category_ens_l2}, we compare the normalized enstrophy $\Omega$ for the flowfields obtained from \textit{Source} and the corresponding $L_2$ norm error, each in ascending order. It is found that there exists a clear relationship between the dissipation resulting from the higher enstrophy in the flowfields and the corresponding error. This suggests that the error in predictions obtained from \textit{Source} is predominantly influenced by the numerical dissipation. 


\subsection{Generalizability towards out-of-distribution cases}

A crucial component towards determining the performance of a neural network-based predictive model is its generalizability to new, unseen datasets. 
As such, an evaluation with test data precludes inherent biases of the training dataset. 
Evaluations on out-of-distribution (OOD) datasets allows for estimating the robustness of the model while potentially identifying the model's inherent bias towards certain kind of data. In addition, such practices hint towards the applicability of the model to real-world data, which are likely to be OOD. We have carefully designed experiments for OOD datasets which allow us to perform the following:

\begin{itemize}
  \item \textbf{Three equi-diameter cylinders}. This setup mimics the existing one as shown in Fig. \ref{cyl_layout} with the exception of the geometric symmetry induced by three equi-diameter cylinders at spacing ratios of $L/D$ = [1.5, 3.5, 5.0] with each cylinder having unit diameter. This serves as an interesting problem considering that the model was never trained for symmetry (the equi-diameter cylinders result in symmetric distribution about the x-axis). Additionally, this allows for a direct comparison of the wake category with that obtained by \citet{zheng2016numerical}. For the chosen spacing ratio, the flowfield should exhibit deflected gap wake, anti-phase wake, and fully-developed in-phase wake, respectively. 

  \item \textbf{Two side-by-side cylinders}. This layout chooses two side-by-side equi-diameter cylinders which is contrary to the present training setup to assess the performance of the model for an entirely alternative problem of practical relevance. The spacing ratios chosen for this case are $L/D$ = [1.5, 2.5, 4.0], which allow for direct comparison with \citet{bao2013flow} and should yield flip-flopping (or chaotic) wake, in-phase wake, and anti-phase wake, respectively. 
\end{itemize}

\subsubsection{Test case 1: three cylinders}

\begin{figure}
\hspace{-1.5cm}
    \includegraphics[width=1.25\textwidth]{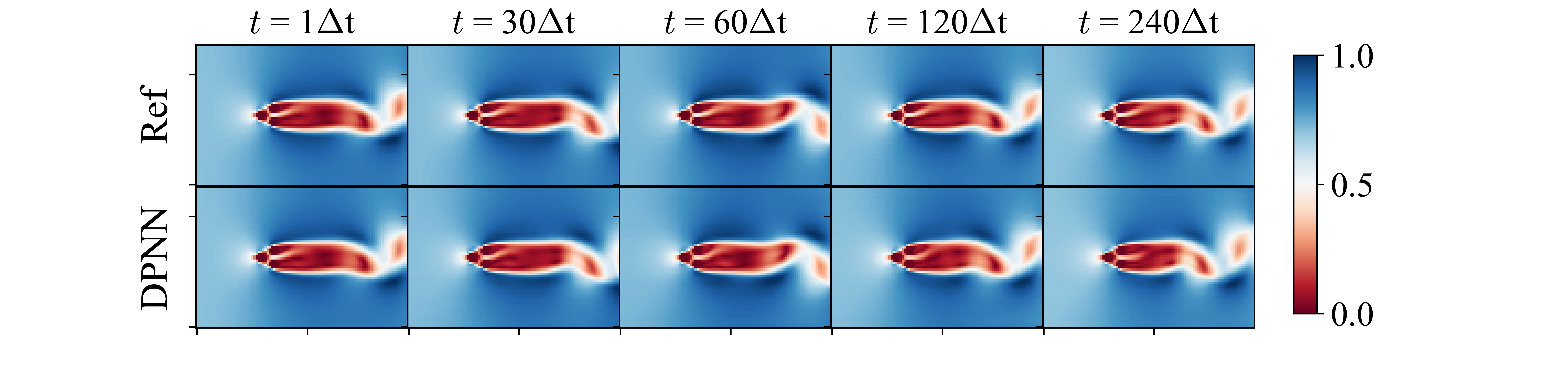}
    \caption{Comparison of velocity flowfield obtained from DPNN and \textit{Reference} solver at multiple time instances for flow past the three-cylinder case at $L/D$ = 1.5}
\label{gen_vel_3cyl}
\end{figure}

\begin{figure}
\hspace{-1.5cm}
    \includegraphics[width=1.25\textwidth]{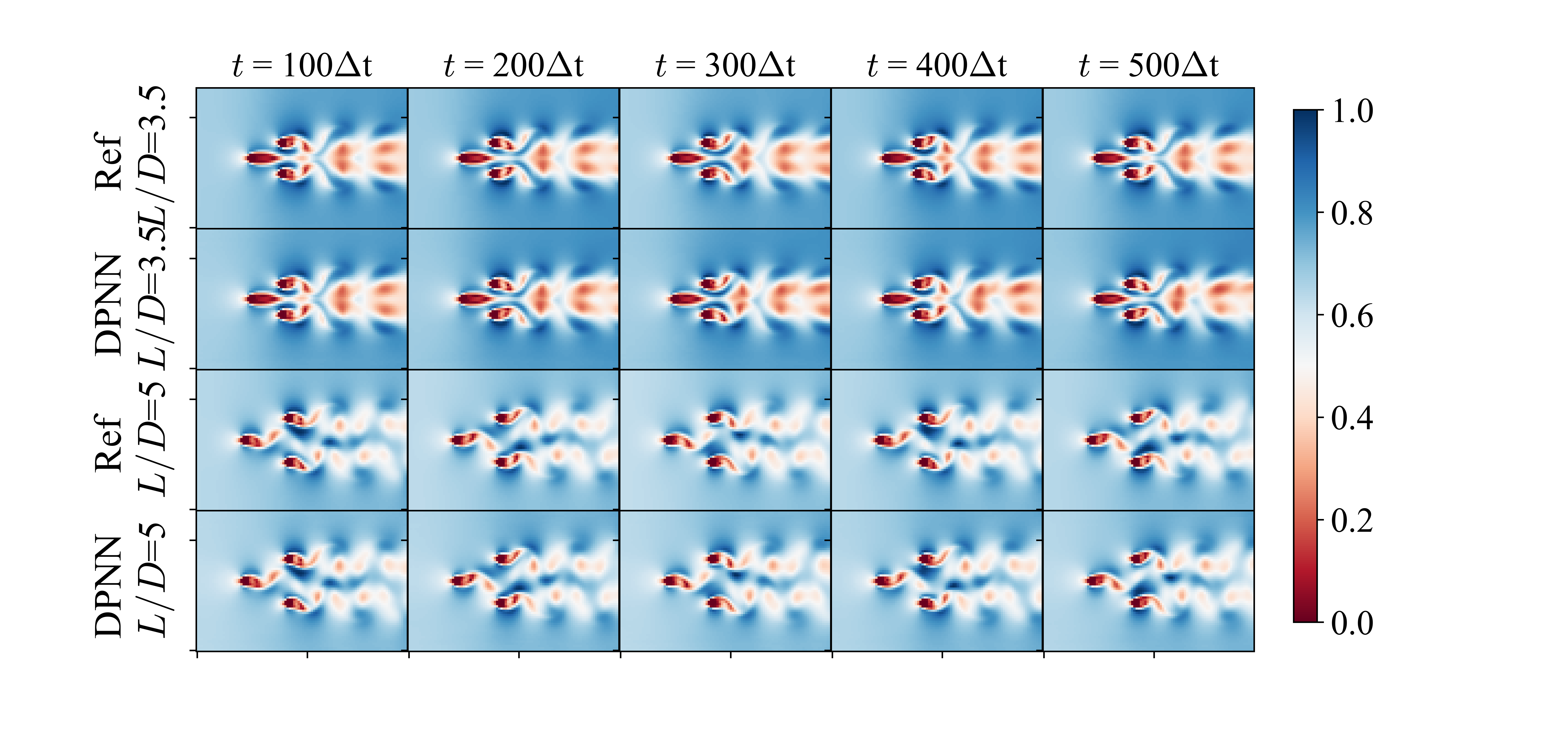}
    \caption{Comparison of velocity flowfield obtained from DPNN and \textit{Reference} solver at multiple time instances for flow past the three-cylinder case at $L/D$ = 3.5 and 5}
\label{gen_vel_3cyl2}
\end{figure}

Figure \ref{gen_vel_3cyl} presents the velocity flowfields obtained from the DPNN approach and its comparisons with the data obtained from the \textit{Reference} solver at multiple time instances, $\textit{i.e.,}$ three equi-diameter cylinders arranged in an equilateral triangle layout for the spacing ratio $L/D$ = 1.5. The \textit{Reference} (or ground truth data) solver depicts a deflected gap flow wake, with the deflected gap flow being deflected towards the narrow wake region (upper cylinder), which bodes well with the observations by \citet{zheng2016numerical}. It is found that the predictions obtained by the present DPNN approach also result in the same wake dynamics. In addition, the vortex shedding cycle result in an excellent agreement with the data obtained from the \textit{Reference} solver, which instils further confidence in the long-term temporal predictions. 


Figure \ref{gen_vel_3cyl2} presents the velocity flowfields obtained from the present approach and its comparison with the data obtained from the \textit{Reference} solver at the spacing ratios $L/D$ = 3.5 and 5, up to a total time of $t=500\Delta t$. The wake exhibited at these spacing ratios clearly depicts anti-phase and fully developed in-phase patterns, respectively, as reported by \citet{zheng2016numerical}. The comparison shows that DPNN is able to preserve the overall wake structures remarkably well for long time horizons, irrespective of either in-phase or fully developed anti-phase wake. 

\subsubsection{Test case 2: two cylinders}

\begin{figure}
\hspace{-1.5cm}
    \includegraphics[width=1.25\textwidth]{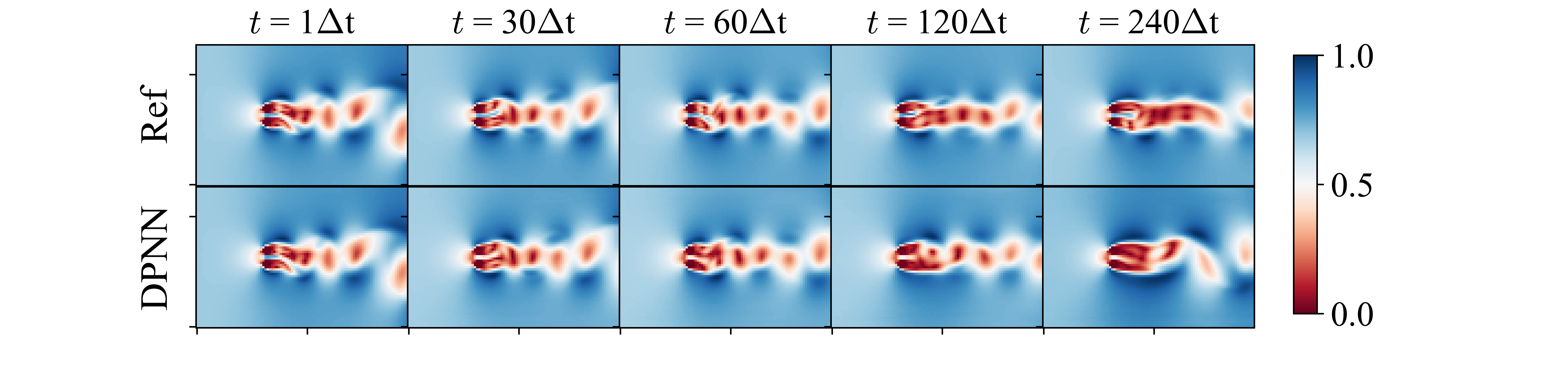}
    \caption{Comparison of velocity flowfield obtained from DPNN and \textit{Reference} solver at multiple time instances for flow past the two cylinder case at $L/D$ = 1.5}
\label{gen_vel_2cyl}
\end{figure}

\begin{figure}
\hspace{-1.5cm}
    \includegraphics[width=1.25\textwidth]{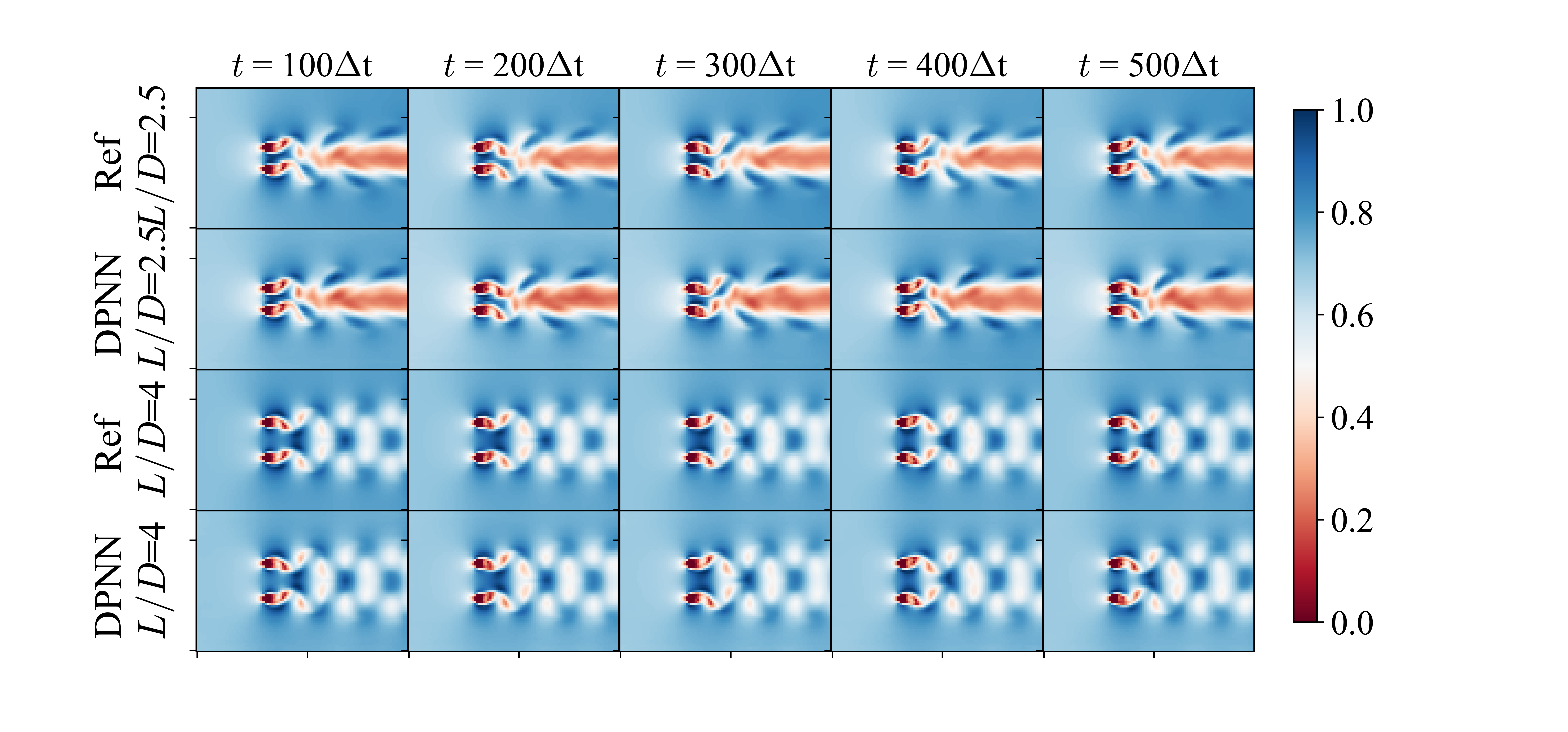}
    \caption{Comparison of velocity flowfield obtained from DPNN and \textit{Reference} solver at multiple time instances for flow past the two cylinder case at $L/D$ = 1.5}
\label{gen_vel_2cyl2}
\end{figure}

Figure \ref{gen_vel_2cyl} highlights the velocity flowfield obtained from the present approach and its comparison with the corresponding ground truth data at multiple time instances, $\textit{i.e.,}$ two side-by-side cylinders at a spacing ratio $L/D$ = 1.5. The data obtained from the \textit{Reference} solver exhibits flip-flopping wake pattern wherein the gap flow switches direction in a chaotic manner and augers well with the observation by \citet{bao2013flow}. This can be clearly seen in Fig. \ref{gen_vel_2cyl}, \textit{e.g.,} the gap flow at $t=1\Delta t$  (directed towards the lower cylinder) switches direction at $t=30\Delta t$ (directed towards the upper cylinder). Addressing such a tumultuous shift poses a significant challenge for any predictive framework (\cite{vlachas2018data,doan2021short}). For instance, while DPNN accurately predicts the first two shifts, its effectiveness diminishes when it comes to long-term forecasting, particularly struggling with the accurate reconstruction of chaotic wake dynamics. Nevertheless, considering the fact that the proposed framework yields satisfactory performance for predicting the chaotic dynamics for mid-time horizons, this acts as a promising direction for future research. 

Figure \ref{gen_vel_2cyl2} presents the velocity flowfield obtained from the present DPNN approach and its comparison with the \textit{Reference} data for spacing ratios $L/D$ = 2.5 and 4 for multiple time instances. The wake dynamics exhibit in-phase and anti-phase wake, respectively, which bodes well with the observations made by \citet{bao2013flow}. It is found that the predictions from the present approach compare really well with the data obtained from the \textit{Reference} solver, especially for the anti-phase wake at $L/D$=4 where it is seen to preserve the symmetry of the flow about the centerline $x$-axis. For the in-phase wake obtained at $L/D$=2.5, the flowfields diverge at long-time predictions \textit{i.e.,} $t=500\Delta t$. Therefore, given these observations, it is encouraging to note that the current DPNN approach is successful in faithfully reconstructing wake dynamics for periodic flows (both in-phase and anti-phase) over extended time horizons. Although the accuracy of predictions for chaotic wakes is limited to mid-time horizons, the progress made so far offers a solid foundation for further refinement and optimization.

\section{Conclusions}

This work implements a differentiable physics-assisted neural network (DPNN) approach as a coarse-grained surrogate for unsteady incompressible flow past arbitrarily shaped bodies for a low Reynolds number of around 100. The bodies under consideration are arranged in a staggered equilateral triangle manner at multiple spacing ratio 1.2 $\leq$ $L/D$ $\leq$ 5.5 to produce wakes of varying nature. The main conclusions from the work can be summarized as follows:

\begin{enumerate}
	\item The DPNN-based predictive framework outperforms the standard data-driven supervised learning approach and the \textit{Source} as a robust coarse-grained surrogate for unsteady fluid flows past arbitrary bodies. This has been qualitatively and quantitatively verified using multiple performance measuring indices that indicate accurate reconstruction of local (boundary layer) and global (wake dynamics) fluid phenomenon and for both physical quantities and Fourier space or reduced feature space. 
	
	\item The framework has also been evaluated for out-of-distribution test samples for cases that assess the ability of the predictive framework to preserve symmetry (\textit{i.e.,} anti-phase wake), stable gap flow (\textit{i.e.,} deflected gap wake), long-term temporal stability, chaotic dynamics ($\textit{i.e.}$, flip-flopping wake). It is found that the framework is largely wake agnostic, \textit{i.e.,} it renders a good prediction irrespective of the wake category. However, for the chaotic or flip-flopping wake, the framework can only accurately maintain the chaotic switching of the gap flow up to a certain mid-time range. 
	
	\item The $L_2$ errors computed on the predictions obtained from DPNN and \textit{Source} are lowest for the deflected gap wake owing to its stable gap flow as well as long vortex formation length, the latter of which occupies the bulk of the computational space. The DPNN-based predictions seem to suffer from the onset of chaotic dynamics, which later manifest into inaccurate wake dynamics. The flowfields from \textit{Source} seem to suffer from large dissipation due to the underlying coarse computational grid, which results in the continuous decay of kinetic/rotational energy. 
	
	\item The Strouhal number distributions with the spacing ratio for the arbitrary bodies results in good agreement with the literature based on flow past equi-diameter bodies. Also, the transition in the wake category from a single bluff-body wake to a deflected gap wake and eventually to a chaotic wake occurs at nearly the same spacing ratios. Moreover, we discover similar categories and patterns in the wake exhibited by the arbitrary bodies, except that in-phase and anti-phase wakes yield in quasi in/anti-phase wakes. Additionally, the arbitrary nature of the bodies promotes an early vortex shedding by the upstream cylinder and precludes quasi in-phase wakes from occurring at all. 
	
	\item The enstrophy calculated based on the flowfields obtained from DPNN exhibit clear strata of test samples belonging to unique wake categories. This is particularly noticeable for single bluff-body, deflected gap, and chaotic wakes, whereas these delineations tend to overlap for the fully developed quasi in/anti-phase wake categories. 

\end{enumerate}

In view of these discussions, we postulate that such a strategy that incorporates both a low-fidelity solver as well as the neural network as a hybrid predictive framework derives merit from each component. Such a strategy paves the way for the low-fidelity solver to incorporate as much of the underlying physics as possible while the network learns the difference between low-fidelity solver and high-fidelity data. This specific attribute is key that makes such a framework more promising than purely data-driven supervised learning. 

The present approach also shows potential merits as a generic reconstruction strategy for near-body boundary fluid properties. This is especially true for non body-conformal sharp interface immersed boundary approaches for high Mach and Reynolds numbers flows \citep{brahmachary2021role} that are prone to suffer from heat flux reconstruction at the body boundary. Moreover, the present learning framework would also be interesting avenue for flows over rough surfaces \citep{lee2022predicting,jouybari2021data} with practical relevance. In addition,  the generalizability of the approach opens up the potential directions for learning the kinematics of dispersed spherical particles \citep{ozaki2022prediction}.


\section*{Funding}

This work was supported by the European Research Council Consolidator Grant \textit{SpaTe} (CoG-2019-863850)

\section*{Data availability statement} The data and code supporting this study are openly available in disclosed upon publication.

\section*{Declaration of interests} The authors report no conflict of interest.

\bibliographystyle{elsarticle-harv} 
\bibliography{cas-refs}

\begin{thebibliography}{60}
\expandafter\ifx\csname natexlab\endcsname\relax\def\natexlab#1{#1}\fi
\providecommand{\url}[1]{\texttt{#1}}
\providecommand{\href}[2]{#2}
\providecommand{\path}[1]{#1}
\providecommand{\DOIprefix}{doi:}
\providecommand{\ArXivprefix}{arXiv:}
\providecommand{\URLprefix}{URL: }
\providecommand{\Pubmedprefix}{pmid:}
\providecommand{\doi}[1]{\href{http://dx.doi.org/#1}{\path{#1}}}
\providecommand{\Pubmed}[1]{\href{pmid:#1}{\path{#1}}}
\providecommand{\bibinfo}[2]{#2}
\ifx\xfnm\relax \def\xfnm[#1]{\unskip,\space#1}\fi
\bibitem[{Bao et~al.(2010)Bao, Zhou and Huang}]{bao2010numerical}
\bibinfo{author}{Bao, Y.}, \bibinfo{author}{Zhou, D.}, \bibinfo{author}{Huang,
  C.}, \bibinfo{year}{2010}.
\newblock \bibinfo{title}{Numerical simulation of flow over three circular
  cylinders in equilateral arrangements at low reynolds number by a
  second-order characteristic-based split finite element method}.
\newblock \bibinfo{journal}{Computers \& Fluids} \bibinfo{volume}{39},
  \bibinfo{pages}{882--899}.
\bibitem[{Bao et~al.(2013)Bao, Zhou and Tu}]{bao2013flow}
\bibinfo{author}{Bao, Y.}, \bibinfo{author}{Zhou, D.}, \bibinfo{author}{Tu,
  J.}, \bibinfo{year}{2013}.
\newblock \bibinfo{title}{Flow characteristics of two in-phase oscillating
  cylinders in side-by-side arrangement}.
\newblock \bibinfo{journal}{Computers \& Fluids} \bibinfo{volume}{71},
  \bibinfo{pages}{124--145}.
\bibitem[{Bearman(2011)}]{bearman2011circular}
\bibinfo{author}{Bearman, P.}, \bibinfo{year}{2011}.
\newblock \bibinfo{title}{Circular cylinder wakes and vortex-induced
  vibrations}.
\newblock \bibinfo{journal}{Journal of Fluids and Structures}
  \bibinfo{volume}{27}, \bibinfo{pages}{648--658}.
\bibitem[{Belbute-Peres et~al.(2020)Belbute-Peres, Economon and
  Kolter}]{belbute2020combining}
\bibinfo{author}{Belbute-Peres, F.D.A.}, \bibinfo{author}{Economon, T.},
  \bibinfo{author}{Kolter, Z.}, \bibinfo{year}{2020}.
\newblock \bibinfo{title}{Combining differentiable pde solvers and graph neural
  networks for fluid flow prediction}, in: \bibinfo{booktitle}{international
  conference on machine learning}, \bibinfo{organization}{PMLR}. pp.
  \bibinfo{pages}{2402--2411}.
\bibitem[{Brahmachary(2019)}]{brahmachary2019finite}
\bibinfo{author}{Brahmachary, S.}, \bibinfo{year}{2019}.
\newblock \bibinfo{title}{Finite volume/immersed boundary solvers for
  compressible flows: development and applications}.
\newblock Ph.D. thesis. Ph. D. thesis dissertation, Indian Institute of
  Technology Guwahati.
\bibitem[{Brahmachary et~al.(2021)Brahmachary, Natarajan, Kulkarni, Sahoo,
  Ashok and Kumar}]{brahmachary2021role}
\bibinfo{author}{Brahmachary, S.}, \bibinfo{author}{Natarajan, G.},
  \bibinfo{author}{Kulkarni, V.}, \bibinfo{author}{Sahoo, N.},
  \bibinfo{author}{Ashok, V.}, \bibinfo{author}{Kumar, V.},
  \bibinfo{year}{2021}.
\newblock \bibinfo{title}{Role of solution reconstruction in hypersonic viscous
  computations using a sharp interface immersed boundary method}.
\newblock \bibinfo{journal}{Physical Review E} \bibinfo{volume}{103},
  \bibinfo{pages}{043302}.
\bibitem[{Brandstetter et~al.(2022)Brandstetter, Worrall and
  Welling}]{brandstetter2022message}
\bibinfo{author}{Brandstetter, J.}, \bibinfo{author}{Worrall, D.},
  \bibinfo{author}{Welling, M.}, \bibinfo{year}{2022}.
\newblock \bibinfo{title}{Message passing neural pde solvers}.
\newblock \bibinfo{journal}{arXiv preprint arXiv:2202.03376} .
\bibitem[{Brunton et~al.(2020)Brunton, Noack and
  Koumoutsakos}]{brunton2020machine}
\bibinfo{author}{Brunton, S.L.}, \bibinfo{author}{Noack, B.R.},
  \bibinfo{author}{Koumoutsakos, P.}, \bibinfo{year}{2020}.
\newblock \bibinfo{title}{Machine learning for fluid mechanics}.
\newblock \bibinfo{journal}{Annual review of fluid mechanics}
  \bibinfo{volume}{52}, \bibinfo{pages}{477--508}.
\bibitem[{Chen et~al.(2021)Chen, Cakal, Hu and Thuerey}]{chen2021numerical}
\bibinfo{author}{Chen, L.W.}, \bibinfo{author}{Cakal, B.A.},
  \bibinfo{author}{Hu, X.}, \bibinfo{author}{Thuerey, N.},
  \bibinfo{year}{2021}.
\newblock \bibinfo{title}{Numerical investigation of minimum drag profiles in
  laminar flow using deep learning surrogates}.
\newblock \bibinfo{journal}{Journal of Fluid Mechanics} \bibinfo{volume}{919},
  \bibinfo{pages}{A34}.
\bibitem[{Chen et~al.(2020)Chen, Ji, Alam, Williams and Xu}]{chen2020numerical}
\bibinfo{author}{Chen, W.}, \bibinfo{author}{Ji, C.}, \bibinfo{author}{Alam,
  M.M.}, \bibinfo{author}{Williams, J.}, \bibinfo{author}{Xu, D.},
  \bibinfo{year}{2020}.
\newblock \bibinfo{title}{Numerical simulations of flow past three circular
  cylinders in equilateral-triangular arrangements}.
\newblock \bibinfo{journal}{Journal of Fluid Mechanics} \bibinfo{volume}{891},
  \bibinfo{pages}{A14}.
\bibitem[{Cheng et~al.(2020)Cheng, Fang, Pain and Navon}]{cheng2020data}
\bibinfo{author}{Cheng, M.}, \bibinfo{author}{Fang, F.}, \bibinfo{author}{Pain,
  C.C.}, \bibinfo{author}{Navon, I.}, \bibinfo{year}{2020}.
\newblock \bibinfo{title}{Data-driven modelling of nonlinear spatio-temporal
  fluid flows using a deep convolutional generative adversarial network}.
\newblock \bibinfo{journal}{Computer Methods in Applied Mechanics and
  Engineering} \bibinfo{volume}{365}, \bibinfo{pages}{113000}.
\bibitem[{Choi et~al.(2008)Choi, Jeon and Kim}]{choi2008control}
\bibinfo{author}{Choi, H.}, \bibinfo{author}{Jeon, W.P.}, \bibinfo{author}{Kim,
  J.}, \bibinfo{year}{2008}.
\newblock \bibinfo{title}{Control of flow over a bluff body}.
\newblock \bibinfo{journal}{Annu. Rev. Fluid Mech.} \bibinfo{volume}{40},
  \bibinfo{pages}{113--139}.
\bibitem[{Chorin(1968)}]{chorin1968numerical}
\bibinfo{author}{Chorin, A.J.}, \bibinfo{year}{1968}.
\newblock \bibinfo{title}{Numerical solution of the navier-stokes equations}.
\newblock \bibinfo{journal}{Mathematics of computation} \bibinfo{volume}{22},
  \bibinfo{pages}{745--762}.
\bibitem[{Constant et~al.(2017)Constant, Favier, Meldi, Meliga and
  Serre}]{constant2017immersed}
\bibinfo{author}{Constant, E.}, \bibinfo{author}{Favier, J.},
  \bibinfo{author}{Meldi, M.}, \bibinfo{author}{Meliga, P.},
  \bibinfo{author}{Serre, E.}, \bibinfo{year}{2017}.
\newblock \bibinfo{title}{An immersed boundary method in openfoam: verification
  and validation}.
\newblock \bibinfo{journal}{Computers \& Fluids} \bibinfo{volume}{157},
  \bibinfo{pages}{55--72}.
\bibitem[{Davydzenka and Tahmasebi(2022)}]{davydzenka2022high}
\bibinfo{author}{Davydzenka, T.}, \bibinfo{author}{Tahmasebi, P.},
  \bibinfo{year}{2022}.
\newblock \bibinfo{title}{High-resolution fluid--particle interactions: a
  machine learning approach}.
\newblock \bibinfo{journal}{Journal of Fluid Mechanics} \bibinfo{volume}{938},
  \bibinfo{pages}{A20}.
\bibitem[{Doan et~al.(2021)Doan, Polifke and Magri}]{doan2021short}
\bibinfo{author}{Doan, N.A.K.}, \bibinfo{author}{Polifke, W.},
  \bibinfo{author}{Magri, L.}, \bibinfo{year}{2021}.
\newblock \bibinfo{title}{Short-and long-term predictions of chaotic flows and
  extreme events: a physics-constrained reservoir computing approach}.
\newblock \bibinfo{journal}{Proceedings of the Royal Society A}
  \bibinfo{volume}{477}, \bibinfo{pages}{20210135}.
\bibitem[{Duraisamy et~al.(2019)Duraisamy, Iaccarino and
  Xiao}]{duraisamy2019turbulence}
\bibinfo{author}{Duraisamy, K.}, \bibinfo{author}{Iaccarino, G.},
  \bibinfo{author}{Xiao, H.}, \bibinfo{year}{2019}.
\newblock \bibinfo{title}{Turbulence modeling in the age of data}.
\newblock \bibinfo{journal}{Annual review of fluid mechanics}
  \bibinfo{volume}{51}, \bibinfo{pages}{357--377}.
\bibitem[{Fukami et~al.(2019)Fukami, Fukagata and Taira}]{fukami2019super}
\bibinfo{author}{Fukami, K.}, \bibinfo{author}{Fukagata, K.},
  \bibinfo{author}{Taira, K.}, \bibinfo{year}{2019}.
\newblock \bibinfo{title}{Super-resolution reconstruction of turbulent flows
  with machine learning}.
\newblock \bibinfo{journal}{Journal of Fluid Mechanics} \bibinfo{volume}{870},
  \bibinfo{pages}{106--120}.
\bibitem[{Guillaume and LaRue(1999)}]{guillaume1999investigation}
\bibinfo{author}{Guillaume, D.}, \bibinfo{author}{LaRue, J.},
  \bibinfo{year}{1999}.
\newblock \bibinfo{title}{Investigation of the flopping regime with two-,
  three-and four-cylinder arrays}.
\newblock \bibinfo{journal}{Experiments in Fluids} \bibinfo{volume}{27},
  \bibinfo{pages}{145--156}.
\bibitem[{Hasegawa et~al.(2020)Hasegawa, Fukami, Murata and
  Fukagata}]{hasegawa2020machine}
\bibinfo{author}{Hasegawa, K.}, \bibinfo{author}{Fukami, K.},
  \bibinfo{author}{Murata, T.}, \bibinfo{author}{Fukagata, K.},
  \bibinfo{year}{2020}.
\newblock \bibinfo{title}{Machine-learning-based reduced-order modeling for
  unsteady flows around bluff bodies of various shapes}.
\newblock \bibinfo{journal}{Theoretical and Computational Fluid Dynamics}
  \bibinfo{volume}{34}, \bibinfo{pages}{367--383}.
\bibitem[{He et~al.(2016)He, Zhang, Ren and Sun}]{he2016deep}
\bibinfo{author}{He, K.}, \bibinfo{author}{Zhang, X.}, \bibinfo{author}{Ren,
  S.}, \bibinfo{author}{Sun, J.}, \bibinfo{year}{2016}.
\newblock \bibinfo{title}{Deep residual learning for image recognition}, in:
  \bibinfo{booktitle}{Proceedings of the IEEE conference on computer vision and
  pattern recognition}, pp. \bibinfo{pages}{770--778}.
\bibitem[{Huang et~al.(2017)Huang, Liu, Van Der~Maaten and
  Weinberger}]{huang2017densely}
\bibinfo{author}{Huang, G.}, \bibinfo{author}{Liu, Z.}, \bibinfo{author}{Van
  Der~Maaten, L.}, \bibinfo{author}{Weinberger, K.Q.}, \bibinfo{year}{2017}.
\newblock \bibinfo{title}{Densely connected convolutional networks}, in:
  \bibinfo{booktitle}{Proceedings of the IEEE conference on computer vision and
  pattern recognition}, pp. \bibinfo{pages}{4700--4708}.
\bibitem[{Issa(1986)}]{issa1986solution}
\bibinfo{author}{Issa, R.I.}, \bibinfo{year}{1986}.
\newblock \bibinfo{title}{Solution of the implicitly discretised fluid flow
  equations by operator-splitting}.
\newblock \bibinfo{journal}{Journal of computational physics}
  \bibinfo{volume}{62}, \bibinfo{pages}{40--65}.
\bibitem[{Jasak and Tukovic(2015)}]{jasak2015immersed}
\bibinfo{author}{Jasak, H.}, \bibinfo{author}{Tukovic, Z.},
  \bibinfo{year}{2015}.
\newblock \bibinfo{title}{Immersed boundary method in foam: theory,
  implementation and use}.
\bibitem[{Jouybari et~al.(2021)Jouybari, Yuan, Brereton and
  Murillo}]{jouybari2021data}
\bibinfo{author}{Jouybari, M.A.}, \bibinfo{author}{Yuan, J.},
  \bibinfo{author}{Brereton, G.J.}, \bibinfo{author}{Murillo, M.S.},
  \bibinfo{year}{2021}.
\newblock \bibinfo{title}{Data-driven prediction of the equivalent sand-grain
  height in rough-wall turbulent flows}.
\newblock \bibinfo{journal}{Journal of Fluid Mechanics} \bibinfo{volume}{912},
  \bibinfo{pages}{A8}.
\bibitem[{Kingma and Ba(2014)}]{kingma2014adam}
\bibinfo{author}{Kingma, D.P.}, \bibinfo{author}{Ba, J.}, \bibinfo{year}{2014}.
\newblock \bibinfo{title}{Adam: A method for stochastic optimization}.
\newblock \bibinfo{journal}{arXiv preprint arXiv:1412.6980} .
\bibitem[{Kochkov et~al.(2021)Kochkov, Smith, Alieva, Wang, Brenner and
  Hoyer}]{kochkov2021machine}
\bibinfo{author}{Kochkov, D.}, \bibinfo{author}{Smith, J.A.},
  \bibinfo{author}{Alieva, A.}, \bibinfo{author}{Wang, Q.},
  \bibinfo{author}{Brenner, M.P.}, \bibinfo{author}{Hoyer, S.},
  \bibinfo{year}{2021}.
\newblock \bibinfo{title}{Machine learning--accelerated computational fluid
  dynamics}.
\newblock \bibinfo{journal}{Proceedings of the National Academy of Sciences}
  \bibinfo{volume}{118}, \bibinfo{pages}{e2101784118}.
\bibitem[{Krizhevsky et~al.(2017)Krizhevsky, Sutskever and
  Hinton}]{krizhevsky2017imagenet}
\bibinfo{author}{Krizhevsky, A.}, \bibinfo{author}{Sutskever, I.},
  \bibinfo{author}{Hinton, G.E.}, \bibinfo{year}{2017}.
\newblock \bibinfo{title}{Imagenet classification with deep convolutional
  neural networks}.
\newblock \bibinfo{journal}{Communications of the ACM} \bibinfo{volume}{60},
  \bibinfo{pages}{84--90}.
\bibitem[{Kutz(2017)}]{kutz2017deep}
\bibinfo{author}{Kutz, J.N.}, \bibinfo{year}{2017}.
\newblock \bibinfo{title}{Deep learning in fluid dynamics}.
\newblock \bibinfo{journal}{Journal of Fluid Mechanics} \bibinfo{volume}{814},
  \bibinfo{pages}{1--4}.
\bibitem[{Lam and Cheung(1988)}]{lam1988phenomena}
\bibinfo{author}{Lam, K.}, \bibinfo{author}{Cheung, W.}, \bibinfo{year}{1988}.
\newblock \bibinfo{title}{Phenomena of vortex shedding and flow interference of
  three cylinders in different equilateral arrangements}.
\newblock \bibinfo{journal}{Journal of Fluid Mechanics} \bibinfo{volume}{196},
  \bibinfo{pages}{1--26}.
\bibitem[{Lee et~al.(2022)Lee, Yang, Forooghi, Stroh and
  Bagheri}]{lee2022predicting}
\bibinfo{author}{Lee, S.}, \bibinfo{author}{Yang, J.},
  \bibinfo{author}{Forooghi, P.}, \bibinfo{author}{Stroh, A.},
  \bibinfo{author}{Bagheri, S.}, \bibinfo{year}{2022}.
\newblock \bibinfo{title}{Predicting drag on rough surfaces by transfer
  learning of empirical correlations}.
\newblock \bibinfo{journal}{Journal of Fluid Mechanics} \bibinfo{volume}{933},
  \bibinfo{pages}{A18}.
\bibitem[{Lee and You(2019)}]{lee2019data}
\bibinfo{author}{Lee, S.}, \bibinfo{author}{You, D.}, \bibinfo{year}{2019}.
\newblock \bibinfo{title}{Data-driven prediction of unsteady flow over a
  circular cylinder using deep learning}.
\newblock \bibinfo{journal}{Journal of Fluid Mechanics} \bibinfo{volume}{879},
  \bibinfo{pages}{217--254}.
\bibitem[{Li et~al.(2020)Li, Yang, Zhang, He, Deng and Shen}]{li2020using}
\bibinfo{author}{Li, B.}, \bibinfo{author}{Yang, Z.}, \bibinfo{author}{Zhang,
  X.}, \bibinfo{author}{He, G.}, \bibinfo{author}{Deng, B.Q.},
  \bibinfo{author}{Shen, L.}, \bibinfo{year}{2020}.
\newblock \bibinfo{title}{Using machine learning to detect the turbulent region
  in flow past a circular cylinder}.
\newblock \bibinfo{journal}{Journal of Fluid Mechanics} \bibinfo{volume}{905},
  \bibinfo{pages}{A10}.
\bibitem[{Ling et~al.(2016)Ling, Kurzawski and Templeton}]{ling2016reynolds}
\bibinfo{author}{Ling, J.}, \bibinfo{author}{Kurzawski, A.},
  \bibinfo{author}{Templeton, J.}, \bibinfo{year}{2016}.
\newblock \bibinfo{title}{Reynolds averaged turbulence modelling using deep
  neural networks with embedded invariance}.
\newblock \bibinfo{journal}{Journal of Fluid Mechanics} \bibinfo{volume}{807},
  \bibinfo{pages}{155--166}.
\bibitem[{List et~al.(2022)List, Chen and Thuerey}]{list2022learned}
\bibinfo{author}{List, B.}, \bibinfo{author}{Chen, L.W.},
  \bibinfo{author}{Thuerey, N.}, \bibinfo{year}{2022}.
\newblock \bibinfo{title}{Learned turbulence modelling with differentiable
  fluid solvers: physics-based loss functions and optimisation horizons}.
\newblock \bibinfo{journal}{Journal of Fluid Mechanics} \bibinfo{volume}{949},
  \bibinfo{pages}{A25}.
\bibitem[{McKinley et~al.(1993)McKinley, Armstrong and
  Brown}]{mckinley1993wake}
\bibinfo{author}{McKinley, G.H.}, \bibinfo{author}{Armstrong, R.C.},
  \bibinfo{author}{Brown, R.}, \bibinfo{year}{1993}.
\newblock \bibinfo{title}{The wake instability in viscoelastic flow past
  confined circular cylinders}.
\newblock \bibinfo{journal}{Philosophical Transactions of the Royal Society of
  London. Series A: Physical and Engineering Sciences} \bibinfo{volume}{344},
  \bibinfo{pages}{265--304}.
\bibitem[{Mittal and Iaccarino(2005)}]{mittal2005immersed}
\bibinfo{author}{Mittal, R.}, \bibinfo{author}{Iaccarino, G.},
  \bibinfo{year}{2005}.
\newblock \bibinfo{title}{Immersed boundary methods}.
\newblock \bibinfo{journal}{Annu. Rev. Fluid Mech.} \bibinfo{volume}{37},
  \bibinfo{pages}{239--261}.
\bibitem[{Morimoto et~al.(2022)Morimoto, Fukami, Zhang and
  Fukagata}]{morimoto2022generalization}
\bibinfo{author}{Morimoto, M.}, \bibinfo{author}{Fukami, K.},
  \bibinfo{author}{Zhang, K.}, \bibinfo{author}{Fukagata, K.},
  \bibinfo{year}{2022}.
\newblock \bibinfo{title}{Generalization techniques of neural networks for
  fluid flow estimation}.
\newblock \bibinfo{journal}{Neural Computing and Applications} ,
  \bibinfo{pages}{1--23}.
\bibitem[{Ozaki and Aoyagi(2022)}]{ozaki2022prediction}
\bibinfo{author}{Ozaki, H.}, \bibinfo{author}{Aoyagi, T.},
  \bibinfo{year}{2022}.
\newblock \bibinfo{title}{Prediction of steady flows passing fixed cylinders
  using deep learning}.
\newblock \bibinfo{journal}{Scientific Reports} \bibinfo{volume}{12},
  \bibinfo{pages}{447}.
\bibitem[{Papaioannou et~al.(2006)Papaioannou, Yue, Triantafyllou and
  Karniadakis}]{papaioannou2006three}
\bibinfo{author}{Papaioannou, G.V.}, \bibinfo{author}{Yue, D.K.},
  \bibinfo{author}{Triantafyllou, M.S.}, \bibinfo{author}{Karniadakis, G.E.},
  \bibinfo{year}{2006}.
\newblock \bibinfo{title}{Three-dimensionality effects in flow around two
  tandem cylinders}.
\newblock \bibinfo{journal}{Journal of Fluid Mechanics} \bibinfo{volume}{558},
  \bibinfo{pages}{387--413}.
\bibitem[{Pawar et~al.(2019)Pawar, Rahman, Vaddireddy, San, Rasheed and
  Vedula}]{pawar2019deep}
\bibinfo{author}{Pawar, S.}, \bibinfo{author}{Rahman, S.},
  \bibinfo{author}{Vaddireddy, H.}, \bibinfo{author}{San, O.},
  \bibinfo{author}{Rasheed, A.}, \bibinfo{author}{Vedula, P.},
  \bibinfo{year}{2019}.
\newblock \bibinfo{title}{A deep learning enabler for nonintrusive reduced
  order modeling of fluid flows}.
\newblock \bibinfo{journal}{Physics of Fluids} \bibinfo{volume}{31},
  \bibinfo{pages}{085101}.
\bibitem[{Pino et~al.(2023)Pino, Schena, Rabault and
  Mendez}]{pino2023comparative}
\bibinfo{author}{Pino, F.}, \bibinfo{author}{Schena, L.},
  \bibinfo{author}{Rabault, J.}, \bibinfo{author}{Mendez, M.A.},
  \bibinfo{year}{2023}.
\newblock \bibinfo{title}{Comparative analysis of machine learning methods for
  active flow control}.
\newblock \bibinfo{journal}{Journal of Fluid Mechanics} \bibinfo{volume}{958},
  \bibinfo{pages}{A39}.
\bibitem[{Raissi et~al.(2019)Raissi, Perdikaris and
  Karniadakis}]{raissi2019physics}
\bibinfo{author}{Raissi, M.}, \bibinfo{author}{Perdikaris, P.},
  \bibinfo{author}{Karniadakis, G.E.}, \bibinfo{year}{2019}.
\newblock \bibinfo{title}{Physics-informed neural networks: A deep learning
  framework for solving forward and inverse problems involving nonlinear
  partial differential equations}.
\newblock \bibinfo{journal}{Journal of Computational physics}
  \bibinfo{volume}{378}, \bibinfo{pages}{686--707}.
\bibitem[{Ronneberger et~al.(2015)Ronneberger, Fischer and
  Brox}]{ronneberger2015u}
\bibinfo{author}{Ronneberger, O.}, \bibinfo{author}{Fischer, P.},
  \bibinfo{author}{Brox, T.}, \bibinfo{year}{2015}.
\newblock \bibinfo{title}{U-net: Convolutional networks for biomedical image
  segmentation}, in: \bibinfo{booktitle}{Medical Image Computing and
  Computer-Assisted Intervention--MICCAI 2015: 18th International Conference,
  Munich, Germany, October 5-9, 2015, Proceedings, Part III 18},
  \bibinfo{organization}{Springer}. pp. \bibinfo{pages}{234--241}.
\bibitem[{Srinivasan et~al.(2019)Srinivasan, Guastoni, Azizpour, Schlatter and
  Vinuesa}]{srinivasan2019predictions}
\bibinfo{author}{Srinivasan, P.A.}, \bibinfo{author}{Guastoni, L.},
  \bibinfo{author}{Azizpour, H.}, \bibinfo{author}{Schlatter, P.},
  \bibinfo{author}{Vinuesa, R.}, \bibinfo{year}{2019}.
\newblock \bibinfo{title}{Predictions of turbulent shear flows using deep
  neural networks}.
\newblock \bibinfo{journal}{Physical Review Fluids} \bibinfo{volume}{4},
  \bibinfo{pages}{054603}.
\bibitem[{Stachenfeld et~al.(2021)Stachenfeld, Fielding, Kochkov, Cranmer,
  Pfaff, Godwin, Cui, Ho, Battaglia and
  Sanchez-Gonzalez}]{stachenfeld2021learned}
\bibinfo{author}{Stachenfeld, K.}, \bibinfo{author}{Fielding, D.B.},
  \bibinfo{author}{Kochkov, D.}, \bibinfo{author}{Cranmer, M.},
  \bibinfo{author}{Pfaff, T.}, \bibinfo{author}{Godwin, J.},
  \bibinfo{author}{Cui, C.}, \bibinfo{author}{Ho, S.},
  \bibinfo{author}{Battaglia, P.}, \bibinfo{author}{Sanchez-Gonzalez, A.},
  \bibinfo{year}{2021}.
\newblock \bibinfo{title}{Learned coarse models for efficient turbulence
  simulation}.
\newblock \bibinfo{journal}{arXiv preprint arXiv:2112.15275} .
\bibitem[{Sumner(2010)}]{sumner2010two}
\bibinfo{author}{Sumner, D.}, \bibinfo{year}{2010}.
\newblock \bibinfo{title}{Two circular cylinders in cross-flow: A review}.
\newblock \bibinfo{journal}{Journal of fluids and structures}
  \bibinfo{volume}{26}, \bibinfo{pages}{849--899}.
\bibitem[{Sun et~al.(2020)Sun, Gao, Pan and Wang}]{sun2020surrogate}
\bibinfo{author}{Sun, L.}, \bibinfo{author}{Gao, H.}, \bibinfo{author}{Pan,
  S.}, \bibinfo{author}{Wang, J.X.}, \bibinfo{year}{2020}.
\newblock \bibinfo{title}{Surrogate modeling for fluid flows based on
  physics-constrained deep learning without simulation data}.
\newblock \bibinfo{journal}{Computer Methods in Applied Mechanics and
  Engineering} \bibinfo{volume}{361}, \bibinfo{pages}{112732}.
\bibitem[{Thuerey et~al.(2021)Thuerey, Holl, Mueller, Schnell, Trost and
  Um}]{thuerey2021pbdl}
\bibinfo{author}{Thuerey, N.}, \bibinfo{author}{Holl, P.},
  \bibinfo{author}{Mueller, M.}, \bibinfo{author}{Schnell, P.},
  \bibinfo{author}{Trost, F.}, \bibinfo{author}{Um, K.}, \bibinfo{year}{2021}.
\newblock \bibinfo{title}{Physics-based Deep Learning}.
\newblock \bibinfo{publisher}{WWW}.
\newblock \URLprefix \url{https://physicsbaseddeeplearning.org}.
\bibitem[{Tong et~al.(2015)Tong, Cheng and Zhao}]{tong2015numerical}
\bibinfo{author}{Tong, F.}, \bibinfo{author}{Cheng, L.}, \bibinfo{author}{Zhao,
  M.}, \bibinfo{year}{2015}.
\newblock \bibinfo{title}{Numerical simulations of steady flow past two
  cylinders in staggered arrangements}.
\newblock \bibinfo{journal}{Journal of Fluid Mechanics} \bibinfo{volume}{765},
  \bibinfo{pages}{114--149}.
\bibitem[{Um et~al.(2020)Um, Brand, Fei, Holl and Thuerey}]{um2020solver}
\bibinfo{author}{Um, K.}, \bibinfo{author}{Brand, R.}, \bibinfo{author}{Fei,
  Y.R.}, \bibinfo{author}{Holl, P.}, \bibinfo{author}{Thuerey, N.},
  \bibinfo{year}{2020}.
\newblock \bibinfo{title}{Solver-in-the-loop: Learning from differentiable
  physics to interact with iterative pde-solvers}.
\newblock \bibinfo{journal}{Advances in Neural Information Processing Systems}
  \bibinfo{volume}{33}, \bibinfo{pages}{6111--6122}.
\bibitem[{Vlachas et~al.(2018)Vlachas, Byeon, Wan, Sapsis and
  Koumoutsakos}]{vlachas2018data}
\bibinfo{author}{Vlachas, P.R.}, \bibinfo{author}{Byeon, W.},
  \bibinfo{author}{Wan, Z.Y.}, \bibinfo{author}{Sapsis, T.P.},
  \bibinfo{author}{Koumoutsakos, P.}, \bibinfo{year}{2018}.
\newblock \bibinfo{title}{Data-driven forecasting of high-dimensional chaotic
  systems with long short-term memory networks}.
\newblock \bibinfo{journal}{Proceedings of the Royal Society A: Mathematical,
  Physical and Engineering Sciences} \bibinfo{volume}{474},
  \bibinfo{pages}{20170844}.
\bibitem[{Wan and Sapsis(2018)}]{wan2018machine}
\bibinfo{author}{Wan, Z.Y.}, \bibinfo{author}{Sapsis, T.P.},
  \bibinfo{year}{2018}.
\newblock \bibinfo{title}{Machine learning the kinematics of spherical
  particles in fluid flows}.
\newblock \bibinfo{journal}{Journal of Fluid Mechanics} \bibinfo{volume}{857},
  \bibinfo{pages}{R2}.
\bibitem[{Williamson(1996)}]{williamson1996vortex}
\bibinfo{author}{Williamson, C.H.}, \bibinfo{year}{1996}.
\newblock \bibinfo{title}{Vortex dynamics in the cylinder wake}.
\newblock \bibinfo{journal}{Annual review of fluid mechanics}
  \bibinfo{volume}{28}, \bibinfo{pages}{477--539}.
\bibitem[{Wu et~al.(2018)Wu, Xiao and Paterson}]{wu2018physics}
\bibinfo{author}{Wu, J.L.}, \bibinfo{author}{Xiao, H.},
  \bibinfo{author}{Paterson, E.}, \bibinfo{year}{2018}.
\newblock \bibinfo{title}{Physics-informed machine learning approach for
  augmenting turbulence models: A comprehensive framework}.
\newblock \bibinfo{journal}{Physical Review Fluids} \bibinfo{volume}{3},
  \bibinfo{pages}{074602}.
\bibitem[{Xiao et~al.(2015)Xiao, Fang, Pain and Hu}]{xiao2015non}
\bibinfo{author}{Xiao, D.}, \bibinfo{author}{Fang, F.}, \bibinfo{author}{Pain,
  C.}, \bibinfo{author}{Hu, G.}, \bibinfo{year}{2015}.
\newblock \bibinfo{title}{Non-intrusive reduced-order modelling of the
  navier--stokes equations based on rbf interpolation}.
\newblock \bibinfo{journal}{International Journal for Numerical Methods in
  Fluids} \bibinfo{volume}{79}, \bibinfo{pages}{580--595}.
\bibitem[{Xie et~al.(2017)Xie, Girshick, Doll{\'a}r, Tu and
  He}]{xie2017aggregated}
\bibinfo{author}{Xie, S.}, \bibinfo{author}{Girshick, R.},
  \bibinfo{author}{Doll{\'a}r, P.}, \bibinfo{author}{Tu, Z.},
  \bibinfo{author}{He, K.}, \bibinfo{year}{2017}.
\newblock \bibinfo{title}{Aggregated residual transformations for deep neural
  networks}, in: \bibinfo{booktitle}{Proceedings of the IEEE conference on
  computer vision and pattern recognition}, pp. \bibinfo{pages}{1492--1500}.
\bibitem[{Xie et~al.(2018)Xie, Franz, Chu and Thuerey}]{xie2018tempogan}
\bibinfo{author}{Xie, Y.}, \bibinfo{author}{Franz, E.}, \bibinfo{author}{Chu,
  M.}, \bibinfo{author}{Thuerey, N.}, \bibinfo{year}{2018}.
\newblock \bibinfo{title}{tempogan: A temporally coherent, volumetric gan for
  super-resolution fluid flow}.
\newblock \bibinfo{journal}{ACM Transactions on Graphics (TOG)}
  \bibinfo{volume}{37}, \bibinfo{pages}{1--15}.
\bibitem[{Zheng et~al.(2016)Zheng, Zhang and Lv}]{zheng2016numerical}
\bibinfo{author}{Zheng, S.}, \bibinfo{author}{Zhang, W.}, \bibinfo{author}{Lv,
  X.}, \bibinfo{year}{2016}.
\newblock \bibinfo{title}{Numerical simulation of cross-flow around three equal
  diameter cylinders in an equilateral-triangular configuration at low reynolds
  numbers}.
\newblock \bibinfo{journal}{Computers \& Fluids} \bibinfo{volume}{130},
  \bibinfo{pages}{94--108}.
\bibitem[{Zravkovich(1987)}]{zravkovich1987effects}
\bibinfo{author}{Zravkovich, M.}, \bibinfo{year}{1987}.
\newblock \bibinfo{title}{The effects of flow interference between two circular
  cylinders in various arrangements}.
\newblock \bibinfo{journal}{Journal of Fluids and Structures}
  \bibinfo{volume}{1}, \bibinfo{pages}{239--261}.

\end{thebibliography}

\appendix

\section{Comprehensive evaluations across all testing samples}
\label{compre}


\begin{figure}
\centering
\includegraphics[scale=0.5, angle=0]{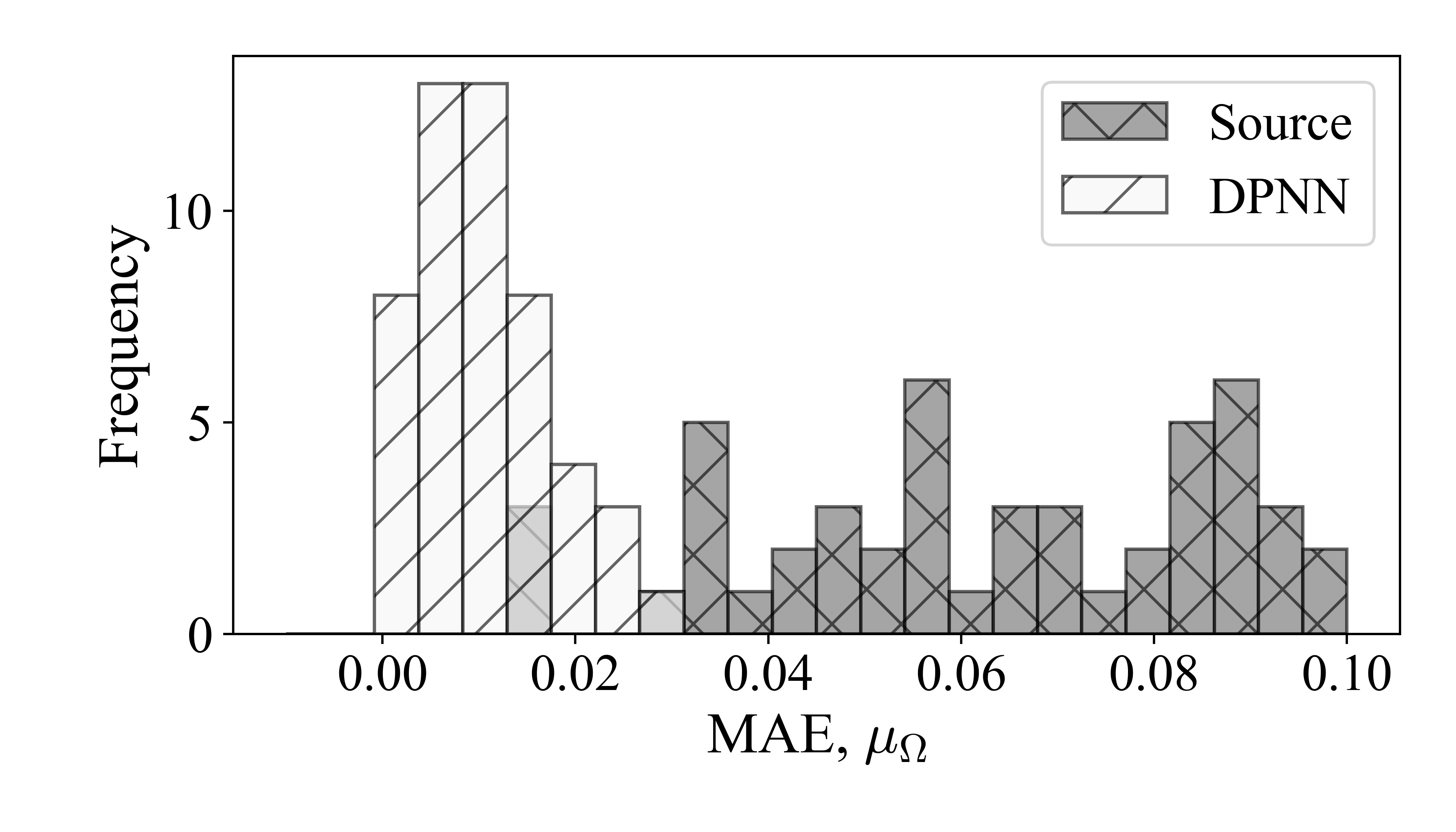}
\caption{Comparison of the mean absolute error based on the enstrophy for 300 frames for all the 50 testing samples via histogram}
\label{hist_ens}
\end{figure}

Sections \ref{qual} and \ref{quan} evaluates representative testing samples via quantities of interest such as $KE$, $\Omega$, etc. To present a comprehensive assessment of the present predictive framework for all testing samples, we compute the  mean absolute error $\mu_{\mathrm{\phi}}$ for the predictions based on the kinetic energy $KE$ and enstrophy $\Omega$ as defined below,

\begin{equation}
\label{eq_mu}
\mu_{\mathrm{\phi}} = \frac{1}{n_f} \mathlarger{\sum_{j=1}^{n_f}}   |\phi_{\mathrm{Ref}} - \phi_{\mathrm{baseline}} |  
\end{equation}

\begin{figure}
\centering
\includegraphics[scale=0.5, angle=0]{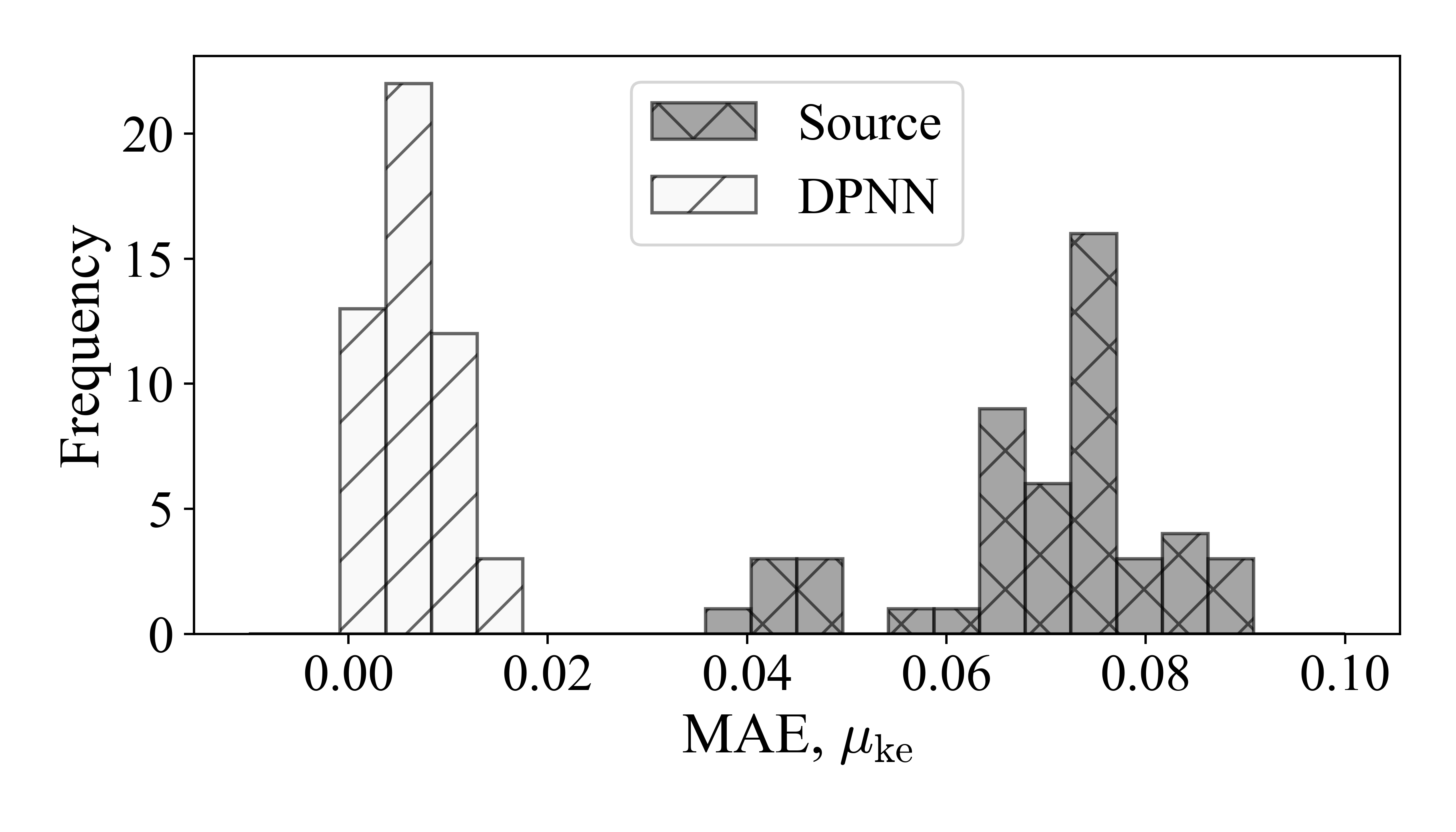}
\caption{Comparison of the mean absolute error based on the kinetic energy for 300 frames for all the testing samples via histogram}
\label{hist_KE}
\end{figure}

 Figure \ref{hist_ens} illustrates the mean absolute error evaluated based on enstrophy $\mu_{\Omega}$ for $n_f$=300 frames for all 50 individual testing samples. Upon observation, it becomes evident that the flowfields generated by our current DPNN methodology consistently result in lower $\mu_{\Omega}$ for all the testing samples. This observation also holds true for the mean absolute error based on kinetic energy  $\mu_{KE}$, as shown in Fig. \ref{hist_KE}. This suggests that, without exception, the predictions derived from the current DPNN methodology are consistently accurate and significantly outperform the \textit{Source}. 

\section{Verification of wake flow regimes}
\label{wake_regime}

In this section, we undertake a few additional verification test cases for flow past three equi-diameter cylinders placed in an equilateral position, as adopted in \citep{chen2020numerical} for multiple  spacing ratio $L/D$. This exercise allows for qualitative reproduction of the expected wake flow regimes for the given spacing ratio, as reported in the literature and further affirms the correctness of the numerical setup employed in the flow solver. In addition, this also sheds light towards the procedure adopted to categorize the flowfield into a distinct wake. Naturally, for the sake of fairness, we adopt the same conditions in terms of Reynolds number $Re$ and spacing ratio, $L/D$ \textit{i.e.,} $Re$=100 at $L/D$ of (1.2, 1.5, 2.25, 3.5, 5). For the aforementioned spacing ratios, the expected wake flow should exhibit single bluff-body wake, deflected gap flow, flip-flopping flow, anti-phase flow, and fully-developed in-phase flow, respectively.

\begin{figure}
\centering
\includegraphics[scale=1.75, angle=0]{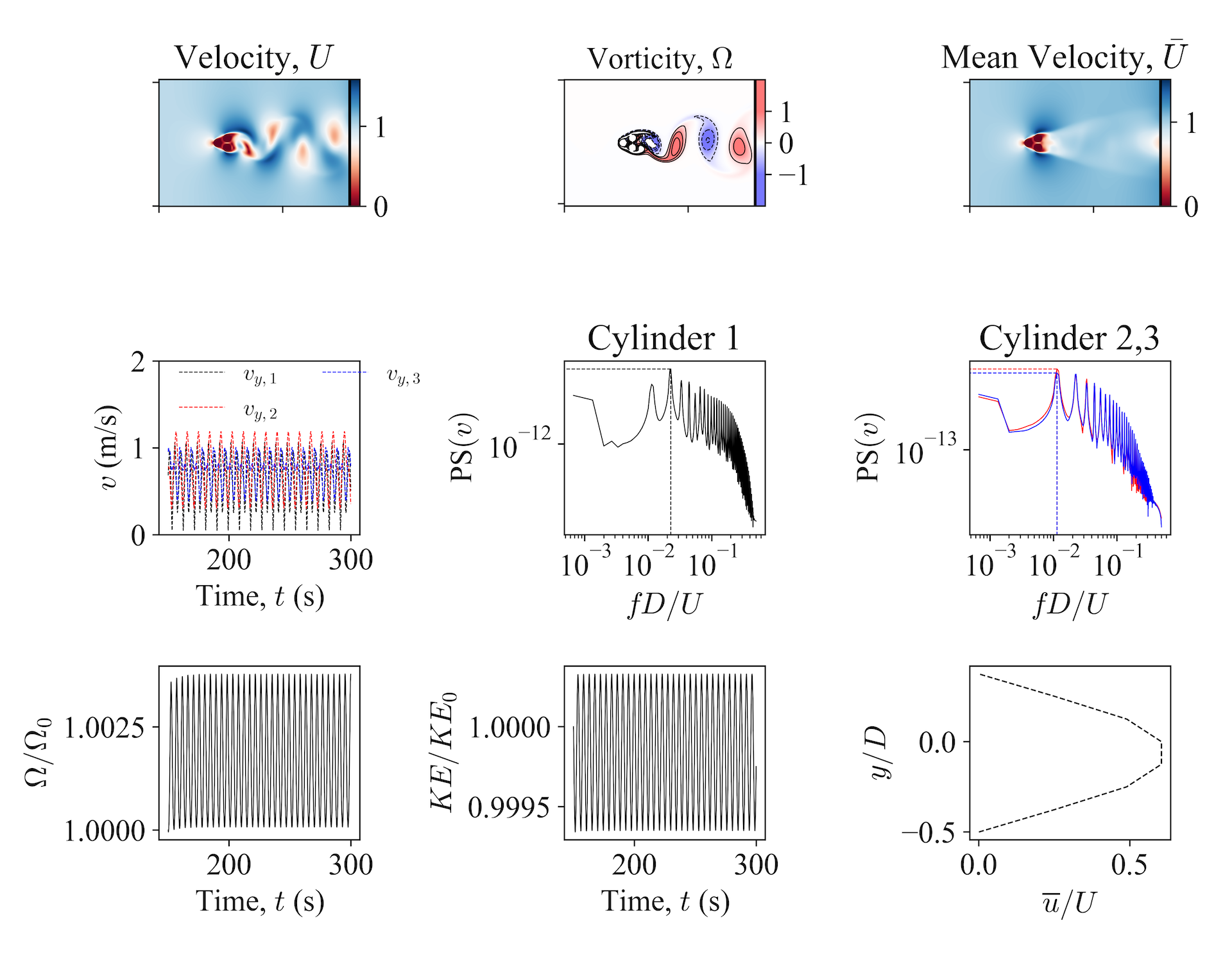}
\caption{Single bluff-body wake obtained from the \textit{Reference} solver at $L/D$ = 1.2 depicted using (a) instantaneous velocity $U$ flowfield, (b) instantaneous vorticity $\omega$ flowfield, (c) time-averaged velocity $\overline{U}$ flowfield (d) lateral velocities with time based on the probe placed just downstream of the cylinders, (e) PSD based on lateral velocity downstream of upstream cylinder 1, (f) PSD based on lateral velocity in the wake of the downstream cylinders 2,3, (g) scaled enstrophy $\Omega/\Omega_0$ with time, (h) scaled kinetic energy $KE/KE_0$ with time, (i) mean velocity at the gap $\overline{u}/U$ }

\label{LD_12}
\end{figure}

Figure \ref{LD_12} presents the characteristics of the flowfield obtained at the spacing ratio $L/D$=1.2, which also agrees with \citep{chen2020numerical}. It can be found from Fig. \ref{LD_12}(a)-(b) that the velocity and corresponding vorticity flowfields depict a single bluff-body wake, \textit{i.e.,} the cluster of bodies shed vortex as if a single bluff-body. The time-averaged mean velocity flowfield shown in Fig. \ref{LD_12}(c) is descriptive of the zone of influence of the downstream wake. The lateral $v$-velocity measured just downstream of the three cylinders exhibits periodicity in the flow with time. The power spectral density (or PSD) calculated based on the lateral $v$-velocities is shown in Figs. \ref{LD_12}(e)-(f) for upstream cylinder and downstream cylinders, respectively. It is found that the downstream cylinders exhibit identical dominant frequencies of vortex shedding $f$. In addition, Figs. \ref{LD_12}(g)-(h) potray the scaled enstrophy $\Omega / \Omega_0$ and scaled kinetic energy $KE/KE_0$, that indicates periodicity of the flowfield. Lastly, Fig. \ref{LD_12}(i) indicates the mean gap flow based on the time-averaged streamwise velocity normalized by freestream velocity. It is found that at such small $L/D$, the gap flow between the two downstream cylinders is weak and exhibits a parabolic profile. With increasing $L/D$, the shear layer emitted from the upstream cylinder is captured by the gap.

\begin{figure}
\centering
\includegraphics[scale=1.75, angle=0]{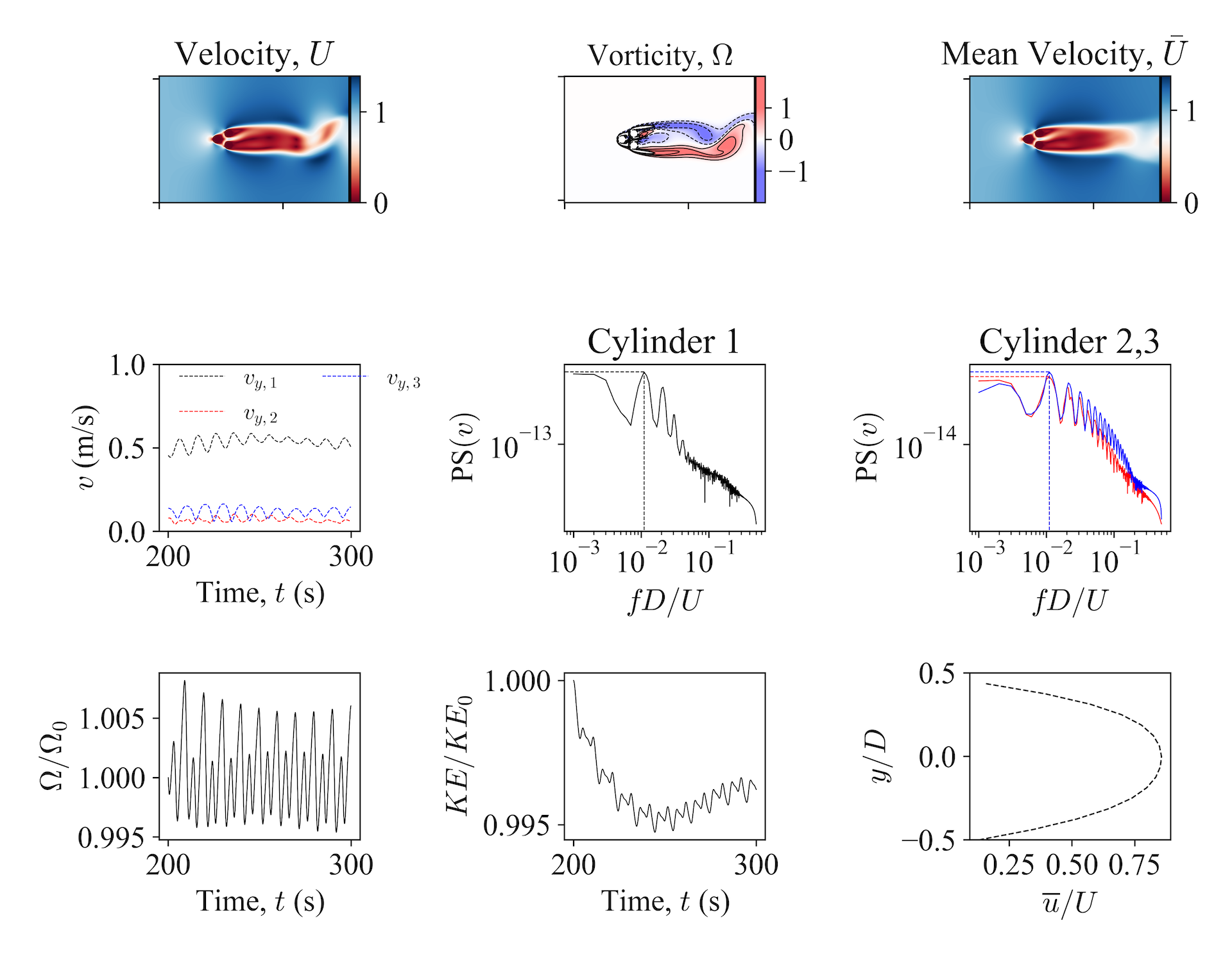}
\caption{deflected gap wake obtained from the \textit{Reference} solver at $L/D$ = 1.5 depicted using (a) instantaneous velocity $U$ flowfield, (b) instantaneous vorticity $\omega$ flowfield, (c) time-averaged velocity $\overline{U}$ flowfield (d) lateral velocities with time based on the probe placed just downstream of the three cylinders, (e) PSD based on lateral velocity downstream of the cylinder 1, (f) PSD based on lateral velocity in the wake of the downstream cylinders 2,3, (g) scaled enstrophy $\Omega/\Omega_0$ with time, (h) scaled kinetic energy $KE/KE_0$ with time, (i) mean velocity at the gap $\overline{u}/U$ }
\label{LD_15}
\end{figure}

Figure \ref{LD_15} depicts the attributes of the deflected gap wake obtained at a spacing ratio $L/D$ = 1.5, as also reported in the independent work by \citep{chen2020numerical} and \citep{zheng2016numerical}. Figures \ref{LD_15}(a)-(b) shows the instantaneous velocity $U$ and vorticity $\omega$ flowfields, which is typical of the deflected gap wake in terms of the long vortex formation length with the gap flow pointing towards the narrow wake region (\textit{i.e.,} upper downstream cylinder as reported in \citep{chen2020numerical} in Figure 3 of their work). The temporal stability of the gap flow is clearly shown in terms of the time-averaged velocity field in Fig. \ref{LD_15}(c). The lateral velocities measured just downstream of the three cylinders are plotted in Fig. \ref{LD_15}(d), which fluctuate about a mean position. The PSD evaluated based on these lateral velocities in the immediate wake of the three cylinders are shown in Figs. \ref{LD_15}(e)-(f), where the dominant frequency of the two downstream cylinders are found to be identical. Figures \ref{LD_15}(g)-(h) present the temporal variation of the scaled enstrophy as well as kinetic energy, with minor dissipation. Lastly, Fig. \ref{LD_15}(i) portrays the parabolic mean velocity at the gap. The spacing ratio $L/D$=1.5 is not large enough for the detached shear layer from the upstream cylinder to be fully trapped within the gap in this case, and they follow the freestream side of the two downstream cylinders.

\begin{figure}
\centering
\includegraphics[scale=1.75, angle=0]{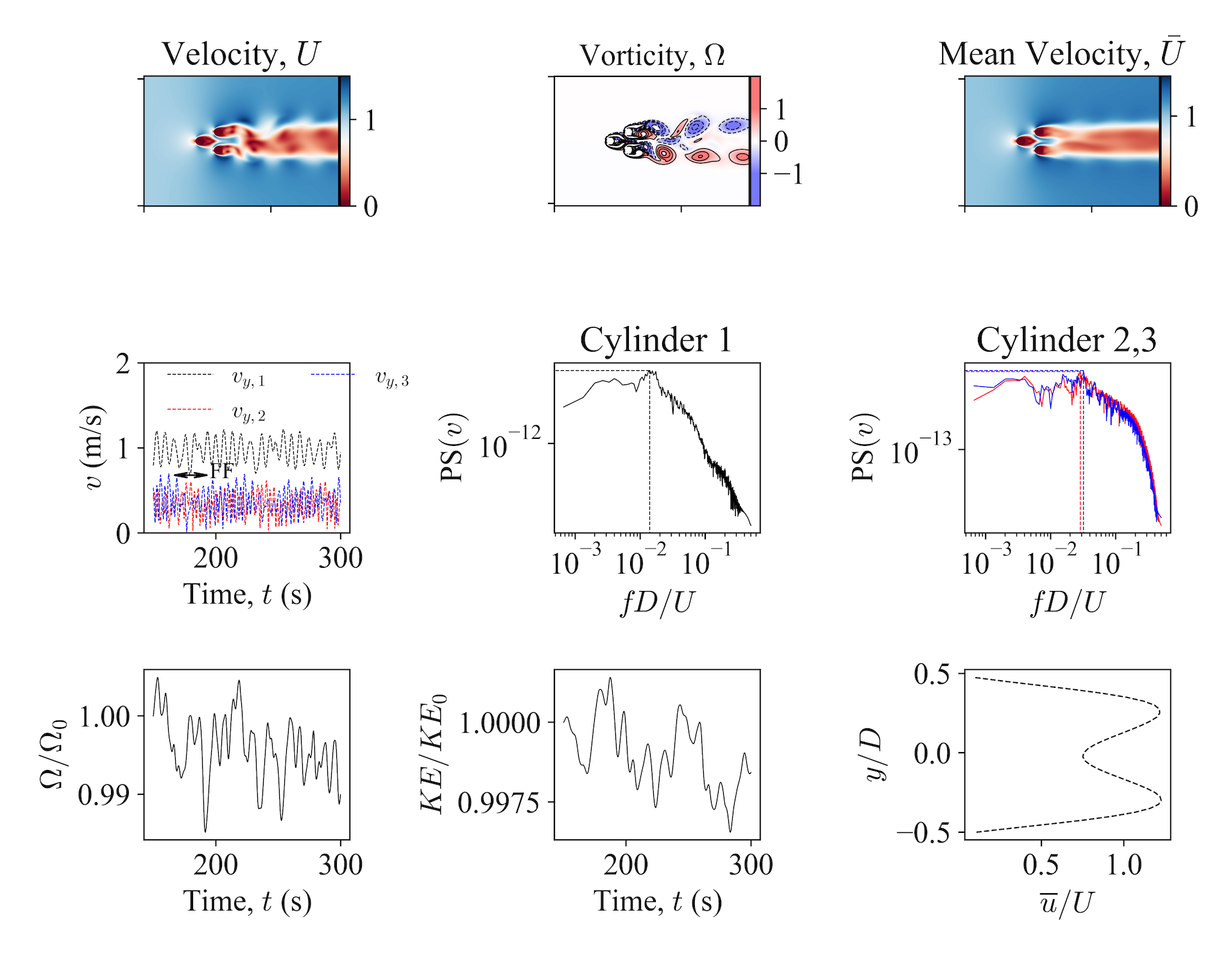}
\caption{Flip-flopping chaotic wake obtained from the \textit{Reference} solver at $L/D$ = 2.25 depicted using (a) instantaneous velocity $U$ flowfield, (b) instantaneous vorticity $\omega$ flowfield, (c) time-averaged velocity $\overline{U}$ flowfield (d) lateral velocities with time-based on probe placed just downstream of the cylinders, (e) PSD based on lateral velocity downstream of upstream cylinder 1, (f) PSD based on lateral velocity in the wake of the downstream cylinders 2,3, (g) scaled enstrophy $\Omega/\Omega_0$ with time, (h) scaled kinetic energy $KE/KE_0$ with time, (i) mean velocity at the gap $\overline{u}/U$ }
\label{LD_225}
\end{figure}

Figure \ref{LD_225} presents the features exhibited by a chaotic wake for a spacing ratio $L/D$ = 2.25, as also reported by \citep{chen2020numerical}. This category of wake exhibits irregular switching of the gap flow direction, which is indicated in the temporal variation of the lateral velocities measured just downstream of the three cylinders shown in Fig. \ref{LD_225} (d). This transition in the gap is chaotic in nature and also results in an altered frequency of vortex shedding by the two downstream cylinders (as shown in Fig. \ref{LD_225} (f)). The irregularity in the flow is also present in the temporal variation of the scaled enstrophy and kinetic energy, as seen in Figs. \ref{LD_225}(g-h). The shear layer shed from the upstream cylinder now enters the gap region along the inner side of the two downstream cylinders and cases two peaks in the time-averaged gap flow (see Fig. \ref{LD_225}(i)).

\begin{figure}
\centering
\includegraphics[scale=1.75, angle=0]{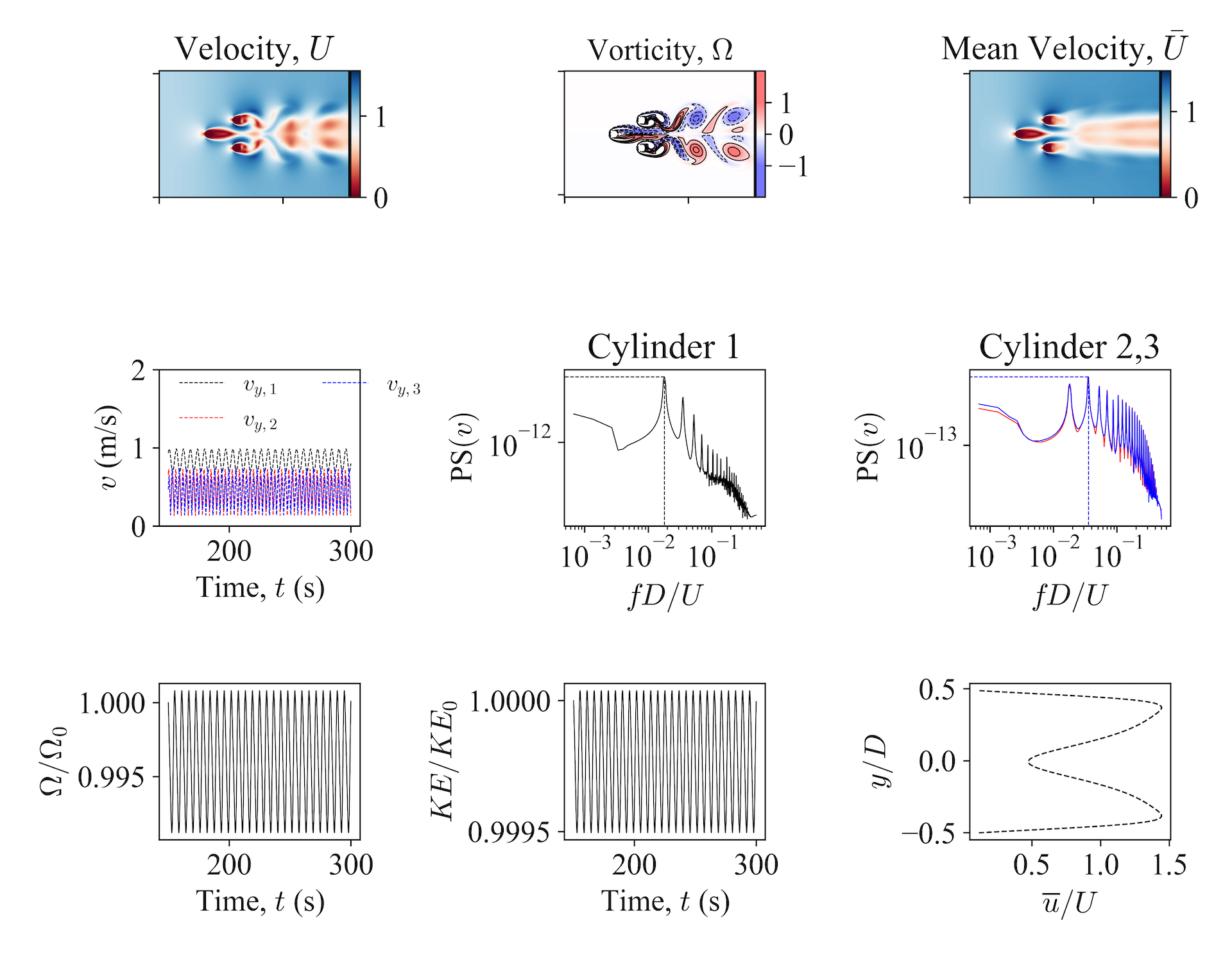}
\caption{Anti-phase wake obtained from the \textit{Reference} solver at $L/D$ = 3.5 depicted using (a) instantaneous velocity $U$ flowfield, (b) instantaneous vorticity $\omega$ flowfield, (c) time-averaged velocity $\overline{U}$ flowfield (d) lateral velocities with time based on the probe placed just downstream of the cylinders, (e) PSD based on lateral velocity downstream of upstream cylinder 1, (f) PSD based on lateral velocity in the wake of the downstream cylinders 2,3, (g) scaled enstrophy $\Omega/\Omega_0$ with time, (h) scaled kinetic energy $KE/KE_0$ with time, (i) mean velocity at the gap $\overline{u}/U$ }
\label{LD_35}
\end{figure}

Figure \ref{LD_35} indicates the characteristics of the anti-phase wake obtained at a spacing ratio $L/D$ = 3.5, as also reported by \citep{chen2020numerical} and \citep{zheng2016numerical}. The vorticity flowfield shown in Fig. \ref{LD_35} (b) indicates the counter-rotating vortices while maintaining symmetrical distribution along the centerline $x$-axis, resulting in an identical dominant frequency of vortex shedding by the two downstream cylinders. Besides, this category of wake exhibits a uniform variation of the lateral velocity as well as the scaled enstrophy and kinetic energy, as also seen in the distribution of scaled enstrophy and kinetic energy in Figs. \ref{LD_35}(g-h). Further, Fig. \ref{LD_35}(i) indicates greater penetration of the shear layer emitted from the upstream cylinder, as evident from the prominent peaks in the gap flow profile.

\begin{figure}
\centering
\includegraphics[scale=1.75, angle=0]{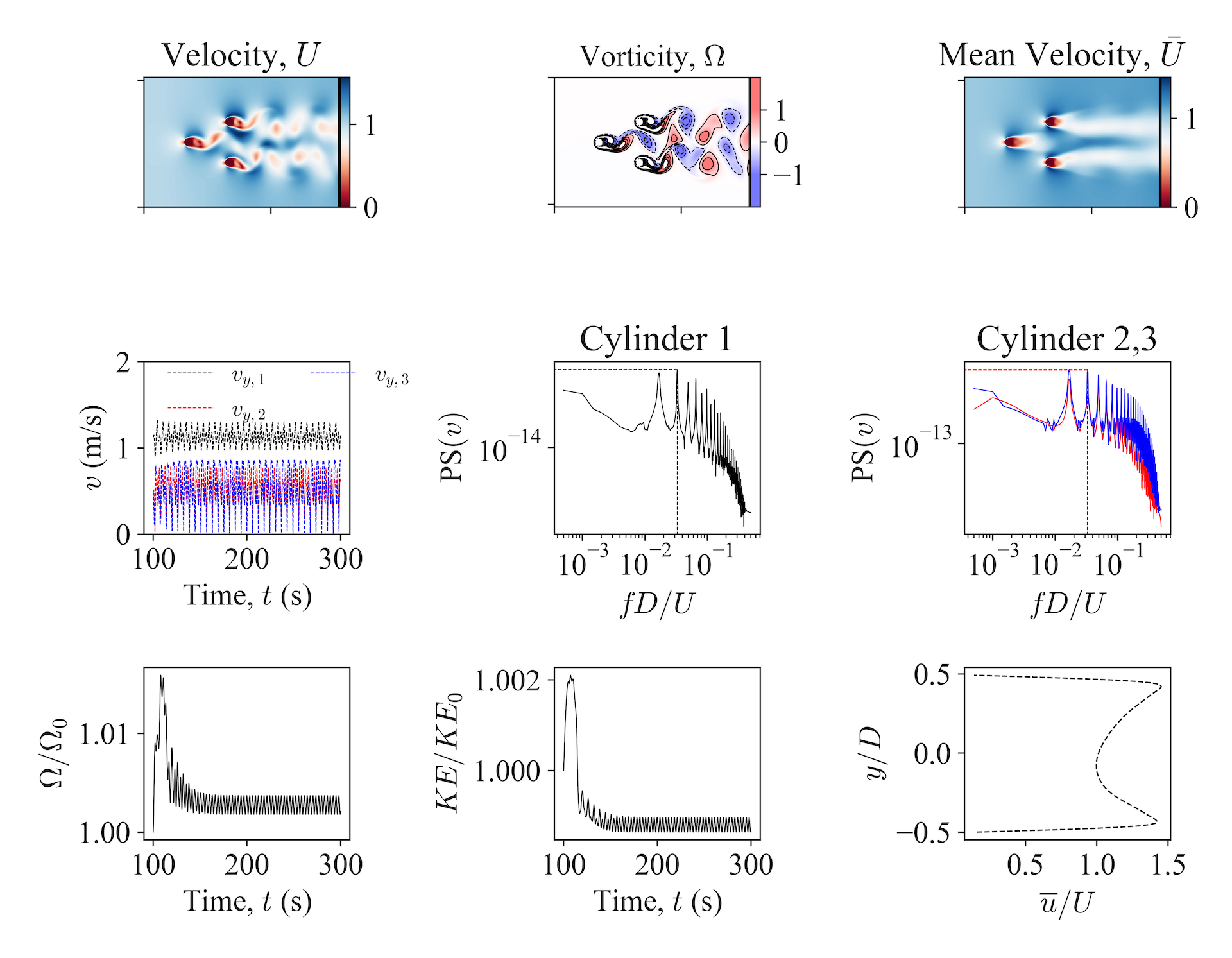}
\caption{Fully developed in-phase wake obtained from the \textit{Reference} solver at $L/D$ = 5 depicted using (a) instantaneous velocity $U$ flowfield, (b) instantaneous vorticity $\omega$ flowfield, (c) time-averaged velocity $\overline{U}$ flowfield (d) lateral velocities with time based on probe placed just downstream of the cylinders, (e) PSD based on lateral velocity downstream of upstream cylinder 1, (f) PSD based on lateral velocity in the wake of the downstream cylinders 2,3, (g) scaled enstrophy $\Omega/\Omega_0$ with time, (h) scaled kinetic energy $KE/KE_0$ with time, (i) mean velocity at the gap $\overline{u}/U$ }
\label{LD_5}
\end{figure}

Figure \ref{LD_5} presents the features of a fully developed in-phase wake, obtained at a spacing ratio $L/D$ = 5, as also confirmed by \citep{chen2020numerical} and \citep{zheng2016numerical}. For this category of wake, the vortices are also shed by the upstream cylinder and are swallowed by the gap between the two downstream cylinders and later merge with the vortices shed by the two downstream cylinders. Besides, such a category of wake results in a uniform variation of the lateral velocity as well as scaled enstrophy, and kinetic energy, resulting in an identical dominant frequency of vortex shedding by the two downstream cylinders.  

\begin{figure}
\centering
\includegraphics[scale=3, angle=0]{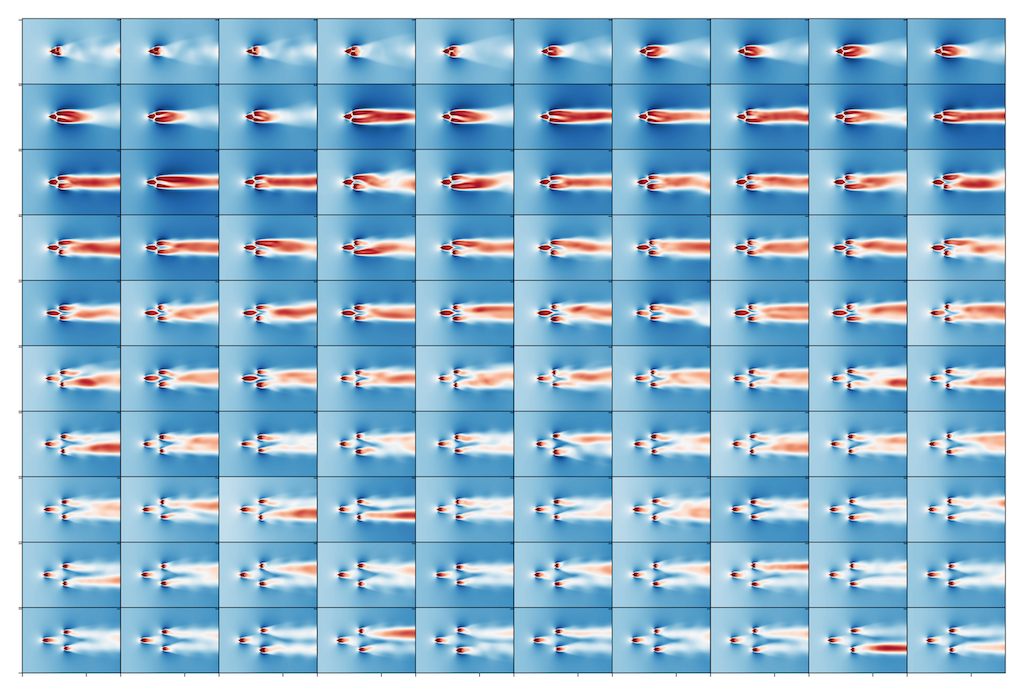}
\caption{Time averaged velocities fields obtained from the \textit{Reference} solver for all datasets arranged in the order of ascending spacing ratio}
\label{mean_flow}
\end{figure}

Figure \ref{mean_flow} presents the time-averaged mean velocity flowfields of all the 100 datasets, arranged in increasing $L/D$ values (left to right, top to bottom), for the time range 150s $\leq$ $t$ $\leq$ 300s. It can be found that the first few datasets exhibit a single bluff-body wake. At slightly higher spacing ratios, the wake transitions to \textit{wake} interference as the downstream cylinders are fully submerged in the wake of the upstream cylinder. At moderate spacing ratios, the wake pertains to \textit{proximity} + \textit{wake} interference due to partial immersion of the downstream cylinders within the wake of the upstream cylinders. At the high spacing ratio values, the wake of the upstream cylinders has minimal interference with the downstream cylinders and hence corresponds to the no-interference wake category. 

\begin{figure}
\centering
\includegraphics[scale=0.525, angle=0]{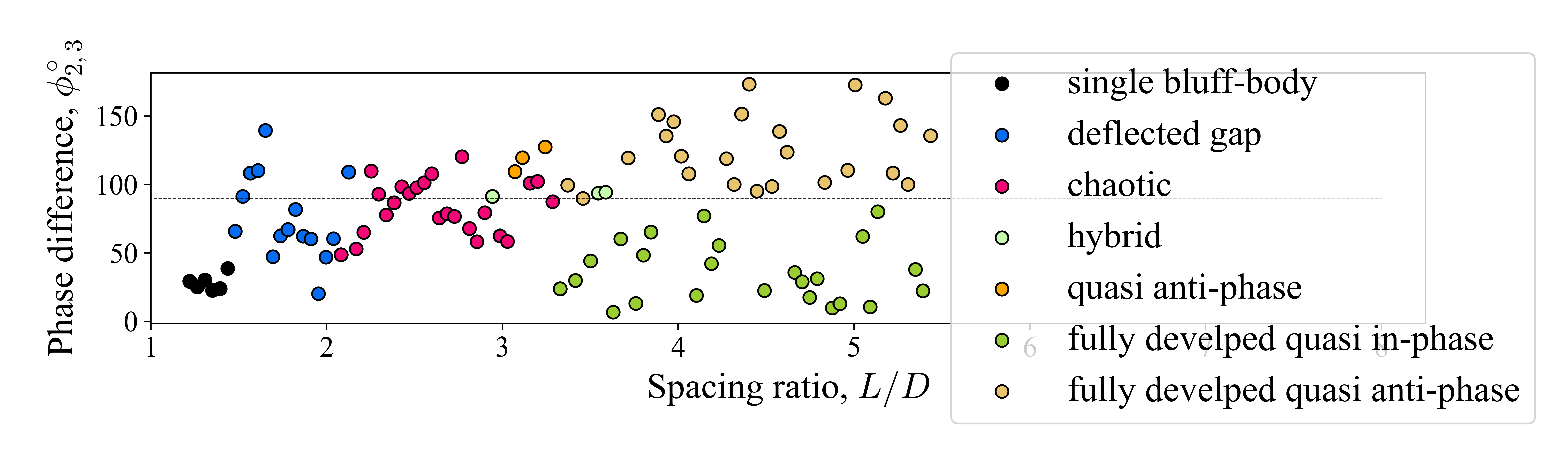}
\caption{Comparison of phase difference based on the $v$-velocity probed in the wake of the two downstream cylinders with respect to spacing ratio $L/D$ obtained from the \textit{Reference} solver}
\label{category_phase}
\end{figure}

\begin{figure}
\centering
\includegraphics[scale=2.4, angle=0]{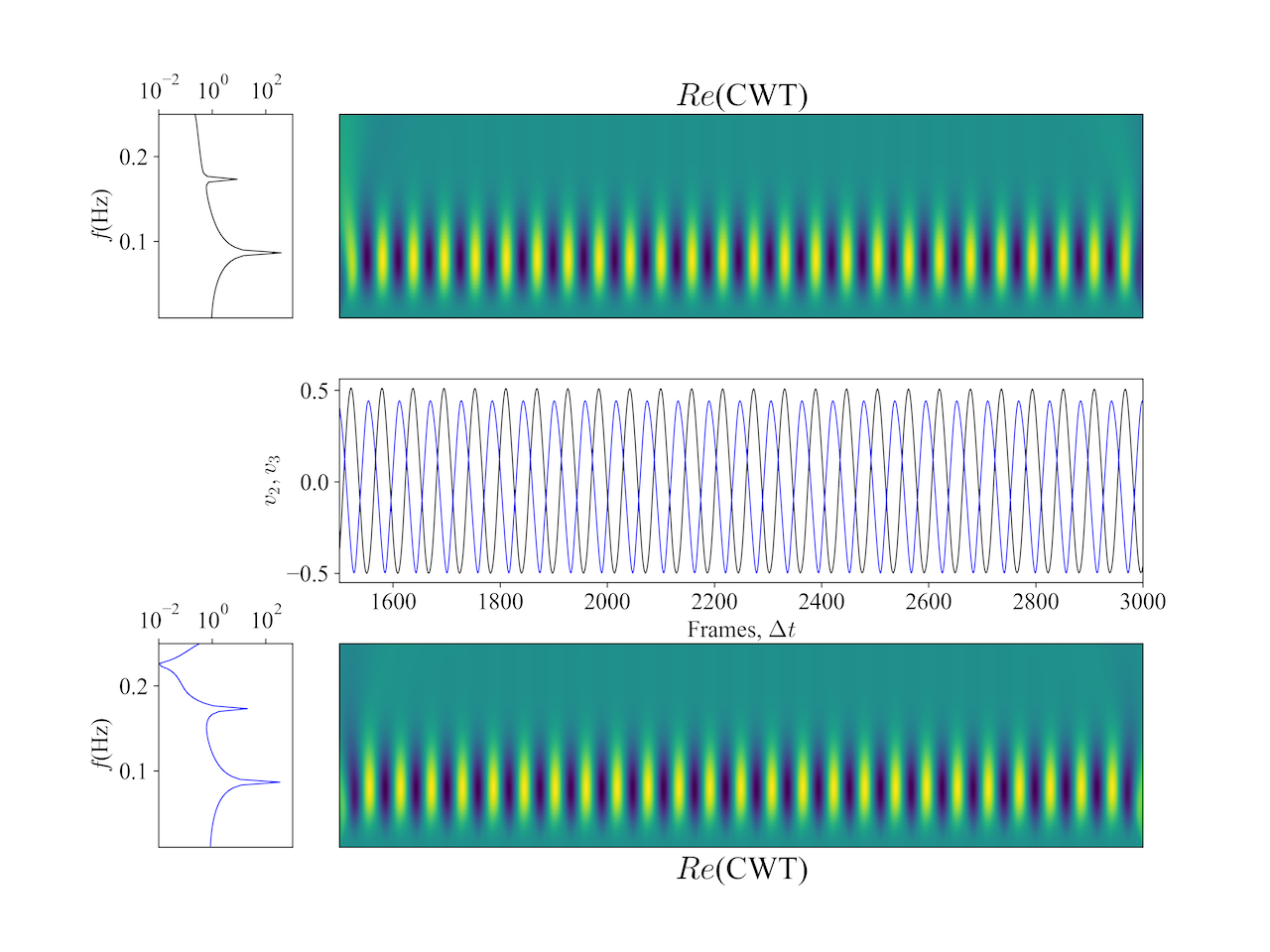}
\caption{Continuous wavelet transformation and the associated frequency distribution obtained using the \textit{Reference} solver for the $v$-velocity distributions of the 83rd dataset. The upper row corresponds to the $v$-velocity obtained from the wake of the upper downstream cylinder (black), whereas the lower row corresponds to the lower downstream cylinder (blue). }
\label{phase_diff2}
\end{figure}

\begin{figure}
\centering
\includegraphics[scale=2.4, angle=0]{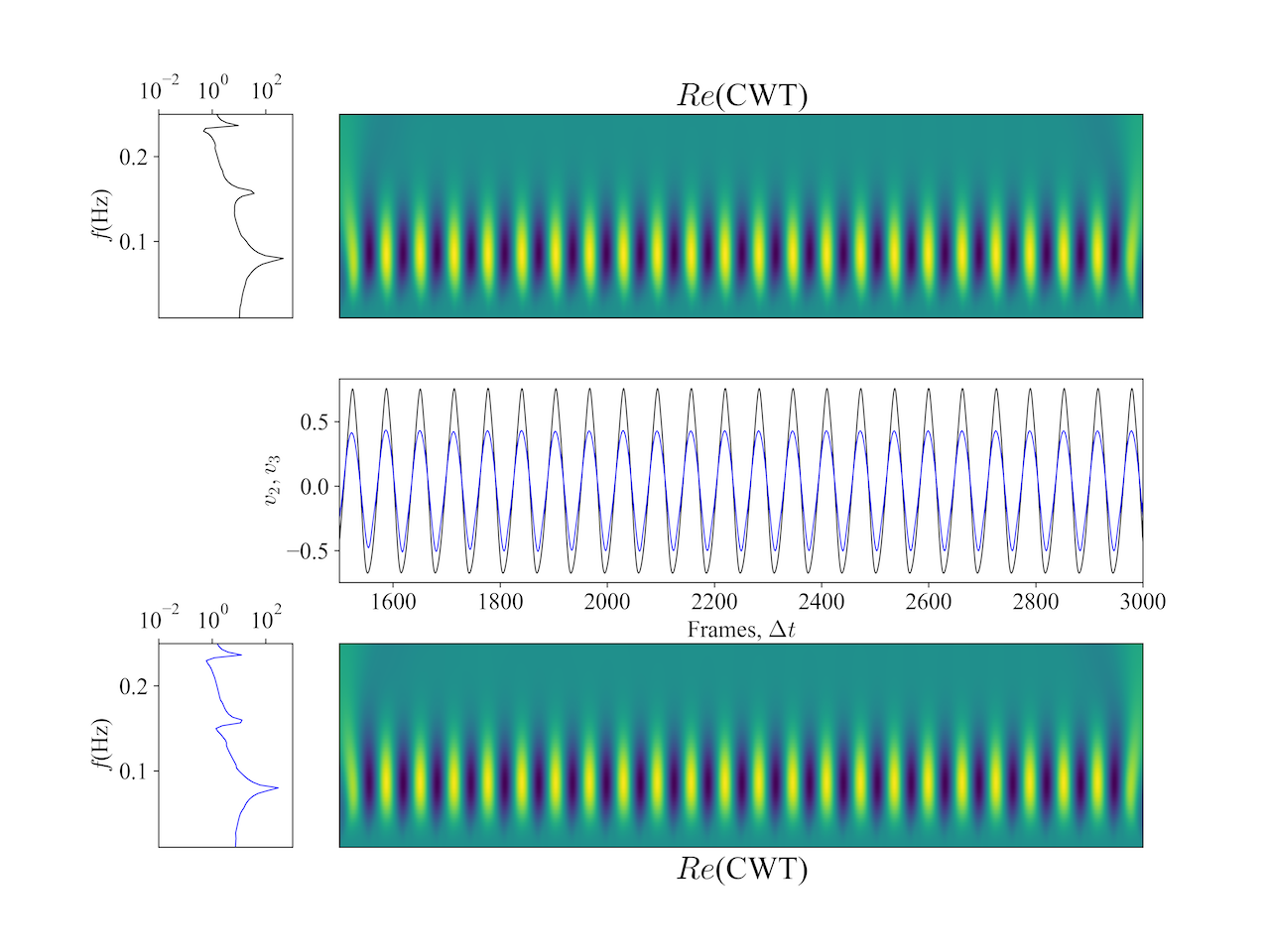}
\caption{Continuous wavelet transformation and the associated frequency distribution obtained using the \textit{Reference} solver for the $v$-velocity distributions of the 95th dataset. The upper row corresponds to the $v$-velocity obtained from the wake of the upper downstream cylinder (black), whereas the lower row corresponds to the lower downstream cylinder (blue). }
\label{phase_diff3}
\end{figure}

\begin{figure}
\centering
\includegraphics[scale=0.4, angle=0]{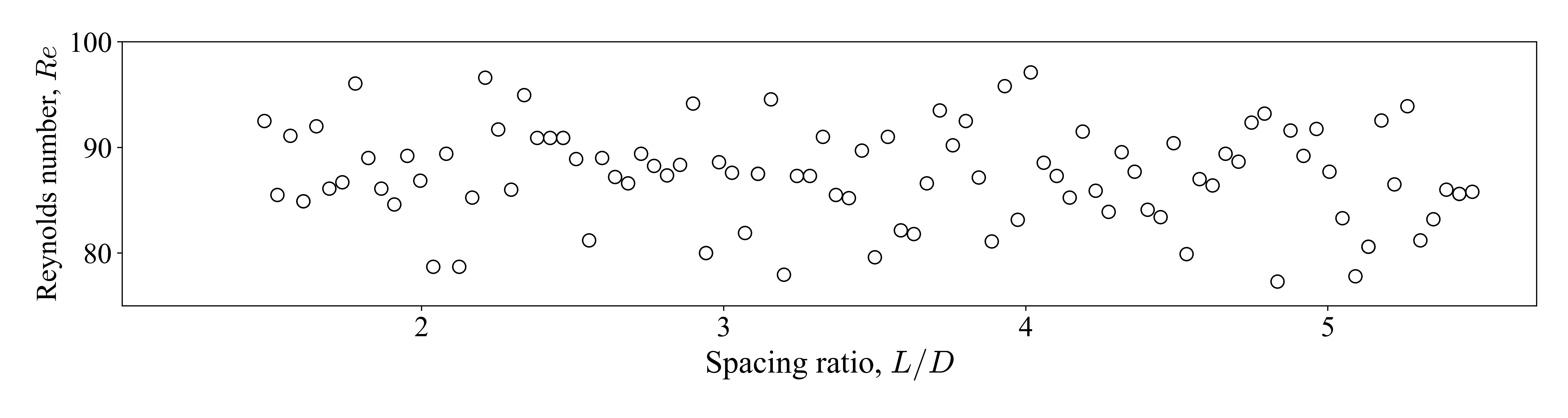}
\caption{Comparison of the effective Reynolds number $Re$ for all 100 individual datasets based on the characteristic length scale of the upstream cylinder (cylinder height $H_{\mathrm{c1}}$).}
\label{re_all}
\end{figure}

Besides probing the velocity fields, we analyze the temporal distribution of the cross-stream velocities probed in the immediate wake of the two downstream cylinders. This enables us to further categorize the wake into in-phase (or quasi in-phase) and anti-phase (quasi anti-phase). Specifically, we use cross-correlation to evaluate the phase difference between the two downstream cylinders using their respective cross-stream $v$-velocities. Figure \ref{category_phase} presents the phase difference $\phi_{2,3}$ of the 100 training datasets in terms of their spacing ratio $L/D$. It can be observed that there is a sudden jump in the $\phi_{2,3}$ between the single bluff-body wake and deflected gap wake. This result bodes well with the observations reported in \cite{chen2020numerical}. In addition, there exists a clear difference in the $\phi_{2,3}$ values for the quasi anti-phase wake and fully developed quasi anti-phase wake flow, which is in line with the expectations. It is also found that unlike the reported observations in \cite{chen2020numerical}, the present study shows a transition from quasi anti-phase wake to fully developed quasi in-phase wake due to early shedding of the vortices by the upstream cylinder. Consequently, the arbitrary nature of the bodies promotes such a transition and precludes quasi in-phase wake from appearing at all. Figs. \ref{phase_diff2} and \ref{phase_diff3} show the continuous wavelet transformation on the cross-stream velocities for two representative samples that depict quasi anti-phase and fully developed quasi in-phase wake flow, respectively. Finally, Fig. \ref{re_all} presents the effective Reynolds number $Re$ for all the individual datasets calculated based on the height of the upstream cylinder $H_{\mathrm{c1}}$. It is found that the $Re$ values between 75 $\leq$ $Re$ $\leq$ 100, among the 100 experiments.

\section{Fine tuning neural network architecture and parameters}
\label{NN_arch}

\subsection{Adaptive Learning Rate}
\label{LRate}

\begin{figure}
\centering
\includegraphics[scale=0.6, angle=0]{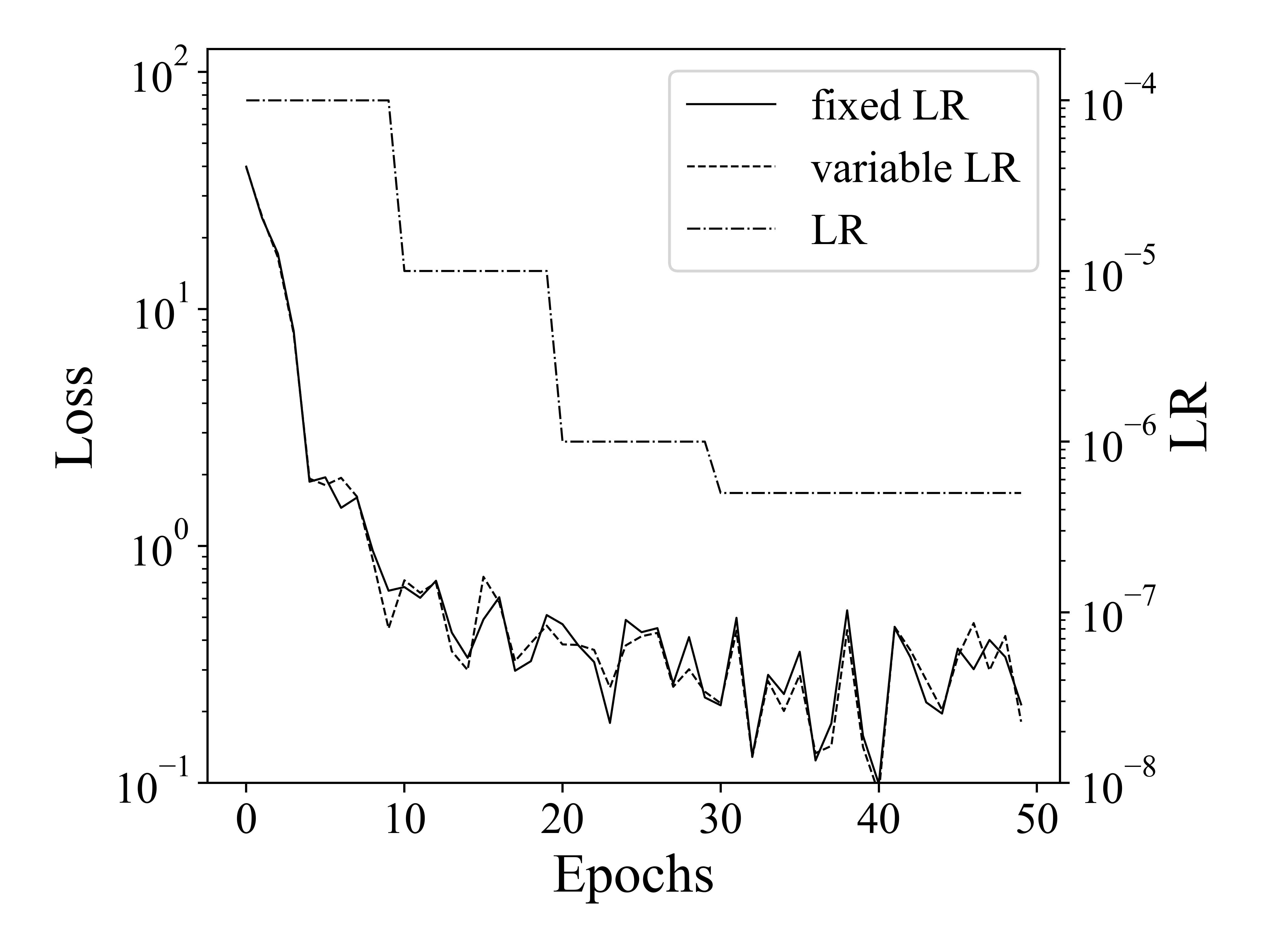}
\caption{Influence of learning rate schedule on the convergence. The dash-dotted line indicates the sequential decay of the learning rate value with epochs.}
\label{LR_curve}
\end{figure}

While choosing a constant learning rate, $\eta$ is straightforward, it may heavily influence the decay of the loss function. For instance, an extremely low $\eta$ may lead to low convergence, whereas for relatively high $\eta$ may result in divergence or oscillatory convergence. This issue can be remedied using a learning rate schedule$, i.e.,$ start with a relatively high $\eta$ and progressively reduce it with training epochs. The influence of a constant and a variable $\eta$ is shown in Fig. \ref{LR_curve}. It is found that while the loss decays with training epochs, the influence of learning rate $\eta$ on the loss is minimal. We incorporate a variable learning rate for the remainder of the study.

\subsection{Residual Network: ResNet}

\begin{figure}
\centering
\includegraphics[scale=0.475, angle=0]{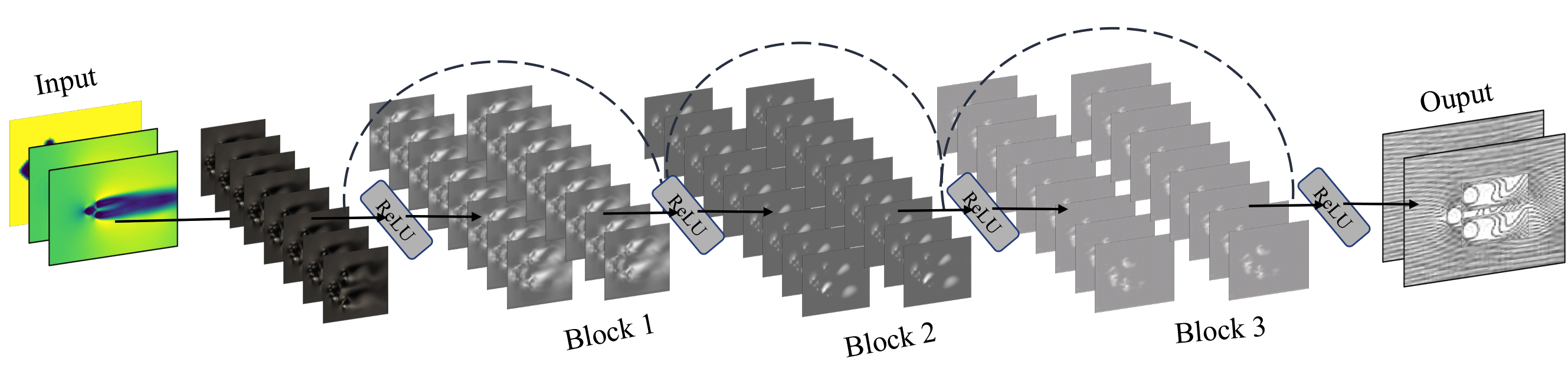}
\caption{A residual network architecture with skip connections}
\label{Resnet}
\end{figure}

Figure \ref{Resnet} describes a typical residual network (or ResNet) architecture \citep{he2016deep} wherein multiple ResNet blocks have been connected in series with skip-connections in between. Each ResNet block comprises two convolutional layers having 32 filters, each with a kernel of size of [5 $\times$5]. The input is first fed to a convolutional layer whose output is then fed to a ResNet block. There are a total of nine ResNet blocks, the output of which is activated by the rectified linear activation function (or ReLU) activation function. Such a deep network results in a total of 516,674 trainable parameters. 

\subsection{Convolutional Neural Network: CNN}

A fully convolutional neural network (or CNN) architecture has been adopted \citep{krizhevsky2017imagenet}. The CNN architecture comprises input, hidden, and output layers. For the present case, we have adopted 11 hidden layers of kernel size  = [5 $\times$ 5]  and filters varying as (4, 8, 16, 32, 64, 128, 64, 32, 16, 8, 4). To preserve the size of the inputs, we employ padding while also using the ReLU activation function. No downsampling or max-pooling was applied throughout the CNN layers. This setting resulted in a total of 546,478 trainable parameters. 

\subsection{U-Net}

A U-Net-based architecture has been adopted from \citep{ronneberger2015u} that comprises an encoder path followed by a decoder path with skip connections between each level of encoding and decoding. Each encoding level consists of two convolutional layers followed by a [2 $\times$ 2] Maxpooling operator for downsampling. The number of filters for each level is chosen as (4, 8, 16, 32), with the final convolutional block comprising 64 filters, each having a fixed kernel size of [5 $\times$ 5]. The output from the [2 $\times$ 2] Upsamping operator is concatenated with the corresponding output from the encoded level. The resulting feature maps are then fed to a convolution layer. The number of filters in each convolution layer in the decoded path is chosen as (32,16,8,4). This architecture resulted in a total of 520,342 trainable parameters.

\subsection{Diluted Residual Networks: DilResNet}

A dilated residual network (or Dil-ResNet) architecture has been adopted from an independent work by \citep{stachenfeld2021learned}, who presented this architecture for 2D as well as 3D turbulent flows with an aim to learn high-resolution turbulent fluid dynamics using low spatial and temporal resolutions. This architecture employs an encoder and decoder ($, i.e.,$ using MaxPooling and Upsampling, respectively), each containing a single convolutional layer without any activation. In addition to the encoder and decoder layer, a processor is connected in between to enable the encode-process-decode paradigm. The processor consists of 4 dilated CNN blocks connected in series and residual connections between them.  Each CNN blocks comprise 7 dilated CNN layers, each having $n_f$=36 number of filters, along with dilation rate $D_\mathrm{R} \in $ {1,2,4,8,4,2,1} and ReLU activation. These values have been kept the same as the original work by \citep{stachenfeld2021learned} to allow for a fair comparison. The DilResNet architecture results in a total of 548,055 trainable parameters. 

\subsection{ResNeXt}

A ResNeXt architecture has been adopted from \cite{xie2017aggregated} that in the original work showed improved performance compared to base ResNet architecture for image classification tasks. The ResNeXt architecture proposed by \citep{xie2017aggregated} comprises a branched design that mimics the \textit{split-transform-merge} strategy on the base ResNet design. The \textit{spitting} is performed once the input is convolved using $n_f$ = 32 filters into a number of branches, each having the same topology. The number of branches (or Cardinality) = 5 is chosen for the present study. Each branch executes \textit{transformation} on the resulting input by 2 layers of convolutions followed by Batch Normalisation and ReLU activation. Finally, the outputs from each branch are concatenated to perform \textit{merge} operation. This is followed by another convolution layer which is merged with a skip connection from the input. This operation is repeated 9 times, resulting in a total of 524,322 trainable parameters.

\subsection{Dense convolution network: DenseNet}

A dense convolution network (or DenseNet) architecture has been adopted from \citep{huang2017densely} that was originally implemented for object recognition tasks. In this architecture, each layer is connected to every other layer in a feed-forward manner, allowing for better feature propagation, supporting feature reuse, and alleviating the vanishing gradient problem. In this architecture, the inputs are initially passed through a convolution layer with $n_f$ = 64 filters, followed by a collection of dense layers (or dense blocks) and transition layers, connected in series. In each dense layer, the inputs pass through a bottleneck layer wherein the first convolution layer expands the number of filters to 4 $\times 32$ and then a second convolution layer containing $n_f$ = 32 filters. To further reduce the number of feature maps, the output from the bottleneck layer is passed through the transition layer. There are a total of three Dense layers and 2 transition layers, resulting in a total of 566,578 trainable parameters.

\end{document}